\documentclass[aps,twocolumn,superscriptaddress,floatfix,prx,longbibliography]{revtex4-1}
\usepackage[utf8]{inputenc}
\usepackage{lipsum}
\usepackage{graphicx}
\usepackage{amsmath,amssymb}
\usepackage{braket}
\usepackage{mathtools}
\usepackage{hyperref}
\usepackage{multirow}
\usepackage{color}
\usepackage[normalem]{ulem}
\usepackage{amsfonts}
\usepackage{bbm}
\usepackage[ruled,vlined]{algorithm2e}
\usepackage{braket}
\newcommand{\<}{\langle}

\renewcommand{\>}{\rangle}

\renewcommand{\vec}[1]{\boldsymbol{#1}} 

\renewcommand{\d}{\partial}
\newcommand{\rrangle}{\rangle\rangle}
\newcommand{\llangle}{\langle\langle}
\newcommand{\Z}{\mathbb{Z}}

\makeatletter

\usepackage{xcolor}

\usepackage[normalem]{ulem} 

\AtBeginDocument{\let\LS@rot\@undefined}
\makeatother

\begin{document}
\newcommand{\linkstate}[3]{(\ensuremath{#1#2;#3})}
\newcommand{\state}[1]{\ensuremath{| {#1} \rangle}}
\newcommand{\mindex}{\ensuremath{\mathbf{m}}}
\newcommand{\mindexq}{\ensuremath{{\{\vec{m_q}\}}}}
\newcommand{\avg}[1]{\ensuremath{\left[ #1 \right] }}

\title{Entanglement and charge-sharpening transitions in U(1) symmetric monitored quantum circuits}

\author{Utkarsh Agrawal}
\affiliation{Department of Physics, University of Massachusetts, Amherst, MA 01003, USA}

\author{Aidan Zabalo}
\affiliation{Department of Physics and Astronomy, Center for Materials Theory, Rutgers University, Piscataway, NJ 08854, USA}

\author{Kun Chen}
\affiliation{Center for Computational Quantum Physics, Flatiron Institute, 162 5th Avenue, New York, NY 10010, USA}

\author{Justin H. Wilson}
\affiliation{Department of Physics and Astronomy, Center for Materials Theory, Rutgers University, Piscataway, NJ 08854, USA}
\affiliation{Department of Physics and Astronomy, Louisiana State University, Baton Rouge, LA 70803, USA}
\affiliation{Center for Computation and Technology, Louisiana State University, Baton Rouge, LA 70803, USA}

\author{Andrew C. Potter}
\affiliation{Department of Physics, University of Texas at Austin, Austin, TX 78712, USA}
\affiliation{Department of Physics and Astronomy, and Quantum Matter Institute,
University of British Columbia, Vancouver, BC, Canada V6T 1Z1}

\author{J. H. Pixley}
\affiliation{Department of Physics and Astronomy, Center for Materials Theory, Rutgers University, Piscataway, NJ 08854, USA}
 \affiliation{Department of Physics, Princeton University, Princeton, NJ 08544, USA}
 \affiliation{Center for Computational Quantum Physics, Flatiron Institute, 162 5th Avenue, New York, NY 10010, USA}

\author{Sarang Gopalakrishnan}
\affiliation{Department of Physics, The Pennsylvania State University, University Park, PA 16802, USA}

\author{Romain Vasseur}
\affiliation{Department of Physics, University of Massachusetts, Amherst, MA 01003, USA}

\begin{abstract}

Monitored quantum circuits can exhibit an entanglement transition as a function of the rate of measurements, stemming from the competition between scrambling unitary dynamics and disentangling projective measurements. We study how entanglement dynamics in non-unitary quantum circuits can be enriched 
in the presence of
charge conservation, 
using a combination of exact numerics and a mapping onto a statistical mechanics model of constrained hard-core random walkers. We uncover a charge-sharpening transition that separates different scrambling phases with volume-law scaling of entanglement,  distinguished by whether measurements can efficiently reveal the total charge of the system. We find that while R\'enyi entropies grow sub-ballistically as $\sqrt{t}$ in the absence of measurement, for even an infinitesimal rate of measurements,  all average R\'enyi entropies grow ballistically with time $\sim t$. We study numerically the critical behavior of the charge-sharpening and entanglement transitions in U(1) circuits, and show that they 
 exhibit emergent Lorentz invariance and can also be diagnosed using scalable local ancilla probes.  Our statistical mechanical mapping technique readily generalizes to arbitrary Abelian groups, and offers a general framework for studying dissipatively-stabilized symmetry-breaking and topological orders.

\end{abstract}
\maketitle


\section{Introduction}

The dynamics of quantum information has become a central theme across multiple branches of physics ranging from condensed matter and atomic physics to quantum gravity~\cite{RevModPhys.80.517,RevModPhys.81.865,RevModPhys.82.277,Calabrese_2009,LAFLORENCIE20161}. Of particular recent interest is the exploration of entanglement dynamics in open quantum systems, motivated by the advent of noisy intermediate-scale quantum simulators~\cite{Preskill2018quantumcomputingin}. The dynamics of an open system
monitored by an external observer or coupled to its environment consists of two competing processes: unitary evolution, which generates entanglement and generically leads to chaotic dynamics~\cite{PhysRevX.7.031016,Nahum2018,PhysRevX.8.021013,Zhou2019,PhysRevX.8.031058,PhysRevX.8.031057}, and non-unitary operations resulting from measurements and noisy couplings to the environment, that tend to irreversibly destroy quantum information 
 stored in the  system by revealing it to the environment.

A minimal model that captures these competing processes consists of a quantum circuit made up of random unitary gates interlaced with local projective measurements. 
Remarkably, this minimal model undergoes a dynamical phase transition as the rate of measurements is increased. This transition occurs in individual quantum trajectories (i.e., the state of the system conditional on a set of measurement outcomes), and separates two phases where typical trajectories have very different entanglement properties~\cite{PhysRevB.98.205136,Skinner2019}. 
%
%
When measurements are frequent enough, they rapidly extract quantum information from any initial state, and collapse it into a weakly entangled pure state. 
Below a critical measurement rate, however, initial product states grow highly entangled over time, while initial mixed states remain mixed for extremely long times. In this ``entangling'' phase, unitary dynamics scrambles quantum information into nonlocal degrees of freedom that can partly evade local measurements. 
These nonlocal degrees of freedom span a decoherence-free subspace in which the dynamics is effectively unitary~\cite{Choi2020,Gullans2019,Li2020b,Fan2020}: this subspace can be regarded as the code space of a quantum error correcting code.
The effective size of the protected subspace vanishes at the measurement-induced phase transition. 
The critical properties of the measurement-induced transition are still under investigation, but key qualitative features of the transition including the emergence of conformal invariance are by now well-established~\cite{Bao2020,Jian2020,Li2020a,2021arXiv210703393Z}. Meanwhile, different variants of this transition have been investigated involving various ensembles of gates, circuit geometries, etc., establishing that such transitions are generic aspects of the quantum trajectories of open systems~\cite{PhysRevB.98.205136,Skinner2019,Li2019,PhysRevB.99.224307,Li2020a,10.21468/SciPostPhys.7.2.024,Gullans2019,Szyniszewski2019,Choi2020,Bao2020,Jian2020,Gullans2020,Zabalo2020,PhysRevResearch.2.023288,Ippoliti2020,Lavasani2020,Sang2020,PhysRevResearch.2.013022,PhysRevB.102.064202,Nahum2020,Turkeshi2020,Fuji2020,Lunt, Lunt2020,Fan2020,2020arXiv200503052V,Li2020b,PhysRevB.103.224210,PhysRevLett.126.060501,2021arXiv210306356L,2020arXiv201204666J,PhysRevLett.126.170503,2021arXiv210106245T,2021arXiv210209164B,2021arXiv210413372B,2021arXiv210407688B,2021arXiv210703393Z,2021arXiv210605881N}.

Given the central role of scrambling in the measurement-induced transition, it is natural to ask how the transition (and the entangling phase) are affected if one constrains the scrambling dynamics by imposing symmetries or conservation laws. 
Even in the absence of measurements, conservation laws severely constrain the scrambling of local operators~\cite{PhysRevX.8.031058,PhysRevX.8.031057,Friedman2019}. Moreover, conservation laws can parametrically slow down entanglement growth. For example, the dynamics of the R\'enyi entropies 
\begin{equation}
S_n = \frac{1}{1-n} \ln {\rm tr} \rho_A^n
\label{eq:Sn}
\end{equation}
(with $\rho_A$ the reduced density matrix of a subsystem $A$) 
in systems with a conserved charge 
have been shown to grow sub-ballistically as $S_{n>1} \sim \sqrt{t}$~\cite{Rakovszky2019,Huang2019,Zhou2020,Rakovszky2020,Znidaric2020}, while the von Neumann entropy remains linear in time $S_1 \sim t$. 
These results suggest that slow hydrodynamic modes might fundamentally alter, not just the nature of the measurement-induced phase transition, but also the nature of the entangling phase.
General arguments and numerical studies of both Clifford and non-interacting fermion circuits give evidence that symmetries enable distinct phases of volume-law-entangled dynamics~\cite{2021arXiv210209164B}, which would be fundamentally impossible in thermal equilibrium. Despite this work, many fundamental questions about the nature and phase structure and universal properties of measurement-induced phases and critical phenomena in volume-law entangled matter with symmetries remain unanswered -- in large part due to the absence of tractable analytic and numerical approaches for analyzing disorder-averaged quantum circuits with generic (computationally-universal) gate sets.
 
In this work we study the many-body dynamics of 
 monitored quantum circuits with a conserved charge (or equivalently a U(1) symmetry), and introduce a systematic tool to analytically perform the proper averaging over random gates and measurement outcomes with well-controlled approximations for general family of monitored circuit models with universal gates. This technique enables efficient numerical analysis of large-scale MRCs, and opens the door to potential analytic analysis using well-developed statistical mechanics and field-theoretic tools~\cite{FieldTheorySharpening}. Using this approach, we show that charge-conserving MRCs exhibit not only a measurement-induced entanglement transition but also a new type of ``charge-sharpening'' transition that separates two distinct entangling phases. These two phases are distinguished by whether it is easy or hard (in a way we will make precise below) for measurements to reveal the charge of the system. In the ``charge-fuzzy'' phase, there are large charge fluctuations even \emph{conditional} on the measurement history: i.e., the history of measurement outcomes does not suffice to fix the charge profile. In the ``charge-sharp'' entangling phase, by contrast, the measurement outcomes fix the charge distribution. The charge-sharp phase is nevertheless highly entangled because of the scrambling of neutral degrees of freedom. 
We find that both the entanglement transition and the charge-sharpening transition are Lorentz invariant, with a dynamical exponent $z=1$. 
%
This ballistic dynamics may seem surprising in light of the diffusive growth of R\'enyi entropies in the absence of measurements. However, we give general arguments that, for any finite measurement rate, $p>0$ all R\'enyi entropies grow linearly in time, $S_{n} \sim t$. 

Our conclusions are supported by exact numerics on Haar random U(1) monitored circuits, and by a replica statistical mechanics model~\cite{Bao2020,Jian2020} obtained from the study of a U(1) symmetric qubit degree of freedom coupled to a $d$-dimensional qudit. Remarkably, in the limit $d\to \infty$, we are able to analyze the replica limit analytically, and the contributions to entanglement from the qubit and qudit degree of freedom decouple. This enables us to study the entanglement dynamics of the qubit directly in this limit by simulating an effective statistical mechanics model of a symmetric exclusion process constrained by the measurements and entanglement cuts. We also discuss scalable probes of the charge-sharpening transition using ancilla qubits, and find evidence for a new universality class for the charge-sharpening transition in the limit of small local Hilbert space dimension.
  
The plan of the rest of the paper is as follows. In Sec.~\ref{overview} we specify the models we have explored, and present some general considerations on their steady state phases and dynamics. In Sec.~\ref{ednumerics} we present numerical results for U(1)-symmetric qubit chains. In Sec.~\ref{statmech_section} we present a tractable limit in which the model can be mapped onto the statistical mechanics of constrained random walkers. In Sec.~\ref{statmech_numerics} we present numerical results for the transfer matrix of this statistical model. Finally in Sec.~\ref{discussion} we summarize our results and discuss their broader implications. In particular, we note that these methods can be readily generalized to general Abelian symmetry groups, as detailed in Appendix~\ref{app:generalizations}, where we leverage duality relations to discuss implications for systems with highly-entangled phases with symmetry-breaking and topological orders.

\section{Overview of Results}\label{overview}

In this section we will introduce a family of U(1)-symmetric circuits, and present some general observations concerning their steady-state phase structure and entanglement dynamics. The numerical evidence supporting these observations will be presented below, in Secs.~\ref{ednumerics} and~\ref{statmech_numerics}.

\subsection{Models}\label{modelsec}

%
Following~\cite{PhysRevX.8.031057}, we consider a one-dimensional chain in which each site hosts a two-level system (``qubit'') and a $d$-level system (``qudit''), i.e., the on-site Hilbert space is $\mathbb{C}^2 \otimes \mathbb{C}^d$
for $d>1$ and $\mathbb{C}^2$ for $d=1$. 
The dynamics will consist of local unitary gates and measurements, which are chosen to conserve the U(1) charge 
\begin{equation}
\mathcal{Q} \equiv \sum_i \mathfrak{q}_i \otimes\mathbb{I}_i,\,\,\, \mathrm{where} \,\,\,   \mathfrak{q}_i = (\sigma^z_i +1)/2 
\label{eqn:Q}
\end{equation}
is acting on the $i$th site of the chain 
of length $L$ and $\mathbb{I}$ is the identity matrix on the qudits. 
These chains evolve under~(i)~unitary two-qubit gates, acting on nearest neighbor sites, which conserve the global charge $\mathcal{Q}$, and~(ii)~single-site projective measurements in which the qubit is measured in its $Z$ basis and the qudit is simultaneously measured in some reference basis~\footnote{Since the unitaries acting on the qudits are random, the randomizing measurement basis is superfluous.}. At each time-step, a given site is measured with probability $p$; for specificity, we assume that when this happens both the qubit \emph{and} the qudit are measured, so the measurement acts on that site as a rank-1 projector.
The symmetry-preserving two-site
unitary gates are arranged in a brickwork geometry and take the form  
\begin{align}
U_{i,i+1} = \begin{pmatrix}
U^0_{d^2\times d^2} &0 & 0  \\
0 & U^1_{2d^2\times 2d^2} &0 \\
0 & 0 & U^2_{d^2 \times d^2}
\end{pmatrix}, \label{eq: Unitary matrix structure}
\end{align}
where $i$ labels a site, $U^{\mathfrak{q}}_{D\times D}$ is a unitary matrix of size $D\times D$ acting on the charge $\mathfrak{q}_1 +\mathfrak{q}_2 = \mathfrak{q} \in \{0,1,2\}$ sector (a local charge is defined to take values $0$ and $1$), and $D$ is the dimension of the Hilbert space of the charge sector. 
Each matrix is drawn independently from the Haar random ensemble of unitary matrices of the appropriate size. 

We present numerical results for this class of circuits in two limits. First, we consider the limit $d = 1$, where there is no qudit degree of freedom, and one simply has a chain of qubits interacting via gates that conserve the charge $\mathcal{Q}$. 
In this limit, we obtain numerical results by direct time evolution. Second, we consider the complementary limit $d = \infty$, in which we can map the problem to a statistical mechanics model and explicitly write down a transfer matrix that generates the observables of interest. The phase diagrams in the two complementary limits are similar. 

The qubit-only ($d=1$) model is directly realizable in existing quantum processors. The $d>1$ model is perhaps less natural experimentally, but could be realized in circuit quantum electrodynamics setups~\cite{blais2021circuit} in which superconducting transmon qubits are coupled to multilevel superconducting cavities (qudits), or by blocking multiple qubits together (e.g. $d=2$ could be realized as a two-leg ladder of qubits). Regardless of experimental implementation, we expect the $d>1$ models to capture the generic universal behavior of phases and transitions, while allowing greater theoretical control in the large-$d$ limit.

\subsection{Observables and averaging}\label{sec: observables and averaging}
%
%

For a given choice of unitary gates and measurement locations $\mathbf{X}$ (in spacetime), the unitary-measurement dynamics can be described in terms of quantum trajectories $ \rho_{\mathbf{m}}(t)=K_{\mathbf{m}} \rho K_{\mathbf{m}}^\dagger$. Here $\mathbf{m}$ denotes a  ``quantum trajectory'', associated with a fixed configuration of measurement locations $\mathbf{X}$, and measurement outcomes ${\cal M}(\mathbf{X})$. We will write $\mathbf{m} = \lbrace\mathbf{X},  {\cal M}(\mathbf{X}) \rbrace$. The Kraus operators $K_{\mathbf{m}}$ consist of random unitary gates and projection operators onto the measurement outcomes $ {\cal M}(\mathbf{X})$, and we have $\sum_{\lbrace {\cal M}(\mathbf{X} ) \rbrace} K_{\mathbf{m}}^\dagger K_{\mathbf{m}} = \mathbb{I}$.


We will be concerned with general properties  of the single-trajectory state $\rho_{\mathbf{m}} \equiv K_{\mathbf{m}} \rho K_{\mathbf{m}}^\dagger$. (As a concrete example, consider the purity $\Pi_{\mathbf{m}} = \mathrm{Tr}(\rho_\mathbf{m}^2) / (\mathrm{Tr} \rho_{\mathbf{m}})^2$.) We then average over quantum trajectories, weighting each set of measurement outcomes by its probability of occurrence, i.e., the Born probability $p_{\mathbf{m}} = \mathrm{Tr}(\rho_{\mathbf{m}})$. Finally, we average the answers across the ensemble of quantum circuits. 

A few comments are in order here. 
\begin{enumerate}
\item For an initially pure state $|\psi\rangle$ the Born probability takes the familiar form $p_{\mathbf{m}} = \Vert K_{\mathbf{m}} |\psi\rangle \Vert^2$, i.e., it is just the norm of the projected state. For a series of measurements interspersed with unitary gates, one can pick each measurement outcome based on the Born probability or \emph{equivalently} apply a set of projectors at random and evaluate the probability of the entire measurement history by computing the norm of the state at the end of the trajectory.
\item There are four different types of average that we will consider here: (i)~the quantum expectation value of an observable $A$ in a single trajectory $\rho_\mathbf{m}$, namely $\mathrm{Tr}(A \rho_\mathbf{m})/\mathrm{Tr}\rho_\mathbf{m}$, which we will write as $\langle A \rangle_\mathbf{m}$; 
(ii)~the (Born-weighted) average of a single-trajectory function, such as purity or entanglement entropy, over quantum trajectories (measurement outcomes); (iii)~the average over unitary gates, chosen with Haar measure; and (iv)~the average over spacetime points where the measurements occur (measurement locations). In much of this work, we will present results for which averages (ii)-(iv) have been done. We will use the notation $[\cdot]$ for this full average. At some points it will be useful to separate these averages. In these cases we will use the explicit notations $\sum_{\lbrace {\cal M}(\mathbf{X}) \rbrace}( \cdot) $, $\mathbb{E}_{U}( \cdot)$, and $\mathbb{E}_{\mathbf{X}}( \cdot)$ for averages over measurement outcomes, gates, and measurement locations respectively. We will also use a shorthand notation $\mathbb{E}_\mathbf{m} \equiv \sum_{\mathbf{m}} p_{\mathbf{m}}(\dots)$ to denote summation over all possible measurement locations $\mathbf{X}$ including appropriate probability factors of $p$ and $1-p$, and over all measurement outcomes $ {\cal M}(\mathbf{X})$, including the associated Born probability factor $p_{\mathbf{m}}$.
\item It is crucial that the quantities of interest to us are \emph{nonlinear} functions of $\rho_{\mathbf{m}}$~\cite{Skinner2019}, such as $\Pi_{\mathbf{m}}$. To see the significance of this, let us compare the quantity $\Pi_{\mathbf{m}}$ to that of some simple expectation value. In $\rho_{\mathbf{m}}$, the expectation value of a local operator $A$ would be $\mathrm{Tr} (A \rho_{\mathbf{m}}) / \mathrm{Tr} \rho_{\mathbf{m}}$. Averaging this over trajectories (measurement outcomes) with the Born probabilities would simply give infinite temperature behavior~\footnote{The ensemble-averaged $\rho(t)$ resulting from maximally-random, local, open-systems dynamics is indistinguishable from an infinite temperature state over distances $\sim t$.}:
$\sum_{\lbrace {\cal M}(\mathbf{X}) \rbrace}  ( p_{\mathbf{m}} \langle A \rangle_\mathbf{m} ) = \sum_{\lbrace {\cal M}(\mathbf{X}) \rbrace} \mathrm{Tr} (A \rho_{\mathbf{m}}) = \mathrm{Tr}(A \rho(t)){\approx \mathrm{Tr}(A)/\mathrm{Tr}(1)}$, where $\rho(t) =\sum_{\lbrace {\cal M}(\mathbf{X}) \rbrace}  K_{\mathbf{m}} \rho K_{\mathbf{m}}^\dagger$ describes the dynamics of the density matrix in the case where the environment does not monitor or keep track of the measurement outcomes.
 By contrast, $\sum_{\lbrace {\cal M}(\mathbf{X}) \rbrace} \Pi_{\mathbf{m}} p_{\mathbf{m}}= \sum_{\lbrace {\cal M}(\mathbf{X}) \rbrace} \mathrm{Tr}(\rho_{\mathbf{m}}^2) / \mathrm{Tr} \rho_{\mathbf{m}}$, which cannot simply be written in terms of $\rho(t)$. The averaged density matrix $\rho(t)$ is blind to measurement transitions; only nonlinear functions of single-trajectory wavefunctions detect it.
\end{enumerate}

\subsection{Results}

\begin{figure}
\includegraphics[width=0.5\textwidth]{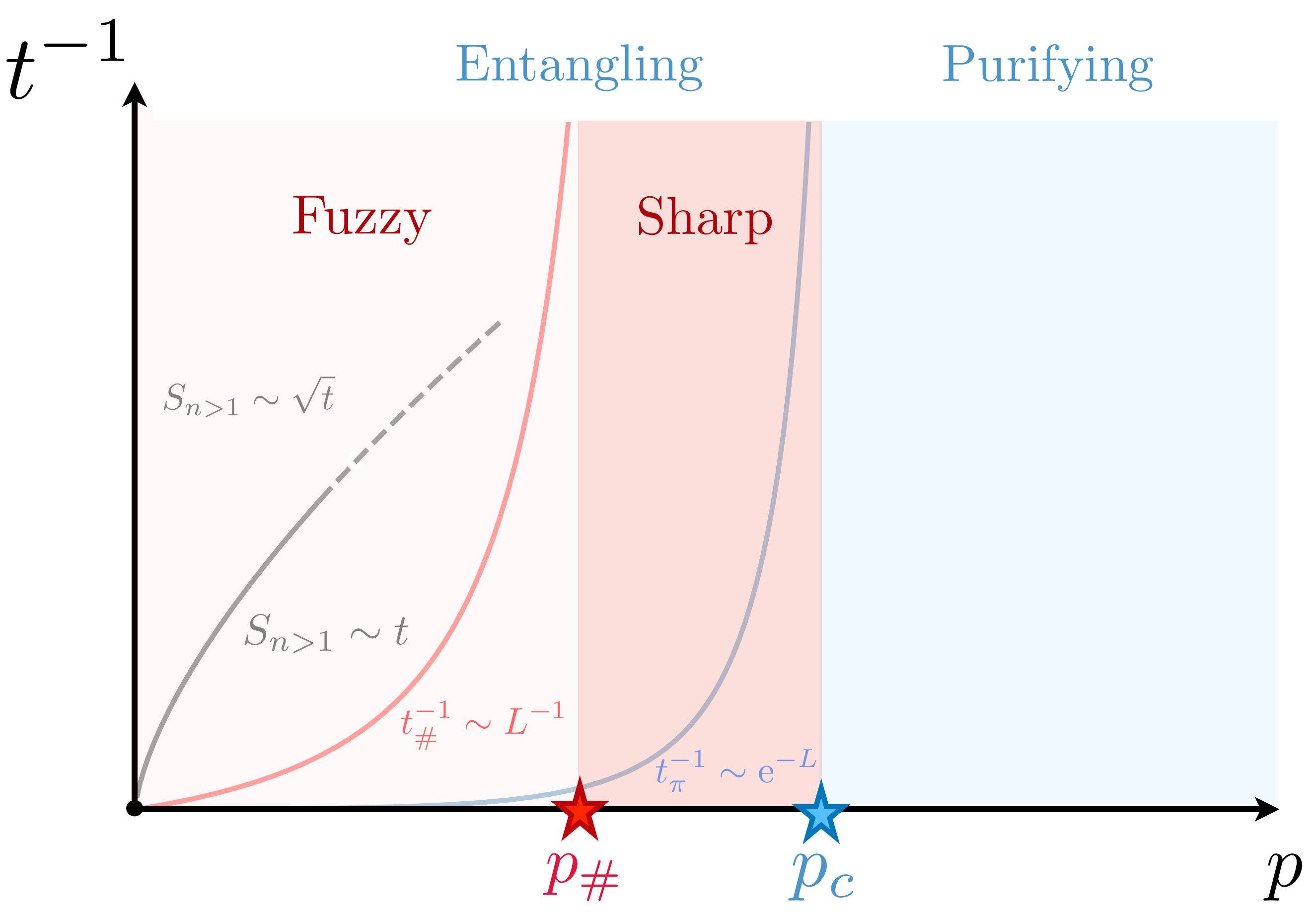}
\caption{ {\bf Phase diagram and crossover time scales in U(1) symmetric monitored quantum circuits.} Our numerical results indicate that there are two distinct phases in the entangling (volume-law) regime $p<p_c$, separated by a charge-sharpening critical point as $p=p_\#$. In the charge-fuzzy phase ($p<p_\#$), we identify three relevant time scales in the dynamics: For a large enough system, first (1) average R\'enyi entropies crossover from diffusive $[S_{n>1}] \sim \sqrt{t}$ to ballistic $\sim t$ scaling over a time scale $\sim p^{-3/2}$, then (2) charge sharpens after the crossover time scale $t_\# \sim L$, and finally (3) the system purifies over a much long time scale $t_\pi \sim {\rm e}^L$.}\label{Fig. phase diagram}
\end{figure}

In the following, we unveil a charge sharpening transition that takes place before the entanglement transition in two distinct models of monitored U$(1)$ symmetric random quantum circuits. Our main results are summarized in Fig.~\ref{Fig. phase diagram} and discussed in more detail below.

\subsubsection{Entanglement transition}

A general feature of unitary-projective circuits is the presence of an \emph{entanglement transition}, separating a phase where initially unentangled states develop volume-law entanglement from one where their entanglement remains area-law at all times. We briefly review the general properties of this transition and discuss how they are modified by the presence of a conservation law.

This transition occurs at some critical measurement rate $p_c$. In the volume-law phase, the half-system entanglement entropy grows linearly in time and saturates on timescales $t \sim L$. At times $t > L$, the entanglement entropy (averaged over circuits and trajectories) reaches a steady state value $[S(L/2)] = c L$, where $c$ is a constant that decreases continuously to zero at $p_c$. At $p = p_c$, we have that $[S_n(L/2)] = \alpha_n \log L$, where $\alpha_n$ (which generally depends on the R\'enyi index $n$) is part of the universal critical data~\cite{Skinner2019, Zabalo2020}. 

An equivalent way to understand the entanglement transition is as a
``purification transition'' for an initially mixed state~\cite{Gullans2019}. For $p < p_c$, an initially mixed state for a system of size $L$ evolves to a pure state on a timescale $t_{\pi} \sim \exp(L)$, whereas for $p > p_c$ purification happens on a timescale that grows sub-linearly in $L$. If one takes the limits $L \rightarrow \infty$, $t/L = \mathrm{constant}$, the purity of an initially mixed state for $p < p_c$ is essentially constant in time: i.e., the steady state is defined to be at early times compared with the slow purification dynamics for $p < p_c$. 

The purity of an initially mixed state on timescales $t \sim L$ can be used to define an effective order parameter for the volume-law phase, as follows~\cite{Gullans2020, Zabalo2020}. Consider evolving a pure state along some trajectory until times $t \sim L$, and then entangling some local degree of freedom with an ancilla qubit. The reduced density matrix of the system is now a rank-2 mixed state. For $p < p_c$, this mixed state remains mixed for an exponentially long time. Therefore, by evolving for another $t \sim L$ and then measuring the entanglement entropy of the ancilla (which is equivalent to measuring the purity of the system density matrix in this setup), one can extract a local order parameter for the volume-law phase~\cite{Gullans2020}. By studying the correlations of this local order parameter---e.g., by coupling in two ancillas at distinct spacetime points and tracking their mutual information---it was established (in the absence of the U$(1)$ symmetry) that the critical theory has an emergent Lorentz invariance with dynamical scaling exponent $z = 1$.

One of our results is to locate and characterize this entanglement transition in the presence of the U$(1)$ conservation law. In the $d \to \infty$ limit, we find that $p_c = 1/2$, exactly as in the absence of a conservation law. Moreover, the entanglement transition corresponds to a percolation transition. For $d = 1$, we find that $p_c = 0.105(3)$ moves to substantially lower measurement rate than in the generic circuit without a conservation law ($p_c^{\mathrm{Haar}}\approx 0.17$~\cite{Zabalo2020}). The correlation length exponent $\nu = 1.32(6)$ is close to the percolation value $4/3$, but the coefficients $\alpha_n$ differ from the percolation value as well as the value for Haar-random circuits without a $U(1)$ symmetry. Finally, we find compelling numerical evidence that the dynamical scaling $z = 1$ holds at this critical point. That this holds regardless of the diffusive ($z = 2$) dynamics of the U$(1)$ conserved charge is perhaps puzzling at first sight. We return below to the resolution of this puzzle.

\subsubsection{Charge-sharpening transition}

In addition to changing the critical properties of the entanglement transition at $p_c$, the conservation law gives rise to a \emph{distinct} ``charge-sharpening'' transition at a measurement rate $p_\#$ inside the volume-law phase. The charge-sharpening transition separates a ``charge-sharp'' phase for $p > p_\#$, in which the measurements along a typical trajectory can rapidly collapse an initial pure superposition (or mixture) of different charge sectors, and a ``charge-fuzzy'' phase where this collapse is parametrically slower occurring on a time scale $t_{\#}\sim L$. Specifically, we can distinguish charge- sharp or fuzzy behavior by the 
variance of the conserved charge $\cal{Q}$ in Eq.~\eqref{eqn:Q} over a single trajectory, 
averaged across trajectories and samples, i.e., 
\begin{align}
[\delta {\cal Q}^2] = [\langle {\cal Q}^2 \rangle_{{\mathbf{m}}} - \langle {\cal Q} \rangle_{{\mathbf{m}}}^2 ],
\label{eq:PQ}
\end{align}
 where the quantity in parentheses is the \emph{quantum} number variance in a given trajectory. In the sharp phase, $[\delta {\cal Q}^2] = 0$ while in the fuzzy phase it remains non-zero at times of order $t \sim L$. 

The dynamics of charge sharpening at small $p$ can be qualitatively understood in terms of a simple classical model, in which one ignores the spatio-temporal correlations between measurements. One can then ask how many independent density measurements $\mathcal{N}_M$ are required to distinguish systems with $N$ particles on $L$ sites from those with $N - 1$ particles on $L$ sites, where $n \equiv N/L = {\cal O}(1)$. Assuming Gaussian density fluctuations (as in the $p = 0$ thermal state) we expect the $N$-particle and $(N-1)$-particle states to become distinguishable when $\mathcal{N}_M \sim L^2$.\footnote{This can seen by the central limit theorem as follows: assuming each measurement outcome to be independent, the statistical error in the outcomes of the measurement of charge density goes as $1/\sqrt{\mathcal{N}_M}$. To distinguish the states with global charge $N$ and $N-1$, we require this error to become smaller than $\sim 1/L$. This gives $\mathcal{N}_M \sim L^2$.}
Since $\mathcal{N}_M = p L t$ in the circuits we consider, sharpening happens on a crossover timescale $t_\# \sim L/p$. For timescales $t \geq t_\#$, we expect that $[\delta \mathcal{Q}^2] \sim \exp(-t/t_\#)$. This follows, e.g., from using the central limit theorem to estimate the probability that an $N$ particle state will give an average density of $n \pm 1/L$ after $p L t$ measurements. This simple model of the volume law phase predicts that a crossover to charge sharpening should take place on a timescale $t_\# \sim L/p$, consistent with our numerical findings (see Sec.~\ref{ednumerics} and~\ref{statmech_numerics}), and parametrically faster than purification. 
Importantly, at any finite $t/L$, $[\delta \mathcal{Q}^2]$ remains non-zero in the fuzzy phase
(albeit exponentially small for $t\gg t_{\#}$). 

By contrast, for $p > p_\#$, charge-sharpening happens on a timescale that is sublinear
(logarithmic) in system size. In the limit $L \to \infty, t/L = \mathrm{constant}$, each trajectory has a definite charge. Thus there is a sharp phase transition at $p_\#$, for which $[\delta \mathcal{Q}^2]$ acts as an order parameter. 
Our numerical results also indicate that charges become devoid of quantum superposition in some regions of space-time exhibiting locally minimal spacing of measurements (see Sec.~\ref{statmech_numerics}).

As with the entanglement transition, one can probe the charge-sharpening transition by coupling an ancilla to the circuit. One entangles the ancilla with the system such that each ancilla state is coupled to a system state with a different value of $\mathcal{Q}$. The system-ancilla entanglement vanishes when $\mathcal{Q}$ sharpens under the circuit dynamics.

\subsubsection{Entanglement dynamics}

We now turn to the dynamics of entanglement at times of order unity. Recall that, absent measurements, the R\'enyi entropies $S_n \sim \sqrt{t}$ for all $n > 1$ in random circuits with U$(1)$-symmetric gates~\cite{Rakovszky2019,Huang2019a,Zhou2020}. This diffusive entanglement dynamics appears to be a generic property of random circuits, so one might expect it to hold throughout the volume-law phase. If it held at the critical point, it would prevent the critical theory from being a conformal field theory (CFT). We now discuss why R\'enyi entropies in fact scale \emph{ballistically} for any non-zero measurement probability, $p > 0$, allowing both the sharpening and entanglement transitions to obey relativistic $z = 1$ dynamic scaling.  

First, we review the argument for diffusive scaling in the absence of measurements~\cite{Rakovszky2019,Huang2019a,Zhou2020}. This phenomenon arises from rare fluctuations that leave a region empty (or maximally filled), as follows. Consider, for concreteness, the dynamics of the initial product state for the qubit $|\psi\rangle = \otimes_{i = 1}^L |+_x\rangle_i$ where $|+_x\rangle = \frac{1}{\sqrt{2}} (|0\rangle + |1\rangle)$. %
Suppose we are interested in the entanglement across a cut at $L/2$ at some later time $t$. 
We can divide the system into three regions: a central region of radius $\ell = \sqrt{D t}$ centered at the entanglement cut, and regions to the left and right. Define 
\begin{equation}
|\psi_{\text{dead}} \rangle = \bigotimes_{i = 1}^{L/2 - \ell} |+_x\rangle_i \bigotimes_{i = L/2 - \ell + 1}^{L/2 + \ell} |0\rangle_i \bigotimes_{i = L/2 + \ell + 1}^L |+_x\rangle_i.
\end{equation}
Initially, $|\langle \psi_{\text{dead}} | \psi \rangle| = 2^{-2\ell}$. After evolving for time $t$, $|(\langle \psi_{\text{dead}}| U^\dagger_t) (U_t | \psi \rangle)| = 2^{-2\ell}$ by unitarity. However, $U_t |\psi_{\text{dead}}\rangle$ is a product state with respect to the cut at $L/2$: by construction, $t$ is not long enough for particles to have diffused to the entanglement cut, and unless there is a $|10\rangle$ or $|01\rangle$ configuration at the cut the gates acting across the cut cannot generate entanglement. The largest Schmidt coefficient of $U_t |\psi\rangle$ is its maximal overlap with \emph{any} product state, so we can lower-bound the largest Schmidt coefficient of $U_t |\psi\rangle$ as $2^{-2\ell} = 2^{-\sqrt{Dt}}$, and therefore $S_\infty \leq 2\ell \ln 2 \sim \sqrt{Dt}$. All R\'enyi entropies with $n > 1$ are dominated by this largest Schmidt coefficient and grow as $\sqrt{t}$. The Von Neumann entropy $S_1$ is dominated instead by \emph{typical} Schmidt coefficients: the number of these grows exponentially in $t$, but they are also exponentially small in $t$ and are therefore subleading for $n > 1$.

We now address how this argument changes when $p > 0$. In a typical trajectory, on a timescale $t$, there are $p \ell t \sim p t^{3/2}$ measurements in the putative dead region near the entanglement cut, and about half the measurements observe a qubit to be in the charge state $|1\rangle$. There are rare circuits with few measurements near the cut, as well as rare histories in a typical circuit where all the measurements yield the same outcome $|0\rangle$. However, both are \emph{at least} exponentially suppressed in $\ell$ and cannot dominate the trajectory-averaged entanglement (since \emph{any} trajectory contributes at most $\sim t$ entanglement, and in any case these atypical trajectories have unusually slow entanglement growth). Therefore, to compute the trajectory-averaged R\'enyi entropies it suffices to consider trajectories with typical measurement locations and typical outcomes. In typical trajectories, one observes a $|1\rangle$ charge after ${\cal O} (1)$ measurements in the region near the cut, so the putative dead region survives only until a time $t \sim p^{-2/3}$. At longer times, the overlap of the wavefunction with dead regions is zero. We conclude that the trajectory-averaged R\'enyi entropies $[S_n]$ grow linearly in time whenever $p > 0$~\footnote{Though other quantities such as $\log [e^{-S_n}]$ are dominated by rare dead-region contributions and do exhibit $\sqrt{t}$ growth due to rare dilute measurement locations. The parametrically strong discrepancy between the average purity and the average R\'enyi entropies is also seen numerically in our statistical model approach (Sec.~\ref{statmech_numerics}).}.
We also note that, the existence of diffusive hydrodynamic modes, which are a purely classical phenomena, does not affect the $z = 1$ dynamical scaling at the volume-to-area law entanglement transition at $p_c$. 
Our 
numerical estimates of the dynamic exponent in Sec.~\ref{ednumerics}
are consistent with this result that $z = 1$ scaling applies and diffusive hydrodynamics decouples also at the charge-sharpening transition at $p_{\#}$.


\section{Numerics on qubit chains} \label{ednumerics}

In this section we present numerical results on a model of random U(1)-conserving gates acting on a chain of qubits (i.e., the $d = 1$ limit of the general model in Sec.~\ref{modelsec}).
Specifically, in the basis of the adjacent qubits
$\{\ket{\downarrow\downarrow},\ket{\uparrow\downarrow},\ket{\downarrow\uparrow},\ket{\uparrow\uparrow}\}$ 
the two-qubit gates at site $i$ take the block diagonal form
\begin{equation}
U_{i,i+1}=
\begin{pmatrix}
  e^{i\phi_0}
  &  &  \\
&  
U^1_{2\times 2}
& \\
   &   & e^{i\phi_1}
\end{pmatrix}
\label{eq:conserving_gate}
\end{equation}
where $\phi_{1}$ and $\phi_{0}$ are chosen at random from the interval
$\left[0,2\pi\right)$ and $U^1_{2\times 2}$ is a Haar-random $2 \times 2$ unitary
matrix that can be parameterized by 4 angles
\begin{equation}
  U^1_{2\times 2}
  (\alpha,\phi,\psi,\chi) = e^{i\alpha}
  \begin{pmatrix}
    e^{i\psi}\cos\phi & e^{i\chi}\sin\phi \\
    -e^{-i\chi}\sin\phi & e^{-i\psi}\cos\phi
  \end{pmatrix}
\end{equation}
where $\alpha$, $\psi$, $\chi$, and $\phi$ are chosen so that $U^1_{2\times 2}$ is uniformly sampled from U$(2)$~\footnote{To obtain a uniform distribution over $U(2)$ we must pick
$\alpha,\psi,\chi \in \left[0,2\pi\right)$ and $\xi \in \left[0,1\right]$ uniformly and compute
$\phi = \arcsin \sqrt{\xi}$.}.
In between layers of gates, projective measurements are performed: with a probability $p$, the qubit is projected onto
${\lvert \uparrow \rangle \mbox{ or } \lvert \downarrow \rangle}$ given by the Born rule.
Utilizing the conservation law, we  work in definite number-sectors 
to reduce the memory load of the exact numerics.
%


The conservation law leads to different charge sectors defined by eigenspaces of $\mathcal{Q}$ in Eq.~\eqref{eqn:Q}.
We will typically focus on charge sectors near the central subspace ($\mathcal{Q} = L/2$). 

\subsection{Entanglement transition}
We begin by locating the entanglement transition point $p_c$ in this model. In order to probe the location of the critical point and the correlation length critical exponent,
we study two quantities that have been identified as good measures of the transition: the tripartite mutual information~\cite{Zabalo2020} and an order parameter defined through the use of an ancilla~\cite{Gullans2020} that is coupled to one charge-$\mathcal{Q}$ sector. In the following it is essential that we use an accurate estimate of $p_c$ to be able to numerically disentangle it from the charge sharpening transition at $p_{\#}$.

\begin{figure*}
    \centering
    \includegraphics{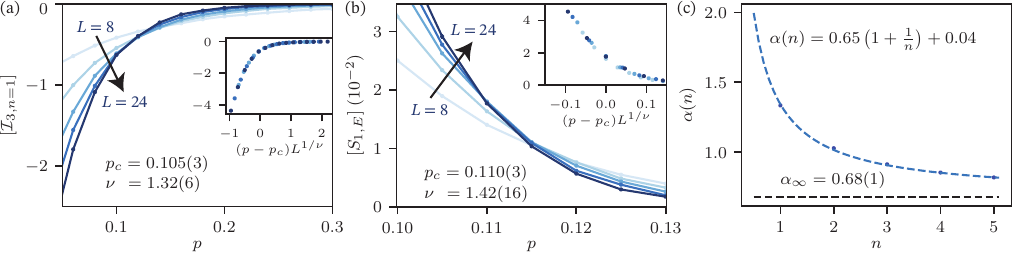}
    \caption{ {\bf Entanglement transition in qubit chains.}
    (a) Data and collapse of the tripartite mutual information, $\mathcal{I}_{3,n=1}$, used to determine
    the critical point of the entanglement transition, $p_c = 0.105(3)$, and the correlation length exponent,
    $\nu = 1.32(6)$. The error bars in $p_c$ and $\nu$ are estimated using the region in parameter space near the minimum such that $\chi^2 < 1.3 \chi_{\mathrm{min}}^2$~\cite{Zabalo2020}, see Fig.~\ref{fil:confidence_blobs}.
    (b) Data and collapse of the entanglement transition order parameter, $\left[S_{1,E}\right]$, used as an
    alternative method to determine the critical point of the entanglement transition, $p_c = 0.110(3)$,
    and the correlation length exponent $\nu = 1.42(16)$.
    {The critical point and the exponent for $\left[S_{1,E}\right]$ are estimated with a finite-size scaling analysis as explained in Appendix D.}
    (c) At the critical point, the bipartite entanglement entropy shows logarithmic scaling with the system size.
    The coefficient of the logarithm has strong R\'enyi index dependence that can be described by a functional form
    $\alpha(n) = 0.65(1)\left(1+\frac{1}{n}\right) + 0.04(1)$. The error bars in $\alpha(n)$ are obtained from the standard deviation of the fit parameters in $\alpha(n) \sim a(1 + \frac{1}{n}) + b$ while the error in the data points themselves take into account the standard deviation of the fit parameters in $S_n(p_c,L) \sim \alpha(n) \ln L$ and the error in $p_c$. 
    This closely resembles the standard result for the groundstate of a CFT, but has an offset slightly larger than zero.
    {Error bars on the data points are obtained from the error on the mean by computing the standard deviation.}
    }
    \label{fig:entanglement-transition}
\end{figure*}

\begin{figure}
    \centering
    \includegraphics{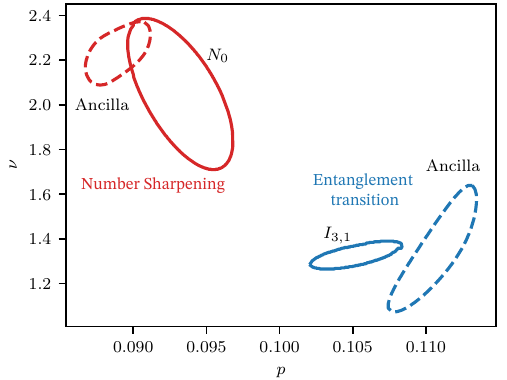}
    \caption{{\bf Two transitions.} Plot of the $68\%$ confidence interval of the critical point $p$ and the correlation length exponent $\nu$ for both the entanglement and charge transitions. The mutual information $I_{3,1}$ (solid blue circle) and the ancilla probe (dashed blue circle) are for the entanglement transition, while the fraction of trajectories $N_0$ (solid red circle) and the ancilla probe (dashed red circle) are for the charge-sharpening phase transition. The two transitions appear to be different with statistical significance, although we cannot exclude systematic finite-size effects that would change this conclusion in the thermodynamic limit.}
    \label{fil:confidence_blobs}
\end{figure}

First, the tripartite mutual information for the R\'enyi index $n$ is defined as
\begin{align}
  \mathcal{I}_{3,n}(A,B,C) \equiv& ~~S_{n}(A) + S_{n}(B) +S_{n}(C)-
  \nonumber\\ 
  &-S_{n}(A\cup B) -S_{n}(A\cup C) -S_{n}(B\cup C)+ \nonumber\\ 
  &+ S_{n}(A\cup B\cup C),
\end{align}
where we have chosen regions $A,B,\mbox{and } C$ to be adjacent regions of size $L/4$ and $S_n(A)$ is the R\'enyi entropy
defined in Eq.~\eqref{eq:Sn}.
For $\mathcal{I}_{3,n}$ at late times ($t=4L$) we apply the finite size scaling hypothesis
$
  \mathcal{I}_{3,n} \sim f(L^{1/\nu}(p-p_c)),
  $
to locate the critical point, where $f(x)$ is a scaling function and $\nu$ is the correlation length exponent.
The data for $n=1$ is shown in Fig.~\ref{fig:entanglement-transition}(a) where we find the data collapses with the minimum $\chi^2$
for the choice of $p_c = 0.105(3)$ and $\nu = 1.32(6)$. 
The error bars are estimated by a region in parameter space near the minimum such that $\chi^2 < 1.3\chi^2_\mathrm{min}$~\cite{Zabalo2020}, see Fig~\ref{fil:confidence_blobs}. 
A similar analysis can be performed for the other R\'enyi entropies and the resulting values of $p_c$ and $\nu$ are similar for all
$n \ge 1$ investigated, 
we find $p_c = 0.103(4), 0.12(2),  0.12(1)$ and $\nu = 1.37(8),  1.47(3), 1.5(2)$ for $n = 2,5,\infty$, respectively.

\begin{figure*}
    \centering
    \includegraphics{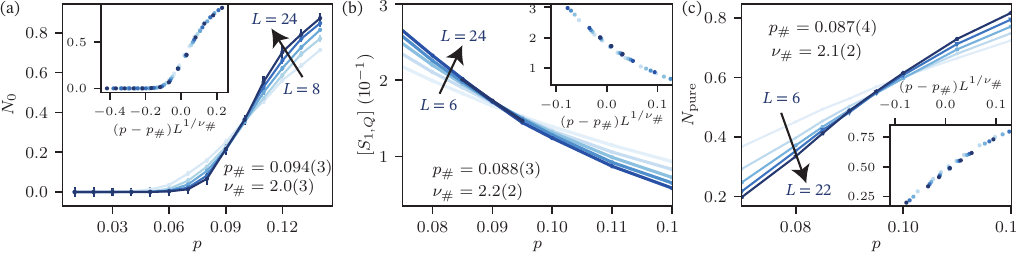}
    \caption{{\bf Charge sharpening transition in qubit chains.}
    (a) Data and collapse of $N_0$, the fraction of trajectories with $\delta\mathcal{Q}^2 < \epsilon$
    used to determine the critical point of the charge sharpening transition,
    $p_\# = 0.094(3)$ and correlation length exponent, $\nu_\# = 2.0(3)$.  The value of
    $\epsilon = 10^{-2}$ and is chosen such that it maximizes the quality of collapse at
    $t/L = 4$. 
    The error bars in $p_\#$ and $\nu_\#$ are estimated using the region in parameter space near the minimum such that $\chi^2 < 1.3 \chi_{\mathrm{min}}^2$~\cite{Zabalo2020}, see Fig.~\ref{fil:confidence_blobs}.
    (b) Data and collapse of the {charge sharpening} transition order parameter, $\left[S_{1,Q}\right]$, used as an
    alternative method to determine the critical point of the {sharpening} transition, $p_{\#} = 0.088(3)$,
    and the correlation length exponent $\nu_{\#} = 2.15(15)$. 
    (c) Data and collapse of the fraction of trajectories where the ancilla qubit is purified $N_{\text{pure}}$ at the charge-sharpening transition. The transition point $p_{\#} = 0.087(4)$ and the correlation length exponent $\nu_{\#} = 2.1(2)$ are consistent with the ancilla probe. 
     {The critical point and the exponent obtained from $\left[S_{1,Q}\right]$ and $N_{\text{pure}}$ are estimated with a finite-size scaling analysis as explained in Appendix D.}
    {Error bars on the data points are obtained from the error on the mean by computing the standard deviation.}
    }
    \label{fig:number-sharpening}
\end{figure*}

\begin{figure}
    \centering
    \includegraphics{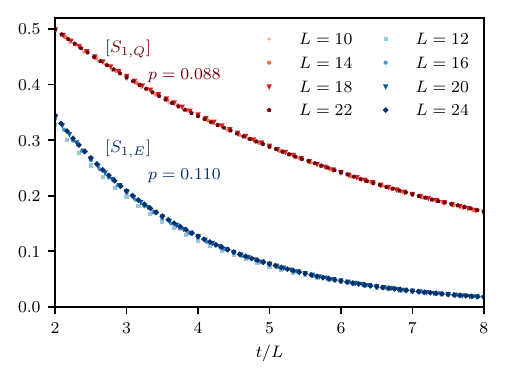}
    \caption{{\bf Dynamical exponent.} Plot of the rescaled time-dependence of the ancilla-circuit entanglement entropy for the entanglement transition at $p=p_c\approx 0.11$ (blue curve) and the charge-sharpening transition at $p=p_{\#}\approx 0.088$ (red curve). The finite size collapse indicates a dynamical exponent $z=1$ for both transitions. }
    \label{fig:dynamic_exponents_QE}
\end{figure}

At the critical point, the bipartite entanglement entropy shows a logarithmic dependence on
the system size and the coefficient of the logarithmic divergence shows strong
R\'enyi index dependence (Fig.~\ref{fig:entanglement-transition}(c)). This behavior can be
described by
\begin{equation}
  S_{n}(p_{c},L) \sim \alpha(n)\ln L, \,\, \alpha(n) = 0.65(1) \left( 1 + \frac{1}{n}\right) + 0.04(1).\label{rendep}
\end{equation}
Apart from the small offset, this R\'enyi index dependence matches the result one expects for the ground state of a CFT~\cite{Calabrese_2009}, $\alpha_{\mathrm{CFT}}(n) = \frac{c}{6}\left(1+\frac{1}{n}\right)$. The coefficients in Eq.~\eqref{rendep} clearly differ from those at the measurement induced transition without a conservation law~\cite{Zabalo2020}.

As an alternative way of locating the entanglement transition, we also study the ``order parameter''~\cite{Gullans2020}.
In order to have the ancilla couple to the system within a particular global charge sector, we consider the qubits at two adjacent sites $i$ and $i+1$ to be in the entangled state with an ancilla: $|\Psi_0\rangle=\frac{1}{\sqrt{2}}(\ket{\uparrow\downarrow}\ket{1} + \ket{\downarrow\uparrow}\ket{0})$ where the ancilla has orthogonal basis states $\ket{1}$ or $\ket{0}$.
We then evolve the system in time $t=2L$ without measurements in order to create a state $\ket{\Psi} = \frac{1}{\sqrt{2}}(\ket{\psi_1}\ket{1} + \ket{\psi_0}\ket{0})$ where $\ket{\psi_{0,1}}$ are orthogonal and in the same charge sector.
We then run the circuit with measurements for an additional time $2L$ and compute the von Neumann entropy of the ancilla, which we denote as $S_{1,E}$ (as it probes the entanglement transition). 

The results for the von Neumann entanglement entropy of the ancilla $S_{1,E}$
are shown in Fig.~\ref{fig:entanglement-transition}(b), and are consistent with $\mathcal I_{3,n}$ data: 
From the scaling ansatz 
$S_{1,E}\sim f_E((p-p_c)L^{1/\nu})$, where $f_E(x)$ is a universal scaling function,
we obtain $p_c = 0.110(3)$ and $\nu=1.4(2)$
in good agreement with $\mathcal{I}_3$.

Summarizing these results, the entanglement transition in U(1)-symmetric circuits has a critical exponent $\nu$ that is consistent with the value for Haar-random circuits with no symmetries, although the nonuniversal $p_c$ has drifted down from the Haar value  $p^\mathrm{Haar}_c \approx 0.17$ (as one might expect since each gate cannot generate as much entanglement).
At $p=p_c$ we extract the dynamical exponent of the entanglement transition using the scaling ansatz $[S_{1,E}]\sim g_E(t/L^z)$, which shows a good quality as seen in Fig.~\ref{fig:dynamic_exponents_QE} for $z=1$ and $g_E(x)$ some universal scaling function. Again, this result is consistent with the non-conserving case. 
While we have focused on $\mathcal{Q}=L/2$ we have checked that for the largest system sizes considered $p_c$ is only very weakly affected for $\mathcal{Q}=L/2-1$ (not shown).

\subsection{Charge sharpening transition}
\label{subsec:charge_sharpeing}
We now turn to estimating $p_{\#}$ in two ways: the charge variance of a state and the entropy of an ancilla entangled with two different number sectors.

First, we compute the variance of the total charge (Eqs.~\eqref{eqn:Q} and \eqref{eq:PQ}).
For a trajectory that lies in a well defined charge sector, $\delta {\cal Q}^2 = 0$, otherwise $\delta {\cal Q}^2 \ne 0$. Therefore, 
we start
with a pure initial
state that is spread out over all of the different $\mathcal{Q}$-sectors, 
$|\psi_0\rangle=\bigotimes_{i=1}^L \frac{1}{\sqrt{2}}\left(\ket{\uparrow}_i + \ket{\downarrow}_i \right)$,
and run the conserving hybrid dynamics to late times to determine if 
the system 
has sharpened into a single charge sector
for some measurement probability $0 \le p_{\#} \le p_c$, where $p_c\approx 0.11$ is the critical point of the entanglement transition.
In this situation, the critical point of the charge sector transition can be determined by the studying the probability, $P(\delta {\cal Q}^2 = 0)$. (Recall that $\delta \mathcal{Q}^2$ is a quantum uncertainty that is a property of each trajectory; the probability distribution $P(\delta \mathcal{Q}^2)$ is over trajectories and circuits, where each trajectory is weighted by its Born probability.)
For large systems, $P(\delta {\cal Q}^2=0) \rightarrow 0$ when the system is distributed over multiple sectors while $P(\delta {\cal Q}^2=0)\rightarrow 1$ when the
system has been constrained to a single sector. In Fig.~\ref{fig:number-sharpening}(a), the fraction $N_0$ of trajectories having a variance
$x=\delta {\cal Q}^2 \le s$ (with $s = 10^{-2}$) is shown for various system sizes and measurement probabilities. The critical point can be identified by the
crossing near $p = 0.1$. 
Performing a finite size scaling analysis, we find the data for different system sizes collapses onto a universal curve
for the critical point $p_{\#} = 0.094(3)$ and correlation length exponent $\nu_{\#} = 2.0(3)$.

Lastly, we consider an ancilla coupled to two different charge sectors, in particular we take $\ket{\Psi} = \ket{\psi_{\mathcal{Q}}}\ket{0}+\ket{\psi_{\mathcal{Q}-1}}\ket{1}$ where $\ket{\psi_{\mathcal{Q}}}$ represents states within the charge sector $\mathcal{Q}$ (while $\ket{1}$ and $\ket{0}$ are states of the ancilla as before).
Since there is no unitary that mixes these sectors, we can say definitively that the reduced density matrix has the form
\begin{equation}
    \rho_\mathrm{anc} = \begin{pmatrix}
      |\braket{\psi_{\mathcal{Q}}(t)|\psi_{\mathcal{Q}}(t)}|^2 & 0 \\ 0 & |\braket{\psi_{\mathcal{Q}-1}(t)
      |\psi_{\mathcal{Q}-1}(t)}|^2
    \end{pmatrix}.
\end{equation}
This formulation is convenient for the numerical algorithm we have developed that conserves charge since if the ancilla were just considered an extra qubit, $\ket{\Psi}$ would be in the conserving sector $M$ for $L+1$ qubits. 
Doing this, we compute the von Neumann entanglement entropy of the ancilla qubit that we denote as $S_{1,\mathcal{Q}}$,
that is shown in
Fig.~\ref{fig:number-sharpening}(b).  Based on the crossing in the data and the ansatz $S_{1,\mathcal{Q}} \sim g_{\mathcal{Q}}((p-p_{\#})L^{1/\nu_{\#}})$
we obtain $p_{\#} = 0.088(3)$ and $\nu_{\#}=2.2(2)$, which matches the $p_{\#}$ and $\nu_{\#}$ found by $P(\delta {\cal Q}^2=0)$.
In addition, we  extract the charge sharpening transition from the probability that the ancilla has fully disentangled from the circuit by computing the fraction of trajectories $N_{\mathrm{pure}}$ that have fully purified the ancilla [Fig.~\ref{fig:number-sharpening}(c)]. From the crossing of $N_{\mathrm{pure}}$ we find a third consistent estimate of $p_{\#}$ and $\nu_{\#}$.
Thus, we have identified the charge sharpening transition across all sectors of $\mathcal{Q}$ with the 
transition in $S_{1,\mathcal{Q}}$ across $\mathcal{Q}=L/2,L/2-1$. The dynamical exponent of the charge sharpening transition is also $z=1$, as $[S_{1,\mathcal{Q}}]\sim g_\mathcal{Q}(t/L)$ at criticality (Fig.~\ref{fig:dynamic_exponents_QE}).


The  two critical points we have identified in this model at $p_{\#}$ and $p_c$ are at least $\sim 3.5$ error bars from each other, providing evidence that a charge sharpening transition occurs \emph{before} full purification. 
This is further exemplified by the estimated confidence intervals for the two transitions for the various probes we have considered as shown in Fig.~\ref{fil:confidence_blobs}.
Moreover, the correlation length exponent for charge-sharpening quantities is distinct from that of the entanglement purification transition, further suggesting that these represent distinct critical points with different universality classes.

We note, however, that the close proximity of the putative two transitions make them challenging to cleanly separate numerically in small scale systems, and acknowledge that this data could in principle be accounted for by large, systematic finite-size errors in the critical exponents that affected the charge and entanglement properties differently\footnote{In the absence of a conservation law, it has been shown that the probes we have used for the entanglement transition ($\mathcal{I}_{3,n}$ and $[S_{1,E}]$) have weak finite size drifts in Clifford circuits~\cite{Zabalo2020} by examining small and large system sizes.}.
In the following sections, we will see that for the model with large-$d$ qudits, the location of the two transitions become clearly distinct.


\section{Statistical mechanics model} \label{statmech_section}
In this section, we show that in the $d \to \infty$ limit, the calculation of entanglement in monitored U(1) circuits can be mapped exactly onto a classical statistical model defined on a square lattice. In this limit, the contributions to the entanglement entropy from the qubit with conserving dynamics and the qudit decouple. The resulting qubit contribution can then be obtained from a constrained symmetric exclusion process. 

Our main goal is to compute averaged R\'enyi entropies $\avg{S_n}$. 
The R\'enyi entropies of a spatial sub-region, $A$, for a fixed quantum trajectory are given by
\begin{align}
S_n(A,\mindex) &= \frac{-1}{n-1} \left(\ln \text{Tr}\left(\rho_{{\mathbf{m}}}^{\otimes n} T_{n,A}\right)-\ln \text{Tr}\rho_\mathbf{m}^{\otimes n}\right), \label{eq: S_n[m]}
\end{align}
where $\rho_{\mathbf{m}} = \ket{\psi_\mindex(t)}\bra{\psi_\mindex(t)}$, $\ket{\psi_\mindex (t)}= K_\mindex \ket{\psi_0} $ is the state (``trajectory") of the system after evolution by time $t$ for a measurement history $\mindex$, and $T_{n,A}$ is a ``SWAP'' operator permuting the $n$ copies of the input state in the entanglement region $A$:
\begin{align}
T_{n,A} = \prod_i |s_{\sigma_i(1)} s_{\sigma_i(2)} ...s_{\sigma_i(n)}\rangle \langle s_1 s_2 ... s_n|\nonumber \\
\sigma_i = \begin{cases}
\text{identity}=e, & i\notin A \\
(12\dots n), & i\in A
\end{cases}\label{eq: boundary DW},
\end{align}
where the index $i$ runs over all physical sites, $\ket{s_i}$ are members of the onsite Hilbert space,  $\sigma_i$ is an element of the permutation group $\mathcal{S}_n$, and $(12\dots n)$ denotes a cyclic permutation of the $n$ copies of $\rho$. The key technical difficulty in this problem is to perform the average over gates, measurement locations and outcomes, and to normalize the state after the projective measurements since entanglement is intrinsically non-linear in the density matrix. 
 To bypass this problem, we follow Refs.~\cite{Jian2020,Bao2020} (see also~\cite{RTN,PhysRevB.100.134203} in the context of random tensor networks) and introduce $k$ replica copies of the system. The average R\'enyi entropy $S_n$ is then written as:
\begin{align}
\avg{S_n} &= \lim_{k\rightarrow 0}\frac{-1}{k(n-1)}\sum_\mindex \left(Z_A(\mindex) - Z_{\emptyset}(\mindex)\right), 
\label{eq: s_n replica copies}
\end{align}
where 
\begin{align}
Z_A(\mindex) &=  \mathbb{E}_U \left[\text{Tr} \left( \left(K_\mindex \ket{\psi_0}\bra{\psi_0} K_\mindex^\dagger\right)^{\otimes nk +1}T_{n,A}^{\otimes{k}}\right)\right]  \nonumber \\
Z_{\emptyset}(\mindex) &= \mathbb{E}_U \left[\text{Tr} \left( \left(K_\mindex \ket{\psi_0}\bra{\psi_0} K_\mindex^\dagger\right)^{\otimes n k +1}\right)\right], \label{eq: Z_A, Z_0}
\end{align}
with $T_{n,A}$ defined in \eqref{eq: boundary DW}, and $\ket{\psi_0}$ is the quantum state of the system at $t=0$. 
As the notation suggests, $Z_{A,\emptyset}$ will correspond to the partition function of an effective statistical model, where $Z_A$ and $Z_{\emptyset}$ only differ with respect to the boundary condition at the top boundary region A (see Appendix~\ref{appendix: stat model technical} Fig. \ref{Fig: stat model}).
We will denote the total number of replicas as $Q=n k+1$ in the subsequent discussion. The additional replica is due to the Born probability factor, which ensures quantum trajectories are weighted appropriately~\cite{Jian2020}.  Also note that since the original non-linear quantity has been converted to a linear quantity defined on $Q$ copies, we are free to do various averages in any order we want.

The rest of this section is organized as follows. We start by giving a very brief overview of the statistical model  for random monitored circuits without any symmetries, following Refs.~\cite{Jian2020,Bao2020}, before moving on to summarize the result for the U(1) symmetric system. We include a detailed and technical derivation of the above model in Appendix~\ref{appendix: stat model technical}. This technical section can be skipped without breaking any continuity. 

\subsection{Statistical model for systems without symmetry}\label{Section: stat model no U(1)}

We briefly review the mapping for random monitored circuits without symmetries to a statistical model~\cite{Jian2020,Bao2020}. We focus on the details required for our subsequent discussion, in particular on the large dimension limit $d\rightarrow \infty$. To calculate Eq.~\eqref{eq: Z_A, Z_0} we need to average over $Q$ copies of the circuit over Haar gates and measurement outcomes (but \emph{not} over measurement locations). Since the random Haar gates are drawn independently, we can individually average $Q$ copies of each gate. 
The combinatorial results of the averaging can be captured as a partition function that can be computed as follows: each unitary gate in the circuit is replaced by a vertex associated with a pair of permutation ``spins" $\sigma_a,\bar{\sigma}_a$, each belonging to the permutation group $\mathcal{S}_Q$. In the $d\rightarrow \infty$ limit, these spins become locked together in a single $\mathcal{S}_Q$ degree of freedom, $\sigma_a$. Vertices from adjacent gates, i.e. those which share an input/output qubit, are connected by links.
The weight associated with a vertex in the partition function is given by $V_a = 1/d^{2Q}$. The weight of the links connecting vertices with elements $\sigma_{a,b}$ is given by
\begin{align}
W_{ab}[\sigma^{-1}_{b}\sigma_a] = \begin{cases}
d^{|\sigma^{-1}_{b}\sigma_a|} & \text{if link (ab) is not measured}\\
d & \text{if link (ab) is measured}
\end{cases},
\end{align}
where $|\sigma^{-1}_{b}\sigma_a|$ is equal to the number of cycles in the cycle decomposition of $\sigma^{-1}_{a}\sigma_b = C_1...C_{|\sigma^{-1}_{b}\sigma_a|}$. Note that the above weights are symmetric under left and right multiplication by elements of $\mathcal{S}_Q$.

We see that in this $d \to \infty$ limit, spins (permutations) connected by unmeasured links are forced to be the same, whereas spins on measured links are effectively decoupled, i.e. a measurements ``break" the links connecting spins, diluting the lattice. This naturally yields a picture of the purification transition in terms of {\em classical percolation} of clusters of aligned permutation  ``spins"~\cite{Skinner2019,Jian2020,Bao2020}, though of course this simple percolation picture is special to $d\rightarrow \infty$: $1/d$ fluctuations are a relevant perturbation to the percolation critical point such that finite $d$ transitions are described by a distinct universality class from percolation~\cite{Jian2020,2021arXiv210703393Z}.


As we saw in Eq.~\eqref{eq: s_n replica copies}, the calculation of $S_n$ requires taking the difference between two partition functions of the model described above but with different boundary condition (see Fig. \ref{Fig: stat model}); in the replica limit, this difference in partition functions becomes equivalent to a difference in free energies $F_{A,\emptyset} = -\log Z_{A,\emptyset}$ (since the partition functions approach unity in the replica limit). The boundary condition for the calculation of $Z_A$ forces a different boundary condition in region $A$, and thus introduces a domain wall (DW) near the top boundary. In the limit $d \to \infty$, the DW is forced to follow a minimal cut, defined as a path cutting a minimum number of unmeasured links (assumed to be unique for simplicity \footnote{DWs are restricted by unitarity to only make certain ``turns'' (See \cite{Zhou2019} for details). E.g, for $p=0$, this leads to a unique DW where the DW follows a ``light cone''. For $p>0$ one can still have many degenerate paths ~\cite{Li2020b}.}). This can be seen as follows: due to the boundary condition in $Z_{\emptyset}$, all vertex elements in $Z_{\emptyset}$ are equal\footnote{Any difference in the vertex elements will lead to the creation of DW which are suppressed as ${\cal O}(1/d)$ and whose contribution goes to zero as $d\rightarrow \infty$.}, and $Z_{\emptyset} = d^{(1-Q)N_m}$, where $N_m$ is the number of measured sites. $Z_A$ would be same as $Z_{\emptyset}$ except for the fact that due to the DW some links contribute different weights to $Z_A$. More precisely, we have $Z_A = d^{(k+1-Q)\ell_{\rm DW}}Z_{\emptyset}$, where $\ell_{\rm DW} = \ell_{\rm DW}(\mathbf{X})$ is equal to the number of unmeasured links that the DW crosses. Since $k+1-Q = (1-n)k<0$ for $n>1$, the DW will follow the path that minimizes $\ell_{\rm DW}$~\footnote{Note that the DW permutation element $(1...n)^{\otimes k}$ has $k+1$ cycles and each cycle can follow an independent path (to the leading order)~\cite{Zhou2019}. However, this subtlety will not change the final result about the minimal cut, but only leads to fluctuations contributing sub-leading logarithmic corrections (for $p>0$).}. Using the expression of $Z_{A,\emptyset}$ in Eq. \eqref{eq: s_n replica copies} and taking the replica limit $k\to 0$, we find $S_n = \ell_{\rm DW}(\mathbf{X})\ln d$, which is valid for each configuration of measurement locations. Averaging over measurement locations, we have 
\begin{equation} \label{eqstatmechnosymm}
\avg{S_n} = \left( \mathbb{E}_{\mathbf{X}} \ell_{\rm DW}(\mathbf{X}) \right) \ln d,
\end{equation}
where $\mathbf{X}$ denotes a configuration of measurement locations (measurement outcomes and Haar gates have been averaged over to get the statistical model). In the language of the statistical model, $\mathbf{X}$ denotes a percolation configuration. $\ell_{\rm DW}$ is the length of the minimal cut from one end of the sub system $A$ to the other end.

In the $d\to \infty$ limit, equation~\eqref{eqstatmechnosymm} is valid for any measurement probability, $p$~\footnote{Note that this description of the $d \to \infty$ differs from that of Ref.~\cite{Jian2020}. There the measurement locations $\sum_{\mathbf{X}}$ were averaged over directly in the partition function, in an annealed way, while we chose here to keep the measurement locations as quenched disorder. Our approach predicts a minimal cut picture consistent with Ref.~\cite{Skinner2019}. We leave a discussion of the validity of the replica trick in this limit to future work. }. For $p=0$, there are no measured links and hence $\ell_{\rm DW} =|A|$, where $|A|$ is the length of subsystem $A$. In fact, $\ell_{\rm DW}$ undergoes a percolation transition at $p_c=1/2$, where $\ell_{\rm DW}$ is extensive in $|A|$ for $p<1/2$, and becomes ${\cal O}(1)$ for $p>1/2$~\cite{Skinner2019}.

\begin{figure*}
\centering
\includegraphics[width=\textwidth]{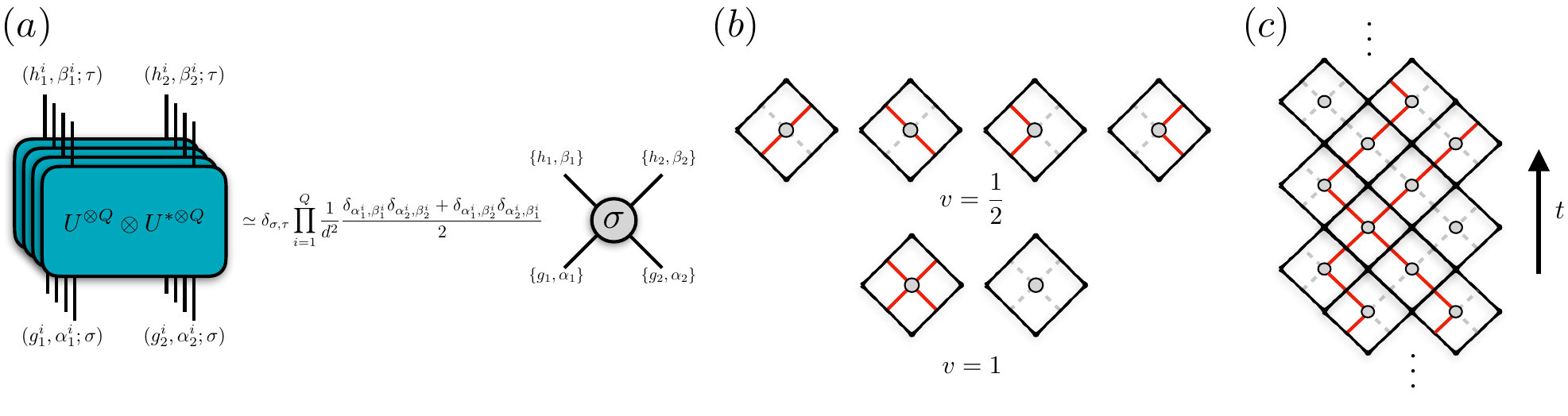} 
\caption{\textbf{Statistical mechanics model.} (a) The average of $\mathcal{U} = U^{\otimes Q} \otimes U^{*\otimes Q}$ over Haar gates is non-zero if and only if the conjugate (bra) replicas are permutations of the non-conjugate (ket) replicas. Hence we can conveniently write each leg in the circuit as a set of $Q$ copies of non-conjugate states combined with a permutation group element (see Eq. \eqref{eq: short hand notation}). In the large $d$ limit, the permutation group elements for in-going and out-going legs become locked together in a single permutation, and the corresponding permutation group element $\sigma$ can be associated with a vertex (one per gate), while the charge states live on links. The U(1) charges ${\alpha, \beta}$  are constrained by charge conservation. (b)The charge dynamics in each replica are given by an effective 6-vertex model with weights $v$, corresponding to a symmetric exclusion process constrained by the measurements and entanglement cut. (c) Example of charge configuration.}\label{Fig: weingarten avg}
\end{figure*}  

\subsection{Statistical Model with U(1) qubits -- Summary} \label{Section: summary of stat model}

Here we provide a concise summary of the statistical model in the case of U(1) circuits, deferring the technical details to Appendix~\ref{appendix: stat model technical}. 

Introducing a U(1) qubit on top of each qudit modifies the above model by introducing an additional degree of freedom (per replica) $\alpha_{ij}$ defined on links, which can take value $0$ or $1$ and correspond to the charge of the U(1) qubits. The weight of each vertex is modified according to the input and output U(1) charges as follows,
\begin{align}
(1,1) &\rightarrow (1,1) \nonumber \\
(0,0) &\rightarrow (0,0) \nonumber \\
(1,0) &\rightarrow \frac{1}{2}\left((1,0) + (0,1)\right) \nonumber \\
(0,1) &\rightarrow \frac{1}{2}\left((1,0) + (0,1)\right),\label{eq: vertex rules}
\end{align} 
where the left-hand side denotes the two input charges and the right-hand side denote output charges. The constants before the output spins are the contribution to the vertex weights. For all other configurations of charges, the weight is equal to 0 thereby enforcing charge conservation. These rules can also be seen as a special case of a 6-vertex model where the states $0,1$ denote two species of links and the weight of the vertex depends on the configuration of the links around the vertex; see Fig.~\ref{Fig: weingarten avg}. Alternatively, those weights can be interpreted as describing hard-core random walkers (symmetric exclusion process), where each state ``$1$'' corresponds to a walker (solid link), with the number of walkers being conserved as a function of  time (vertical direction in the statistical mechanics model). 

We cannot directly average over the measurement outcomes of the U(1) qubits due to the non-local nature of the vertex weights. Hence, we only write a statistical model for a given set of measurement locations $\mathbf{X}$ and outcomes ${\cal M}(\mathbf{X})$ for the U(1) qubits; we collectively denote this set by $\mindex$, as above. The charges in the statistical model at broken links of the percolation sample are pinned by the measurement outcome of the qubit on that link. In other words, for a given configuration $\mindex$, all measured links (broken links in the percolation cluster) carry a fixed value of the local charge $0$ or $1$ determined by the measurement outcome of the qubit, which is fixed in $\mindex$.
 
The statistical model is then given by
\begin{align}
Z(\mindex) = \sum_{\{\alpha\}} \prod_{i\in \text{vertices}} V_i(\{\alpha\}), \label{eq: partition function full model}
\end{align}
where the sum over $\{\alpha\}$ denotes the sum over the set of charges $\alpha$ on all links, $V_i$ is the 6-vertex model weight corresponding to the rules~\eqref{eq: vertex rules},  $\mindex$  represents a percolation configuration combined with a set of values of pinned charges on broken links, corresponding to the measurement outcomes of the qubits on those links. This statistical model has a straightforward physical interpretation: it counts histories of the charge degrees of freedom compatible with a given set of measurement locations and measurement outcomes. 

To calculate $S_n(\mindex)$, we first need to find the minimal cut in the percolation configuration. Recall that $\ell_{\rm DW}$ is the number of unbroken links (not measured) along the cut. There are $2^{\ell_{\rm DW}}$ different charge configurations along the cut; we denote this set of different configurations by ${\{\beta_{\rm DW}\}}$. From the partition function~\eqref{eq: partition function full model}, one can straightforwardly compute the probability of finding configuration $\beta_{\mathrm{DW}}$ along the minimal cut. We denote this $p_{\beta_{\mathrm{DW}}}$. 
%
Taking the replica limit exactly (see Appendix~\ref{appendix: stat model technical}), we find that the R\'enyi entropy is given by
\begin{align}
S_n(\mindex) = \frac{-1}{n-1} \ln \left(\sum_{\{\beta_{\rm DW}\}} p_{\beta_{\rm DW}}^n\right) + \ell_{\rm DW}\ln d . \label{eq: final exp for s_n}
\end{align}
The entropy $S_n$ averaged over all trajectories is then given by,\begin{align}
[S_n] = \sum_\mindex Z(\mindex) S_n(\mindex), \label{eq: averaged s_n}
\end{align}
where $Z(\mindex)$ in Eq. \eqref{eq: partition function full model} can be interpreted as some effective Born probability for observing the trajectory $\mindex$, where unitary gates have been averaged over. In particular, note that $\sum_\mindex Z(\mindex) = \mathbb{E}_{\mathbf{X}} \sum_{\lbrace {\cal M}(\mathbf{X}) \rbrace} Z(\mindex) = 1$.

Note that the second term in Eq. \eqref{eq: final exp for s_n} is the entropy of a pure qudit system. We thus interpret the first term as coming from the qubit sector and treat it as the qubits' contribution to the entanglement entropy. This first term also has an appealing physical interpretation as the classical R\'enyi entropy of qubit configurations along the minimal cut. This is a special feature of the $d \to \infty$ limit. 
From now on, we will use $S^T_n$ to denote total entropy of the qubits and qudits in 
\eqref{eq: final exp for s_n} 
$S_n^d$ for the contribution to the entropy from the qudit sector alone, and $S_n = S^T_n-S^d_n$ which is equal to the first term in~\eqref{eq: final exp for s_n}.

While this expression can, in principle, be computed using Monte Carlo sampling with \emph{no sign problem}, for the one dimensional systems considered here, we find it more convenient to use a disordered transfer matrix
to evolve the initial state up to some time $t$. Specifically, we fix $\mindex$ by randomly generating a percolation configuration, and use the vertex rules described in \eqref{eq: vertex rules} to evolve the system in time. At each broken link (measured qubit) encountered in the evolution, we choose the outcome of the measurement (and hence the fixed value of the charge degree of freedom on that link) with probability equal to the Born probability. This is equivalent to a Monte Carlo sampling for the probability distribution given by $Z(\mindex)$ in Eq.~\eqref{eq: partition function full model}. Many samples are generated and for each sample we calculate the probability distribution $\{p_{\beta_{\rm DW}}\}$. Any physical quantity is then calculated as $\avg{O} = \frac{\sum\displaylimits_{i=1}^{N_s} O(\mindex)}{N_{s}}$, where $N_s$ is the number of samples generated. 

We remark that, in addition to the direct simulation of the transfer matrix techniques we employ in this work, it could also be interesting to investigate further the scaling of the transition using tensor network techniques applied to the transfer matrix of the constrained 6-vertex model~\cite{SCHOLLWOCK201196}.

\section{Numerical results from the statistical mechanics model} \label{statmech_numerics}
In this section, we present numerical results for the U(1) statistical mechanics of constrained symmetric exclusion process described in the previous section, valid in the $d \to \infty$ limit. Unless otherwise stated, we focus on the contribution of the qubit to entanglement, and ignore the qudit contribution $\ell_{\rm DW} \ln d$ which is entirely given by classical percolation physics. We first present late time ($t\sim L$) entanglement data, and present evidence for the existence of the charge-sharpening transition occurring for $p_{\#} = 0.315 \pm 0.01 < p_c = \frac{1}{2}$. We also analyze the time dependence of the R\'enyi entropies, and show that they all grow linearly in time for any $p>0$, in sharp contrast with the $p=0$ behavior.

\subsection{Charge-sharpening transition}\label{section: steady state}

\begin{figure*}
\centering
\includegraphics[width=\textwidth]{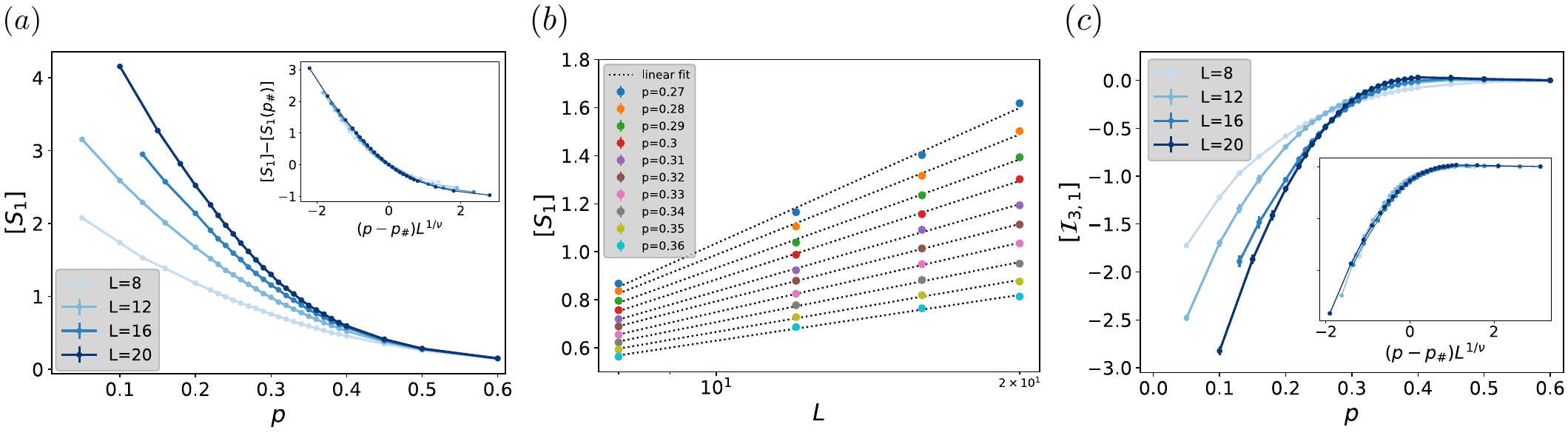}
\caption{\textbf{Entanglement transition in the statistical model qubit contribution.} (a) Plot of the qubit entanglement entropy $S_1$ vs $p$. The \textit{inset} shows the finite size collapse using the ansatz discussed in the main text, with $\nu \sim 1.3$ and $p_{\#}\approx 0.29$. (b) Plot of $S_1$ vs $L$ near the critical point. We find that at the critical point $S_1$ grows logarithmically with $L$. (c) Tripartite mutual information $\mathcal{I}_{3,1}$ vs $p$, showing a crossing around $p_{\#}\approx 0.29$. The \textit{inset} shows the finite size collapse using the same correlation length exponent as in (a). 
{Error bars on the data points are obtained from the error on the mean by computing the standard deviation.}
}\label{Fig. n=1 entropy plot}
\end{figure*}

In the statistical model, the total  entanglement entropy of the subsystem $A$, $S^T_n$, depends on the minimal cut which undergoes a percolation transition at $p_c= \frac{1}{2}$; for $p<p_c$, the length of the minimal cut scales with $L_A$ while for $p>p_c$ the measurement locations percolate and $\ell_{\rm DW}$ becomes ${\cal O}(1)$. Clearly the total entanglement entropy follows the area law for $p>p_c$, and is extensive (and dominated by the qudit contribution) for $p<p_c$. 
As discussed below \eqref{eq: averaged s_n}, $S^T_n$ is given by the sum of two contributions from the qudit and qubit sectors, respectively. In what follows, we will focus on the qubit contribution 
$S_n = S^T_n - S^d_n$, and argue that this quantity undergoes an entanglement transition from volume law to area law for $p=p_{\#}$. We will show that this entanglement transition from the qubit sector coincides with a charge-sharpening transition, which can also be diagnosed in a scalable way using a local ancilla probe, as in Sec.~\ref{ednumerics}.  

\subsubsection{Entanglement transition in the qubit sector}

In this section we look at the R\'enyi entropies $S_n$ at long times, $t>4L$ as a function of $p$. We consider the qubit initial state  $\ket{\psi_0} =\left(\frac{\ket{0}+\ket{1}}{\sqrt{2}} \right)^{\otimes L}$.
To study the behavior of $S_n$ for $p>0$, we numerically run the statistical model \eqref{eq: final exp for s_n} and calculate the half system $S_n$ by averaging $S_n$ over various time steps in the interval of $4$ for $t>4L$. We present results for the $S_1$ in Fig. \ref{Fig. n=1 entropy plot}. 

In analogy with the non-symmetric measurement transition~\cite{Skinner2019,Li2019}, we use the following scaling ansatz for $S_n$ 
\begin{align}
\left[ S_n \right] - S^c_n = f(L/\xi),\label{x}
\end{align}
where $\xi \sim (p-p_{\#})^{-\nu_{\#}}$, and $S^c_n = \left[ S_n(p_\#) \right] \sim \alpha_n \ln L $. 
%
Using both the entanglement entropy scaling and tripartite mutual information as in Sec.~\ref{ednumerics}, we find that the qubit contribution shows an entanglement transition from volume-law to area law at a critical value $p_{\#}$ less than $p_c=1/2$. Finite size collapses are compatible with $p_{\#} = 0.3 \pm 0.02$ and $\nu_{\#} = 1.3\pm 0.2$. We emphasize that this entanglement transition occurs {\em inside} the entangling phase of the total system (the qudit contribution obeys a volume-law scaling in this regime), and occurs only as a subleading contribution to the total entanglement entropy. 

From the point of view of the statistical mechanics model, this transition is especially surprising, as it indicates that the entropy~\eqref{eq: final exp for s_n} of the charge degrees of freedom along the minimal cut does not scale with its length for $p>p_\#$. Instead, our numerical results indicate that measurements are enough to constrain most charges along the cut, so the charges are almost completely ``frozen''~\footnote{Frozen in the sense that there is not much quantum superposition of different charge configurations.} by the measurements near the minimal cut.

\begin{figure*}
\includegraphics[width=\textwidth]{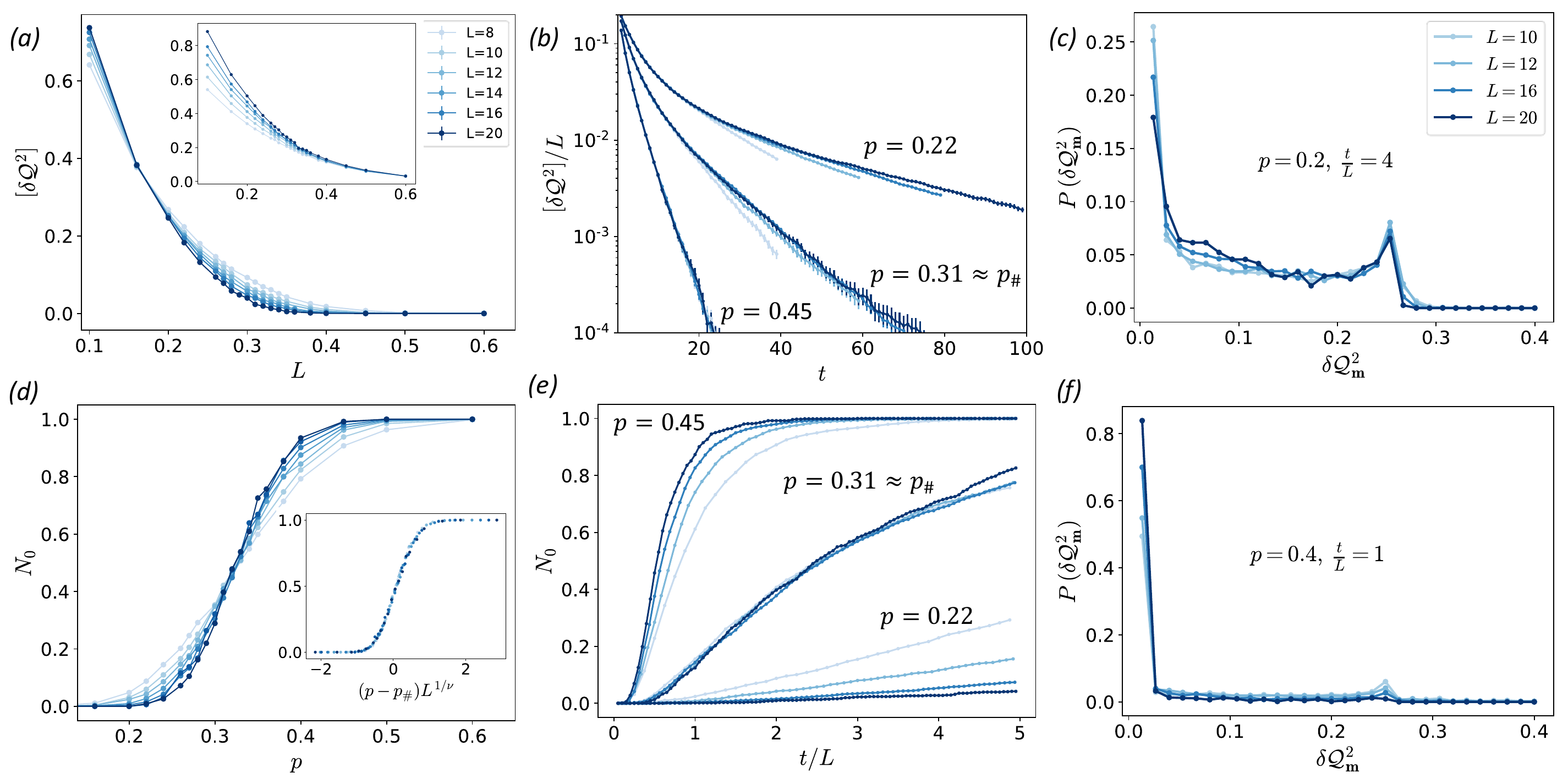}
\caption{\textbf{Charge-sharpening transition in the statistical mechanics model.} \textbf{(a)} Plot of $\left[ \delta\mathcal{Q}^2 \right]$ vs $p$ plotted for $t/L=2$. We find that with increasing $L$ the value of $\left[ \delta\mathcal{Q}^2 \right]$ approaches zero for $p>p_{\#} \approx 0.3$. This is consistent with the entanglement transition in the qubit sector where we observed area-law scaling for $\langle S_n\rangle$ for $p>p_{\#}$. Inset: charge variance in half of the system. \textbf{(b)} Charge variance $\left[ \delta\mathcal{Q}^2 \right]/L$ vs time $t$. For $p\lesssim p_\#$, $\left[ \delta\mathcal{Q}^2 \right]/L$ decays exponentially with a decay rate decreasing with $L$ . For $p>p_{\#}$, the decay rate is same for all $L$ suggesting that $\langle \sigma^2 \rangle/L$ goes to zero faster with increasing system size (faster in units of $t/L$). \textbf{(c)} Histogram of the charge variance in the charge fuzzy phase. \textbf{(d)} Plot of $N_0$ vs $p$ with finite size collapse in the inset. $N_0$ was calculated at $t/L=2$. We find excellent collapse for $p_{\#}=0.315$ and $\nu=1.3$. \textbf{(e)} Time evolution of $N_0$. We clearly see a reversal in trend with system size $L$ around $p_{\#}\approx 0.31$. At the transition $p=p_{\#}$ we find that $N_0\sim h(t/L)$, with $h(x)$ some scaling function, consistent with a dynamical exponent $z=1$. \textbf{(f)} Histogram of the charge variance in the charge sharp phase. The peak at $0.25$ is due to trajectories with superposition of two charge sectors $\mathcal{Q}$ and $\mathcal{Q}+1$. The peak is stronger, and more stable, in the fuzzy phase than in the sharp phase. Error bars in \textbf{(a),(b)} are obtained {from the error of the mean} using standard deviations.}\label{Fig. spin sharpening}
\end{figure*}

\subsubsection{Charge-sharpening transition}\label{sec: charge sharpening stat mech}

Following Sec.~\ref{ednumerics} we probe charge-sharpening by following the dynamics of 
%
the single-trajectory charge variance $ \delta {\cal Q}^2_{{\mathbf{m}}}$ starting from an initial pure state that is a superposition over charge sectors. 
We first discuss the average of this quantity over all trajectories. 
We compute this quantity using the statistical model and plot $ [ \delta {\cal Q}^2 ]  $ as a function $p$ in Fig.~\ref{Fig. spin sharpening}.a. We see that for $p\gtrsim p_{\#} \approx 0.3$, which is the threshold of the area law phase in the qubit sector, $  [ \delta {\cal Q}^2 ] /L$ goes to zero exponentially as a function of time in a way that is independent of $L$. This implies that the time scale $t_\#$ for charge sharpening for $p > p_\#$ (defined as the time it takes for the charge variance to reach a given small value $\epsilon$) scales logarithmically with system size. In contrast, for $p<p_\#$, this charge sharpening time scales as $t_\# \sim L$ (see Appendix~\ref{app:chargesharpeningdynamics}). Fixing $t=2L$, $ [ \delta {\cal Q}^2 ]$  behaves as an order parameter for the charge sharpening transition, coinciding with the entanglement transition in the qubit sector described in the previous section. We observe the same behavior in the bipartite charge variance. 

To extract $p_\#$ it is useful to analyze a quantity that has a discontinuity at the transition. 
To this end, we 
consider $N_0$, the fraction of trajectories with $  \delta {\cal Q}^2_{{\mathbf{m}}}  < \epsilon$ for a given threshold $\epsilon$, as in Sec.~\ref{ednumerics}. 
We check that the results does not depend on $\epsilon$ for small enough values. We plot this quantity in~Fig. \ref{Fig. spin sharpening}.b and find a crossing around $p_{\#} = 0.31$. Note that for all $L$, we chose $\epsilon$ to be small enough so that $N_0$ counts only configurations where the charge is essentially perfectly sharp to numerical accuracy. Defined in this way, $N_0$ approaches 0 in the fuzzy phase, while it goes to 1 in the sharp phase. If we increase the threshold $\epsilon$, instead, we find that $N_0$ behaves more like an order parameter, being fixed to $N_0=1$ in the sharp phase and continuously decreasing in the fuzzy phase. 
It is possible that that the above transition in terms of the fraction of \emph{exactly sharp} trajectories, $N_0$, may be special to the case of perfectly projective measurements, as any slight weakening of the measurements would allow some non-zero quantum fluctuations of charge to persist in a finite space-time volume.
Nevertheless, the transition in $N_0$ provides an upper bound for the ``true" sharpening transition, and can, for example, establish whether the true sharpening transition resides within the volume law phase (for both qubits and qudits). We further explore these questions and the properties of the sharpening transition in a future work~\cite{FieldTheorySharpening}, where we give evidence that the $N_0$-sharpening transition corresponds to a percolation of exactly-sharp regions that occurs within the true charge-sharp phase.

Using a scaling ansatz $N_0 = f\left((p-p_\#)L^{1/\nu_{\#}}\right)$, we find the best collapse for $\nu=1.3\pm 0.15$, consistent with the entanglement data of the qubit. 
We also look at the evolution of $N_0$ with $t/L$ in Fig.~\ref{Fig. spin sharpening}. We find that $N_0$ goes to $1$ for all $p$ at long times but the rate of increase of $N_0$ decreases with $L$ for $p<p_\#$ and increase with $L$ for $p>p_\#$, while remaining constant for $p=p_\#$. We check that the exponent $\nu$ and the critical probability $p_\#$ do not vary much with the time chosen for calculating $N_0$ as long as it is not too large. The crossing value of $N_0$ tends to increase with increasing $t/L$: We focus here on the regime where the thermodynamic limit is taken first so $t/L$ is ``small'' (in practice, $t/L=2$ is small enough to obtain stable results).  We thus conclude that the volume- to area-law transition of $[ S_n ]$ in the qubit sector can be interpreted as a charge sharpening transition wherein starting from a mixed superposition of all charge sectors, the measurements collapse the wave function to one charge sector for $p > p_\#  = 0.315 \pm 0.01$.

\begin{figure}
\includegraphics[width=0.45\textwidth]{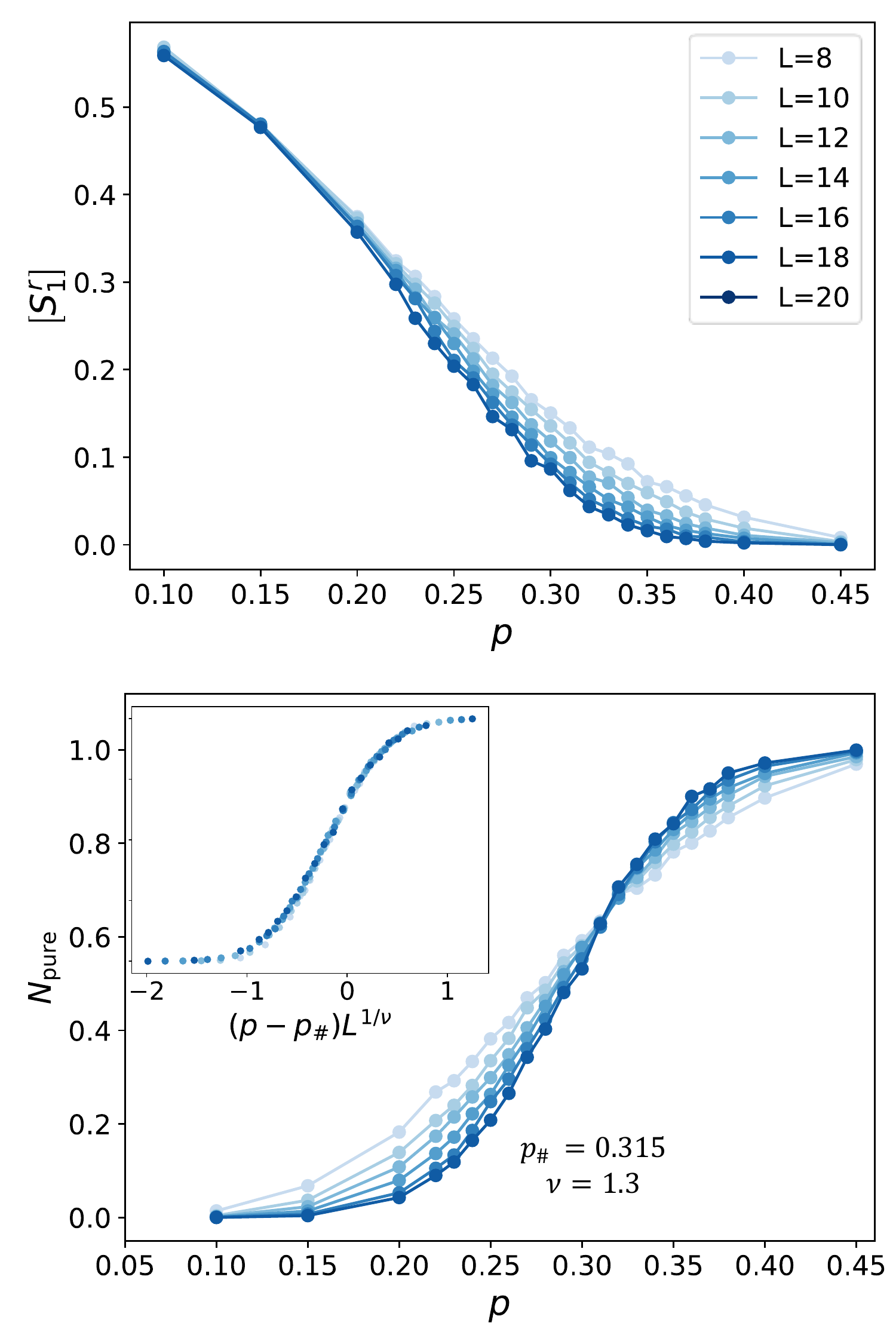}
\caption{{\bf Ancilla probe in the statistical mechanics model. } Top:  entanglement entropy of the ancilla qubit, which behaves as an order parameter for the charge-sharpening transition.
Bottom: Finite size scaling of the number of trajectories where the ancilla qubit is purified $N_\text{pure}$, probing the charge-sharpening transition. 
}\label{Fig. ancilla plot}
\end{figure}

\subsubsection{Local ancilla probe}

As in Sec.~\ref{ednumerics} for the qubit-only ($d=1$) model,  we now present a scalable probe of the charge-sharpening transition by entangling a reference ancilla qubit to different charge sectors $\ket{\Psi_0} = \ket{\psi_{\mathcal{Q}}}\ket{0}+\ket{\psi_{\mathcal{Q}-1}}\ket{1}$. Our numerical protocol is identical to that of Sec.~\ref{ednumerics} (Fig. \ref{Fig. ancilla plot}). Those results are obtained by taking the minimal cut  to be always at the link connecting the ancilla to the system: this is correct in the thermodynamic limit below the percolation threshold $p<p_c=1/2$, and removes spurious finite size effects due to percolation physics.
Our results for this quantity are qualitatively different from the $d=1$ model of Sec.~\ref{ednumerics}, which showed a possible crossing in that quantity, while we observe here a behavior consistent with that of an order parameter for the charge-sharpening transition. This difference might be due to different ``magnetization'' exponents $\beta$, with the $d=1$ model being closer to percolation.
Analogous to the $N_0$ quantity defined above, we introduce $N_\text{pure}$ which is equal to number of trajectories where the reference qubit is purified. We observe a crossing $N_\text{pure}$ near $p_\#\approx 0.315$ and plot the finite size collapse in Fig.~\ref{Fig. ancilla plot}. 

To conclude this section we compare our results for the charge sharpening transition in the two limits of $d=1$ and $d\rightarrow\infty$. In both cases we have found Lorentz invariant critical points with $z=1$ to within numerical accuracy. In the limit of $d\rightarrow\infty$ we have a correlation length exponent $\nu_{\#} = 1.3\pm 0.15$, which is consistent with the percolation universality class that is also found in the qudit sector at $p_c=1/2$. Whereas in the limit of $d=1$ we have $\nu_{\#}\approx 2.0$, which points to a unique universality class that is distinct from both the limit of $d\rightarrow\infty$ and  the entanglement transition at $p_c$.

\subsection{Entanglement dynamics}\label{sec: time evolution}

\begin{figure}
\includegraphics[width=0.5\textwidth]{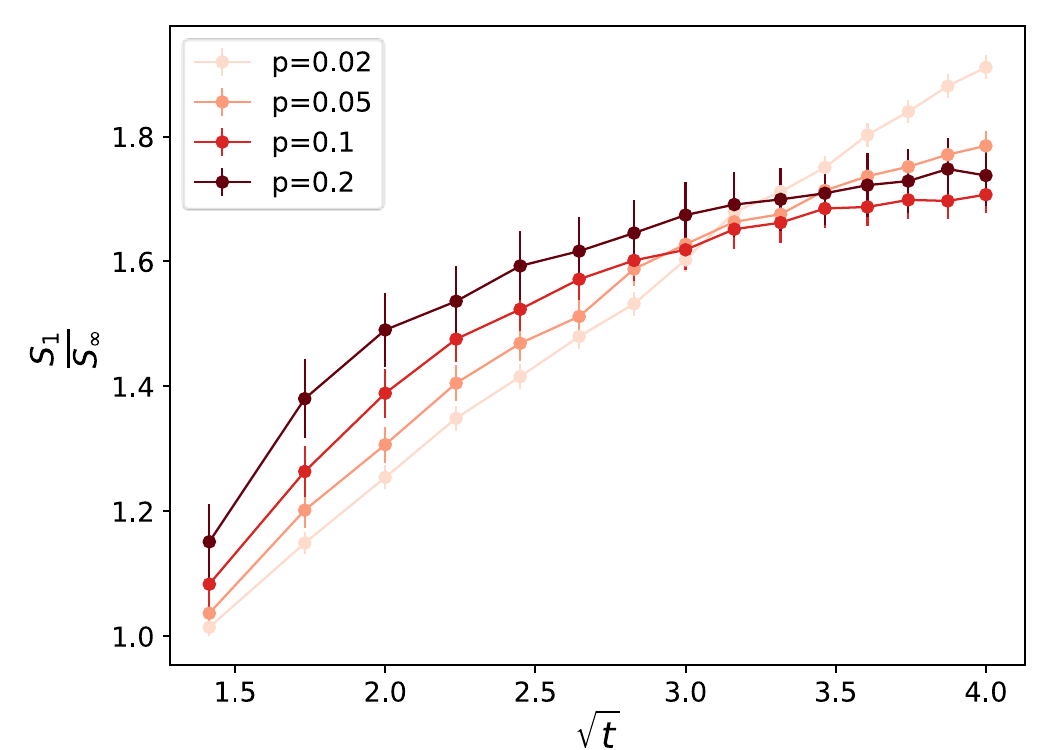}
\caption{\textbf{Entanglement dynamics in the statistical mechanics model.} Ratio $R_s = \frac{S_1}{S_\infty}$ vs $\sqrt{t}$ for $L=12$. We find that $R_s$ saturates implying that $S_\infty$ and $S_1$ are growing at the same rate $\sim t$. As expected, the saturation time is longer at low $p$. Error bars are from error in mean using standard deviations and standard error propagation.}
\label{Fig. time evolution plot}
\end{figure}

Finally, we briefly turn to the dynamics of the R\'enyi entropies $S_n(t)$ using the statistical model. As before, $S_n$ is the contribution of the U(1) qubits to the entanglement entropy: the total entropy $S_n^T$ always grows linearly for $p<p_c=1/2$ due to the qudit sector. The results below should  be interpreted as sub-leading corrections to the growth of the total entanglement entropy arising due to the slow dynamics of the U(1) qubits.

In the absence of measurements, the statistical mechanics model predicts that all R\'enyi entropies scale diffusively $S_{n>1} \sim \sqrt{t}$, whereas $S_{n=1} \sim t$, in agreement with earlier results~\cite{Rakovszky2019,Huang2019a,Zhou2020} (see Appendix~\ref{app:dynamics}).
As argued in Sec.~\ref{overview}, in the presence of measurements, the ``dead regions'' responsible for this unusual behavior survive only until a time $t \sim p^{-2/3}$ in typical trajectories. At long times, the overlap of the wavefunction with dead regions is zero, and we expect the trajectory-averaged R\'enyi entropies $[S_n]$ to grow linearly in time for all $p > 0$.

To confirm this, we plot the ratio $R_s = [S_1]/[S_\infty]$ in Fig. \ref{Fig. time evolution plot}. The 
average von Neumann entropy $[S_1]$ is expected to grow linearly for all $p$. The quantity $R_s$ is thus a measure of the growth of $[S_\infty]$: if $[S_\infty]$ were to increase as $\sqrt{t}$, then we would expect $R_s$ to grow as $\sqrt{t}$ too. This is indeed what we observe at $p=0$. At higher $p$ we find that $R_s$ saturates to a constant value implying that $[S_1] \sim [S_\infty] \sim t$, in agreement with our general argument. Other observables confirming this scaling are presented in Appendix~\ref{app:dynamics}.

\section{Discussion} \label{discussion}		

In this work, we have studied measurement-induced phases and phase transitions in monitored quantum circuits with charge conservation.
We argued that measurements can have a dramatic effect on entanglement growth. While all R\'enyi entropies with index $n>1$ grow diffusively in the absence of measurements, for any $p>0$, the effect of these rare regions are washed out by measurements leading to ballistic scaling $S_n \sim t$ at long times. 

Whereas, in the absence of symmetry, there can only be two possible steady-states, entangling or purifying, charge conservation enriches this dynamical phase diagram.
We uncovered a new type of charge-sharpening transition that separates distinct entangling phases. Even as the dynamics remain scrambling and lead to a volume-law entangled state, the U(1) charge can either be ``fuzzy'' or ``sharp'' depending on the rate of measurements.  This charge-sharpening transition occurs at a critical measurement rate $p_\#$ that is generically smaller than $p_c$, corresponding to the purification transition. 
This new transition is also fundamentally different from the purification entanglement transition, as for any $p>0$, the charge will eventually become sharp with exponentially small corrections
for $t \gg t_{\#}\sim L$ (up to logarithmic corrections) 
for a system of size $L$, 
whereas the purification time diverges exponentially in the system size in the entangling phases. 
The sharpening time scale for U(1) circuits is also parametrically much faster than that in ${\mathbb Z}_2$ symmetric circuits~\cite{2021arXiv210209164B} (linear vs exponential), highlighting the fundamental difference between scrambling of U(1) and ${\mathbb Z}_2$ symmetric modes. Thus the measurement-induced phases inside the volume law for U(1) systems are conceptually very different than those in ${\mathbb Z}_2$ symmetric systems~\cite{2021arXiv210209164B}. The type of sharpening transitions studied here are unique to systems with diffusive modes.

We presented evidence for the existence of this transition using both exact numerical results in a symmetric qubit model ($d=1$), and from the numerical analysis of an emergent statistical mechanics model describing the evolution of charged qubits coupled to large qudits ($d \to \infty$). 
For the model in the $d\rightarrow \infty$ limit, the correlation length exponent $\nu$ of the charge-sharpening transition is consistent with that of percolation. In contrast,
in the qubit-only model we showed that the charge-sharpening correlation length exponent is 
distinct from the that found for the entanglement transition
with 
$\nu_{\#} \approx 2$. Understanding the critical properties of this transition represents a clear challenge for future works. A conceivable scenario could be that the charge-sharpening and entanglement transitions could merge into a single transition below a critical qudit dimension, $d<d_c$. Establishing on firmer grounds the existence of a distinct charge-sharpening transition would also be an important task for future works. 


The statistical mechanics model is also an important step in the understanding of symmetric monitored circuits. 
We were able to take the replica limit analytically which is a crucial step to uncover key properties of measurement-induced phase transitions and is often the most daunting challenge in the studies of monitored circuits~\cite{2021arXiv210209164B}. We find that the contribution of the U(1) degrees of freedoms to the Renyi entropies is related to the entropy of local charge fluctuations along the minimal cut {(eq. \eqref{eq: final exp for s_n})}. Though this mapping is restricted to the $d\rightarrow \infty$ limit, since the permutation degrees of freedom are gapped in the volume-law phase we do not expect them to change the general structure of the phase diagram or the universality class of the sharpening transition for finite $d$. {This is an important distinction with the  $d\to \infty$ percolation limit of the entanglement transition~\cite{Bao2020,Jian2020}, where $1/d$ corrections are relevant and completely modify the simple percolation picture. }

The stat mech approach can also be readily generalized to arbitrary Abelian symmetries (Appendix \ref{app:generalizations}) thus providing a controlled platform for future studies of symmetric circuits, for example ${\mathbb Z}_n$ circuits.
{We emphasize that the change of perspective in treating the measurements as quenched disorder rather than annealed~\cite{Bao2020,Jian2020} is crucial in incorporating symmetric/constrained degrees of freedom in the stat mech model. Recently, this change of perspective turned out to also be useful in other contexts, for example in the study of negativity in non-symmetric circuits~\cite{Weinstein2022}}.

{In the models we consider in this paper, the measurements kill both entanglement and charge fluctuations. This is especially natural for the qubit-only model which is perhaps physically more relevant. In principle, we can consider various modifications of this simple model. For example, in case of the qudits model, one could consider different rate/strength of measurements for the U(1) and neutral degrees of freedom, or measurement-only models where measurements compete against creating and destroying charge fluctuations. A detailed study of such models are left out for future work, although we do not expect any qualitative change to the physics of the charge-sharpening transition discussed in this paper.}

We mostly focused on the global properties of charge dynamics, and defer local properties of the steady-state to future work~\cite{FieldTheorySharpening}. An effective field theory description of the statistical mechanics model introduced above predicts that the local sharpening transition is in a Kosterlitz-Thouless universality class (KT)~\cite{FieldTheorySharpening}. In this picture, the fuzzy phase corresponds to the quasi-long range order and the charge sharp phase is the symmetric phase~\cite{FieldTheorySharpening}. A proper analysis of the replica limit is however crucial to uncover the peculiar nature of this transition, including the dynamical properties distinguishing the phases -- see~\onlinecite{FieldTheorySharpening}. 
It would be interesting to look for signatures of such KT scaling in the qubit model ($d=1$), even though KT criticality is notoriously hard to study in finite size numerics.

The  conservation law has not affected the universality class of the entanglement transition in the limit of $d\rightarrow \infty$. Whereas, in the limit of $d=1$ we have shown that the log-scaling of the Renyi entropy at criticality $S_n \sim \alpha(n)\log L$, has an $\alpha(n)$ that is clearly distinct from the transition with Haar random gates~\cite{Zabalo2020}, which implies the (boundary) universality class is distinct in the presence of a conservation law. Interestingly, we have found that $\nu\approx 1.3$, which is not sensitive enough to discern between percolation, stabilizer dynamics, and the Haar universality class. It will be interesting in future work to probe other  critical exponents of the entanglement transition with a conservation law to discern other uniques properties of this transition. 


It would also be interesting to extend our results to other symmetry groups or kinetic constraints. Our results can be readily generalized to arbitrary Abelian groups (see Appendix~\ref{app:generalizations}). As we discuss in Appendix~\ref{app:generalizations}, if one assumes the  existence of charge-sharp phases for other symmetry groups and spatial dimensionalities, standard duality relations~\cite{fisher2004duality} immediately imply the existence of monitored random circuit classes that exhibit volume-law entangled phases with symmetry-breaking, symmetry-protected topological, and intrinsic topological orders in a typical trajectories. Such trajectory-ordered but volume law entangled phases are clearly forbidden in any equilibrium or closed-system dynamical setting, and are a new feature of non-unitary open system dynamics.

Moreover, it is clear by now that new types of dynamical phases can be obtained in the steady state of monitored quantum circuits, from the combination of different competing (non-commuting) measurements and unitary dynamics~\cite{PhysRevResearch.2.023288,Lavasani2020,Ippoliti2020,Sang2020}. The full phase structure allowed by the microscopic symmetry group and the dynamical symmetries of such monitored quantum circuit appears to be particularly rich~\cite{2021arXiv210209164B}, and remains largely unexplored. We expect non-Abelian symmetries to be especially interesting, as they  could lead to fundamental constraints on the entanglement structure of the steady-state, as in the case of many-body localized systems~\cite{PhysRevB.94.224206}. 

\emph{Acknowledgments.} We thank Michael Gullans, David Huse, Chaoming Jian, Vedika Khemani, and Andreas Ludwig for useful discussions and collaborations on related projects. We acknowledge support from NSF DMR-1653271 (S.G.), NSF DMR-1653007 (A.C.P.), the Air Force Office of Scientific Research under Grant No. FA9550-21-1-0123 (R.V.), and the Alfred P. Sloan Foundation through Sloan Research Fellowships (A.C.P., J.H.P., and R.V.).
A.\ Z.\ and J.H.P.\ are partially supported by Grant No. 2018058 from the United States-Israel Binational Science Foundation (BSF), Jerusalem, Israel. A.\ Z.\ is partially supported through a Fellowship from the Research Discovery Informatics Institute.
The Flatiron Institute is a division of the Simons Foundation.


\begin{appendix}
\section{Mapping to the statistical model with U(1) qubits} \label{appendix: stat model technical}
In this appendix, we present a in detail discussion of the mapping to the statistical model, and derive Eq.~\eqref{eq: final exp for s_n} in the limit $d \to \infty$. To evaluate the quantities in Eq. \eqref{eq: Z_A, Z_0}, we need to calculate the average of $\mathcal{K}\equiv K_\mindex^{\otimes Q}\otimes K_\mindex^{\dagger \otimes Q}$, corresponding to $Q$ copies of the random circuit. 
Each unitary gate in $\mathcal{K}$ is repeated $Q$ times and since they are drawn independently we can individually average them over the random unitary ensemble. 
Let us denote the tensor product of $Q$ copies of a gate by $ \mathcal{U} \equiv U^{\otimes Q}\otimes U^{\dagger\otimes Q}$. We view $\mathcal{U}$ as a super-operator which acts on two sites with each leg containing $Q$ ket states and $Q$ bra states; let $\ket{g^i\alpha^i}\bra{g^{i*}\alpha^{i*}}$ be a basis where $g^i$ is a basis of the qudit Hilbert space ${\mathbb C}^d$, and $\alpha^i$ is the computational basis for the qubit. The index $i$ labels replicas, and runs from $1$ to $Q$. 
Using standard Haar calculus and Weingarten formulas, we find that the action of $\overline{\mathcal{U}}$ on the above basis is non trivial after averaging if and only if $g^{i*} = g^{\sigma(i)}$ and $\alpha^{i*} = \alpha^{\sigma(i)}$, where $\sigma\in \mathcal{S}_Q$ is a permutation. Therefore, we introduce a shorthand notation for writing the relevant members of the basis as 
\begin{align}
(g^i \alpha^i;\sigma) \equiv  \ket{g^i\alpha^i}\bra{g^{\sigma(i)}\alpha^{\sigma(i)}}.\label{eq: short hand notation}
\end{align}
More precisely, each unitary gate in the circuit is replaced by a vertex associated with a pair (corresponding to in- and out-going legs) of permutation ``spins" $\sigma,\bar{\sigma}$, each belonging to the permutation group $\mathcal{S}_Q$. In the $d\rightarrow \infty$ limit, these spins become locked together in a single permutation degree of freedom, $\sigma_a$, that we associate with that vertex. Vertices from adjacent gates, i.e. those which share an input/output qubit and qudits, are connected by links in a way that will be explained below. In the large $d$ limit, the weight associated with a vertex in the partition function is given by $V_a = 1/D^Q$, where $D$ is the size of the block of the relevant symmetry sector. We have $D=d^2$ if all incoming and outgoing charges are the same, and $D=2d^2$ otherwise, see eq.~\eqref{eq: Unitary matrix structure}. 

The results for $\overline{\mathcal{U}}$ to leading orders are summarized in Fig.~\ref{Fig: weingarten avg}; the sub-leading corrections are suppressed as ${\cal O}(1/d^2)$ which we will ignore in rest of the paper. 
The factor of $\frac{\delta_{\alpha_1^i \beta_1^i } \delta_{\alpha_2^i \beta_2^i} + \delta_{\alpha_1^i \beta_2^i} \delta_{\alpha_2^i \beta_1^i }}{2}$ in Fig. \ref{Fig: weingarten avg} enforces U(1) charge conservation, and follows from the size of the different blocks in eq.~\eqref{eq: Unitary matrix structure}. In fact, if we view charge $0$ as vacuum and charge $1$ as a particle, then the dynamics of the U(1) degree of freedom can be understood as a hard core random walk these particles, known as the symmetric exclusion process. Alternatively, it can be seen as a special case of the 6-vertex model (see Fig. \ref{Fig: weingarten avg}.b). 
Though we have focused mainly on the case of U(1) symmetry groups, our approach readily extends to other Abelian groups. In Appendix~\ref{app:generalizations} we provide a general derivation for arbitrary Abelian symmetry groups.


\subsection{Link weights}

\begin{figure}
\centering
\includegraphics[scale=0.4]{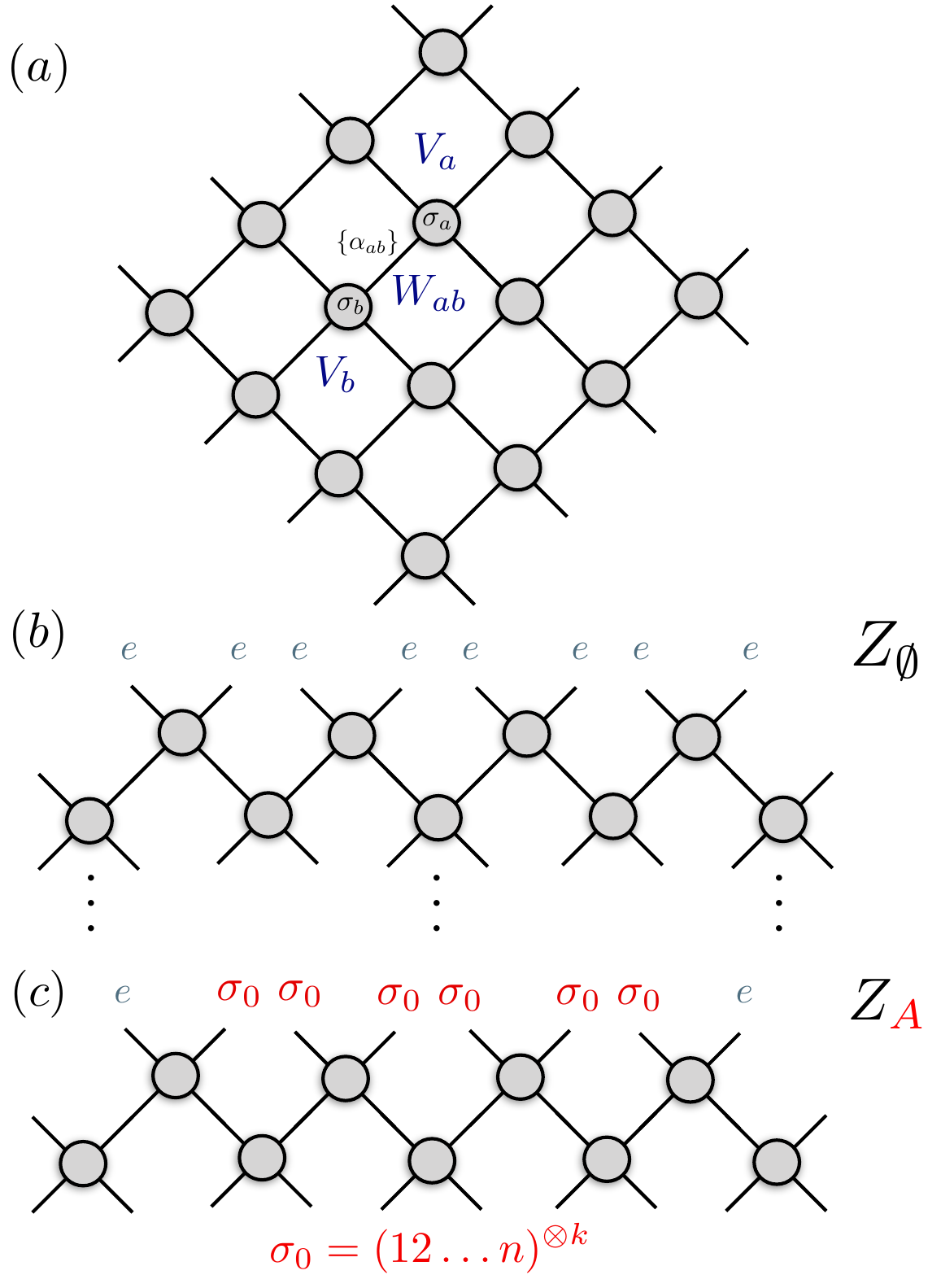}
\caption{{\bf Replicated statistical mechanics model.} (a) The replicated statistical model is defined on a tilted square lattice, with permutation degrees of freedom $\sigma_a \in \mathcal{S}_Q$ living on vertices, and charge degrees of freedom $\lbrace \alpha^i \rbrace_{i=1,\dots,Q}$. The Boltzmann weights have contributions from both vertices $V_a$, see Eq.~\eqref{eq: vertex weight}, and links $W_{ab}$, Eq.~\eqref{eq: link weight}. (b) Fixed boundary conditions in $Z_{\emptyset}$ at the top layer (ending the circuit at a given time $t$), with all permutations fixed to $e$. (c) The partition function $Z_{A}$ differs from $Z_{\emptyset}$ by the boundary condition fixed to $\sigma_0$ in the entanglement interval $A$. This creates a domain wall (DW) that follows a minimal cut in the limit $d\to \infty$.     } \label{Fig: stat model}
\end{figure}

Combining Fig. \ref{Fig: weingarten avg} with the brick wall geometry of the circuit leads to a model described on a square lattice as shown in Fig. \ref{Fig: stat model}. Each vertex has an element from the permutation group $\mathcal{S}_Q$ and each link has $Q$ copies of the elements of the basis of the local Hilbert space. The vertex weights $V_a$ are given by the rule described in Fig. \ref{Fig: weingarten avg}.b. The link weight $W_{ab}$ has two kind of contributions: 1) due to the presence of domain wall (DW) in the permutation group elements $\sigma_{a,b}$ (DW constraint), 2) the state at the link $\langle ab \rangle$ is being measured (measurement constraint). We describe these constraints in detail in the following.

\textit{1) DW constraint.} 
We first consider a link joining two vertices that we label ``1'' and ``2''. 
%
If we integrate out the qudit degrees of freedom, we find the following weight for the links
\begin{align}
W_{12}(\sigma_2^{-1}\sigma_1) &= \sum_{g_1^i,g_2^i} \text{Tr} \left[(g_1^i \alpha_1^i;\sigma_1)(g_2^i \alpha_2^i;\sigma_2)^\dagger\right] \nonumber \\
&=d^{p}\delta_{C_1}...\delta_{C_p}, \label{eq: DW const}
\end{align}
where $C_1...C_p$ is the cycle structure of the permutation element $\sigma_2^{-1}\sigma_1$, $p$ is the number of cycles in that permutation, and $\delta_{C_i}$ is equal to $1$ if all charge states of the replicas within cycle $C_i$ are the same, and otherwise equal to $0$. Note that we cannot sum over the charges $\alpha^i$ as these depends on the states at neighboring links (see Fig. \ref{Fig: weingarten avg}). 

Intuitively we can interpret the above result as follows. Since the link is shared between vertices with different permutations $\sigma_1$ and $\sigma_2$ then we must have the following constraint
\begin{align}
(g^i \alpha^i;\sigma_1) = (g^i \alpha^i;\sigma_2),
\end{align}
which is true if and only if $\sigma_2^{-1}\sigma_1(\{\alpha_i\}) = \{\alpha_i\}$, and $\sigma_2^{-1}\sigma_1(\{g_i\}) = \{g_i\}$. Let us take the simple case where $\sigma_2^{-1}\sigma_1$ is equal to a transposition, say, $(12)$. We can interpret the above equation as saying that $g_1 = g_2$ and $\alpha_1=\alpha_2$. The $g_1=g_2$ condition will reduce the number of allowed basis qudit states from $d^Q$ to $d^{Q-1}$, and since all qudit contributes equally, the weight of the link in this case will be reduced by a factor $1/d$ due to the reduced configurational entropy in the qudit sector. For the qubit degrees of freedom, we cannot sum over all spins due to the non-local charge conservation constraint. The more general case of $\sigma_2^{-1}\sigma_1\in \mathcal{S}_Q$ follows similarly, giving \eqref{eq: DW const}.
An important thing to note is that each transposition in $\sigma_2^{-1}\sigma_1$ reduces the weight by $1/d$ and the weight is strongest when $\sigma_2 = \sigma_1$, that is when there is no DW -- corresponding to a ferromagnetic interaction. Thus at large $d$ limit, it is expensive to have a DW and the system will remain in an ordered phase unless DW are forced, for example, at the entanglement cut (see Fig. \ref{Fig: stat model}). This will play an important role in the subsequent discussion.

\textit{2) Measurement constraint.} If the link $\langle 12 \rangle$ happens to be measured, then the measurement outcomes are same in all replicas, that is, all copies are acted on with the same projection operator. If the projection operator is denoted as $P = P^q \otimes P^d$, then we have the weight
\begin{align}
W_{12}(\sigma_2^{-1}\sigma_1;P) = \text{Tr} \left[ \linkstate{g^i_2}{\alpha^i_2}{\sigma_2} ^\dagger P^{\otimes{Q}} \linkstate{g^i_1}{\alpha^i_1}{\sigma_1} P^{\otimes Q} \right].
\end{align}
Averaging over all measurement outcomes results in
\begin{align}
W_{12} &= \sum_{s=0,1} \sum_{x=1,...,d} W_{12}(\sigma_2^{-1}\sigma_1;P^q_s\otimes P^d_x), \nonumber \\
& = d\sum_{s=0,1} \prod_i \delta_{\alpha^i,s}, \label{eq:measurement const}
\end{align}
where $\delta_{\alpha^i,s}$ ensures that the charge state of the $i^{\rm th}$ replica is compatible with the measurement outcome $s=0,1$ of the qubit. Whenever a measurement occurs, all $Q$ charges on the corresponding link are constrained to be same. This gives the $\delta$ factor in \eqref{eq:measurement const}. For the qudit sector, this leads to a decrease in configurational entropy from $d^Q$ to $d$. An important observation is that the link weights $W_{ab}$ do not depend on the permutation $\sigma_b^{-1}\sigma_a$, a result of crucial importance for the discussion below.
\newline

\subsection{Replicated model}

Combining all these results we can write a statistical model with the partition function given by,
\begin{align}
Z &= \sum_{\mathbf{m}} Z[\mathbf{m}], \\
Z[\mathbf{m}] &=\sum_{\text{configurations}}\left( \prod_{(ab)\in\text{links}} \prod_{v\in\text{vertices}} W_{ab}(\sigma_b^{-1}\sigma_a) V_v \right), \label{eq: formal Z}
\end{align}
where $\sum_{\mathbf{m}} \equiv \sum_{\{ \mathbf{X} \} } p^{N_\mathbf{X}} (1-p)^{LT-N_\mathbf{X}} \sum_{\lbrace {\cal M}(\mathbf{X}) \rbrace}$, where
$\mathbf{X} $ denotes a configuration of measurement locations, $N_\mathbf{X}$ is the number of links being measured (number of bonds in the percolation configuration $\mathbf{X}$), $L$ is the spatial length of the system, $T$ is the number of time steps; and ${\cal M}(\mathbf{X})$ is the set of qubit measurement outcomes at measurement locations $\mathbf{X} $. The sum over configurations is given by
\begin{align}
\sum_{\text{configurations}} &\equiv  \sum_{\{\sigma_v\}\in \mathcal{S}_Q} \sum_{\{\alpha^1\}=0,1}...\sum_{\{\alpha^Q\}=0,1} \nonumber \\
&\equiv \sum_{\{\sigma_v\}\in \mathcal{S}_Q} \sum_{\{\alpha\}},
\end{align}
corresponding to permutation and charge degrees of freedom in each replica. 
The link weights $W_{ab}$ are given by
\begin{align}
W_{ab}(\sigma_b^{-1}\sigma_a) = \begin{cases}
d^{|\sigma^{-1}_b \sigma_a |-Q} \delta_{C_1}...\delta_{C_{|\sigma_b^{-1}\sigma_a |}} & \text{if $(ab)$ not in $\mathbf{X}$} \\ 
d^{1-Q} \delta_{\alpha^i,s_{ab}} & \text{if $(ab)$ in $\mathbf{X}$} \label{eq: link weight}
\end{cases},
\end{align}
with $s_{(ab)}\in {\cal M}(\mathbf{X} )$ the measurement outcome of the qubit on link $(ab)$.  Finally, the vertex weight $V_v$ is given by
\begin{align}
V_v = \prod_{i=1}^Q \frac{\delta_{\alpha_1^i \beta_1^i } \delta_{\alpha_2^i \beta_2^i} + \delta_{\alpha_1^i \beta_2^i} \delta_{\alpha_2^i \beta_1^i }}{2} = \prod_{i=1}^Q V_v^i, \label{eq: vertex weight}
\end{align}
where $\alpha_{1,2}$ and $\beta_{1,2}$ are incoming and outgoing charges (see Fig. \ref{Fig: weingarten avg}). We note that $V_v$ factorizes over the replicas,  that is, $V_v = \prod_{i=1}^{Q} V_v^i$; this will play an important role in factorizing $Z[\mathbf{m}]$ in the discussion below.

Note that we have integrated out the qudit sector from the model. This was possible due to each qudit on a given link being independent of the values at other links. However, this is not possible for the U(1) sector on account of non-local constraints due to the charge conservation. Importantly, the statistical model $Z(\mindex)$ should be thought of as a quenched disordered model where the measurement locations and outcomes (for the qubit) are quenched ``impurities''; averaged quantities in the original problem have become quenched average in the statistical model. From now on, $\mindex = \lbrace \mathbf{X}, {\cal M} (\mathbf{X}) \rbrace$ will denote the system's quantum trajectory with measurement locations + U(1) measurement outcomes fixed, corresponding to a fixed ``disorder'' realization of the statistical model. 
This is unlike the previous-works on the non symmetric problem where the randomness in the measurement locations were absorbed in the statistical model in an annealed way.

\begin{table}
\begin{tabular}{ |c | c|}
\hline
$\alpha^i_l$ & \text{Charge at link $l$ and copy $i$} \\
\hline
$\{\alpha^i\}$ & \text{Set of charges on all links for copy $i$}\\
\hline
$\{\alpha_l\}$ & \text{Set of charges on link $l$ for all copies}\\
\hline
$\{\alpha\}$ & Set of charges on all links and copies\\
\hline
$\delta^{\{l\}}_{\{\alpha\}}$ & All copies of $\alpha$ on set of links $\{l\}$ are equal\\
\hline
$\delta^{\{l\}}_{\{\alpha\},\{{\cal M}\}}$ & All copies of $\{\alpha\}$ on links $\{l\}$ are equal to $\{{\cal M} \}$\\
\hline
\end{tabular}
\caption{Table summarizing the meaning of various notations used in this appendix. }\label{Table: notation}
\end{table}

\subsection{Replica limit} \label{sec: replica limit appendix}

We now proceed to take the replica limit, and will use various notations summarized in Table \ref{Table: notation}. 

We first focus on the partition function $Z_{\emptyset}(\mindex)$, where the links at the top boundary are restricted to be of the form $\linkstate{g^i}{\alpha^i}{e}$. The permutation identity element $e$ represents the fact that we are tracing over all the system and is equal to the Born probability of observing the particular trajectory $\mindex$. As mentioned above, a DW in the statistical model is suppressed by $1/d^{Q-p}$, where $p$ is the number of cycles in the DW. Thus, the leading order contribution to $Z_{\emptyset}$ comes when all vertex elements being equal to $e$. This simplifies $Z_{\emptyset}(\mindex)$ dramatically as we do not need to sum over the permutation elements. We have
\begin{align}
Z_{\emptyset}&(\mindex) = d^{(1-Q)N_m} \nonumber  \left( \sum_{\{\alpha\} = 0,1} \delta^{\mindex}_{\{\alpha\},{\cal M}(\mindex)}\prod_v V_v \right),
\end{align}
where $N_m$ is the number of measured links, $\delta^{\mindex}_{\{\alpha\},{\cal M}(\mindex)}$ is non-zero and equal to $1$ if and only if the charges $\{\alpha\}$ on the measured sites are equal to the measurement outcomes ${\cal M}(\mindex)$ of the qubit, and $V_v$ is given in \eqref{eq: vertex weight}. Intuitively, we should only sum over charge configurations compatible with the measurement outcomes. 
The partition function can be factorized over replicas to give,
\begin{align}
Z_{\emptyset}(\mindex) = d^{(1-Q)N_m}\left( Z_{\emptyset}^{(1)}(\mindex) \right)^Q,
\end{align}
where $Z_{\emptyset}^{(1)}(\mindex)$ is given by
\begin{align}
Z_{\emptyset}^{(1)}(\mindex) = \sum_{\{\alpha\}} \delta^{\mindex}_{\{\alpha\},{\cal M}(\mindex)} \prod_v V_v^{(1)} . \label{eq: partition function}
\end{align}
The superscript $(1)$ denotes the fact that the quantity is for a single replica.

We can similarly factorize $Z_A(\mindex)$ with the caveat that we now have a minimal cut for the permutation degrees of freedom running through the system (see discussion in Section \ref{Section: stat model no U(1)}). A DW between $e$ and $(1...n)^{\otimes k}$ reduces the link weight by $d^{k+1-Q}=d^{-(n-1)k}$~(\ref{eq: link weight}) and the contribution of the cut to the partition function is thus given by $d^{-(n-1)k\ell_{\rm DW}}$, where $\ell_{\rm DW}$ is the length of the minimal cut. There are $k+1$ cycles in the DW; $k$ cycles of the type $(1...n)$ and the last one being an identity on a single copy. Thus we can factorize $Z_A(\mindex)$ as
\begin{align}
Z_A(\mindex) &= d^{(1-Q)N_m-(n-1)k\ell_{\rm DW}} \nonumber \\
&\times   \sum_{\{\alpha^i\} = 0,1}\delta^\mindex_{\{\alpha\},{\cal M}(\mindex)} \prod_{C_a} \delta^{\rm DW}_{\{\alpha^{C_a}\}}\prod_v V_v,
\end{align}
the where $\delta^{\rm DW}_{\{\alpha^{C_a}\}}$ is non-zero (and equal to $1$) if and only if the charges within the cycle $C_a$ are the same on (unmeasured) links on the minimal cut. We can further factorize the above equation to get,
\begin{align}
Z_A(\mindex) = d^{(1-Q)N_m-(n-1)k\ell_{\rm DW}} \left(Z_A^{(n)}(\mindex)\right)^k Z_{\emptyset}^{(1)}(\mindex),
\end{align}
where
\begin{align}
 Z_A^{(n)}(\mindex) &= \sum_{\{\alpha\}} \delta^{\rm DW}_{\{\alpha\}}  \prod_{i=1}^{n} \left( \delta^{\mindex}_{\{\alpha^i\},{\cal M}(\mindex)} \prod_v V_v^{i} \right) \nonumber \\
 &=	\sum_{\beta_1 ... \beta_{\ell_{\rm DW}}} \prod_{i=1}^n \left( \sum_{\{\alpha^{i}\}} \delta^{\rm DW}_{\{\alpha^i\},\{\beta\}} \delta^\mindex_{\{\alpha^i\},{\cal M}(\mindex)}\prod_v V_v^{i} \right) \nonumber\\
 &\equiv \sum_{\{\beta\}} Z_A\left[\mindex;\{\beta\}\right]^n
\end{align}
with $\delta^{\rm DW}_{\{\alpha\},\{\beta\}}$ non-zero (and equal to $1$) if and only if all copies of charges on the unbroken (not measured) links along the minimal cut are equal to $\{\beta\}$. The superscript $(n)$ denotes the fact that we have $n$ charge copies.
Using the above results and \eqref{eq: s_n replica copies}, we find 
\begin{widetext}
\begin{align}
S_n(\mindex) &= \frac{-1}{n-1}\lim_{k\rightarrow 0} d^{(1-Q)N_m} Z_{\emptyset}^{(1)}(\mindex)\frac{d^{-(n-1)k\ell_{\rm DW}}\left(Z_A^{(n)}\right)^{k} - \left(Z_{\emptyset}^{(1)}\right)^{n k}}{k} . 
\end{align}
Remarkably, this factorized form allows us to take the replica limit {\em exactly}:
\begin{align}
S_n(\mindex)&= \frac{-1}{n-1} Z_{\emptyset}^{(1)}(\mindex) \ln \frac{Z_A^{(n)}}{\left(Z_{\emptyset}^{(1)} \right)^n} + \ln d\ \ell_{\rm DW}, \nonumber \\
&= \frac{-1}{n-1} Z_{\emptyset}^{(1)}(\mindex) \ln \left( \sum_{\{\beta\}}\frac{Z_A[\mindex;\{\beta\}]^{n}}{\left(Z_{\emptyset}^{(1)} \right)^n} \right) + \ln d\ \ell_{\rm DW},
\end{align}
\end{widetext}
where $\{\beta\}$ represents all possible configuration of the charge on the unmeasured links along the minimal cut. We can further think of $\frac{Z_A[\mindex;\{\beta\}]}{\left(Z_{\emptyset}^{(1)} \right)}$ as the probability for the charges along the unbroken links of the minimal cut to be equal to $\{\beta\}$ in the statistical model described by the partition function $Z_{\emptyset}^{(1)}$. Denoting this probability by $p_{\{\beta\}}$ we have our final result
\begin{align}
S_n(\mindex) = \frac{-1}{n-1} Z_{\emptyset}^{(1)}(\mindex) \ln \left(\sum_{\{\beta\}} p_{\{\beta\}}^n \right) + \ln d\ \ell_{\rm DW}. \label{eq: final s_n}
\end{align} 

\subsection{$p=0$ limit}
To illustrate the meaning of the statistical model \eqref{eq: final s_n}, we compute $S_n$ for $p=0$. Let us start from the following product state,
\begin{align}
\ket{\psi_0} =  \left( a_0\ket{0} + a_1\ket{1} \right)^{\otimes L}.
\end{align}
In terms of the statistical model, this corresponds to the bottom links being in charge states $1$ or $0$ with probability $a_1^2$ and $a_0^2$ respectively. The minimal cut will be spatial in nature as we are considering late times and $\ell_{\rm DW} = L_A$ since the permutations are fully ordered.
Since the vertex weights \eqref{eq: vertex weight} are SU$(2)$ symmetric, the link charge states are invariant under time evolution. This immediately gives $p_{\{\beta\}} = a_0^{2N_0}a_1^{2N_1}$, where $N_{0,1}$ are the number of links with charge $0,1$ in $\{\beta\}$. Using Eq.~\eqref{eq: final s_n}, we find the following expression for the R\'enyi entropies at late times:
\begin{align}
S_n &= \frac{-1}{n-1}\ln \left( |a_0|^{2n} + |a_1|^{2n} \right)^{L_A} + L_A \ln d. \label{eq: p=0 S_n steady state}
\end{align}
This result is consistent with thermalization to a density matrix $\rho_A = e^{-\mu \mathcal{Q}}/\text{Tr} e^{-\mu \mathcal{Q}}$, where the chemical potential $\mu$ is set by charge conservation
\begin{align}
\langle \mathfrak{q} \rangle = a_1^2 = {\rm Tr} \left[ \mathfrak{q} \rho_A  \right] =  \frac{e^{-\mu}}{1+e^{-\mu}}.
\end{align} 
We check that the R\'enyi entropies are indeed given by $S_n = \frac{-1}{n-1}\ln \text{Tr} \rho_A^n$, since
\begin{align}
S_n &= L_A\left(\ln d + \frac{1}{n-1}\ln \left( \langle \mathfrak{q} \rangle^n + (1- \langle \mathfrak{q} \rangle )^n  \right) \right),
\end{align}
which coincides with~\eqref{eq: p=0 S_n steady state}.
\newline 

\subsection{Charge variance}
In this section we briefly discuss evaluating the charge variance $[\delta \mathcal{Q}^2]$ in the language of the statistical model discussed above. 
The charge variance for fixed measurement locations and outcomes is given by $\delta \mathcal{Q}^2_\mathbf{m} = \langle \mathcal{Q}^2 \rangle_\mathbf{m} - \langle\mathcal{Q}\rangle_\mathbf{m}^2$. As the first term is linear in $\rho_\mathbf{m}$, the average over measurement outcomes will give a trivial answer at infinite temperature (see point 3 in Section \ref{sec: observables and averaging}). We have \begin{align*}
    [\langle \mathcal{Q}^2 \rangle] 
    = \text{Tr} \mathcal{Q}^2/ \text{Tr}\mathbb{I}=L(L+1)/4,
\end{align*}
where charges take value 0 and 1.
Any non-trivial physics is hidden in the second term. Nevertheless, the distribution of the variance over various trajectories is an interesting quantity, and it will be useful to evaluate this quantity using the replica trick. 

We first consider the second term which is given by,\begin{align*}
    [ \langle \mathcal{Q} \rangle^2 ] = \sum_{\mathbf{m}} \mathbb{E}_U \left[ p_{\mathbf{m}}\left(\frac{\text{Tr}\left( {\mathcal{Q}\rho_\mathbf{m}}\right)}{ \text{Tr} \rho_\mathbf{m} }\right)^2 \right].
\end{align*}
We can use the replica trick to re-write the above expression as
\begin{align}
     [ \langle \mathcal{Q} \rangle^2] &= \lim_{k\rightarrow 0} \sum_{\mathbf{m}}  \mathbb{E}_U \left[  \left(\text{Tr}\left( {\mathcal{Q}\rho_\mathbf{m}}\right)\right)^2    \left(\text{Tr}\rho_\mathbf{m} \right)^{2k-1}\right], \notag \\
     &=  \lim_{k\rightarrow 0} \sum_{\mathbf{m}} \mathbb{E}_U [ \text{Tr} \rho_\mathbf{m}^{\otimes (2k + 1)} T ],
\end{align}
where $T$ is an operator acting on the $2k+1$ copies at the top boundary and is given by $T=\mathcal{Q}\otimes\mathcal{Q}\otimes \mathbb{I} \dots \otimes \mathbb{I}$. As discussed above, the above quantity maps to a classical statistical model on averaging over random unitary gates $U$. Since the action of $T$ does not  mix different copies, we can factorize the contribution of different copies as in Section~\ref{sec: replica limit appendix}. The resulting expression, after taking the replica limit, is given by
\begin{align}
    [\langle \mathcal{Q} \rangle^2] = \sum_{\mathbf{m}} Z^{(1)}_\emptyset(\mathbf{m}) \langle \mathcal{Q}_\text{T} \rangle^2_{\mathbf{m},\text{stat}}, \label{eq: <q^2>}
\end{align}
where $\langle \cdot \rangle_\text{stat}$ is the average in the statistical model described by the partition function $Z^{(1)}_\emptyset(\mathbf{m})$ (see eq \ref{eq: partition function}), and the subscript T in $\mathcal{Q}$ is to denote the fact that it is a quantity defined on the top boundary. As mentioned before, there is no simple expression for $[\langle \mathcal{Q}\rangle^2]$, but we can use the statistical mechanics model to evaluate it numerically. Similarly, for $\langle\mathcal{Q}^2\rangle_\mathbf{m}$ the top operator $T$ is given by $T=\mathcal{Q}^2\otimes \mathbb{I}\dots \otimes \mathbb{I}$ and we have
\begin{align}
    [ \langle \mathcal{Q}^2 \rangle] = \sum_{\mathbf{m}} Z^{(1)}_\emptyset(\mathbf{m}) \langle \mathcal{Q}^2_\text{T} \rangle_{\mathbf{m},\text{stat}}. 
\end{align}

\section{Statistical mechanics model for general Abelian symmetries \label{app:generalizations}}
In this section, we generalize the statistical mechanics models to general Abelian groups, and discuss consequences of charge-sharp phases under duality transformations. Notably, our results suggest the existence of volume-law entangled phases with symmetry protected and intrinsic topological order.

\subsection{Haar average}

Consider a general Abelian group $G$, with $\alpha\in\{1\dots |R|\}$ labeling the different combinations of total charge for pairs of sites (e.g. for the U(1) model $\alpha\in\{-1,0,+1\}$, for $\mathbb{Z}_N$ $\alpha\in\{0\dots N-1\}$ etc...). We can decompose a symmetric two-site unitary into a direct sum of reps: $U=\sum_\alpha U_\alpha P_\alpha$, where $P_\alpha$ is a projector onto the $\alpha^\text{th}$-charge-sector subspace.
The main object in the statistical mechanics model is the unitary average of $\mathbb{E}_U\left[U^Q\otimes U^{*Q}\right]$
 where $Q$ is the number of replicas, which decomposes into a direct sum of all $R^{2Q}$ charge-sector combinations. Since Haar-averaging requires that each $U_\alpha$ is ``paired" with a complex-conjugated partner $U^*_\alpha$ of the same total charge, only terms in which the $U^{*Q}$ charge-sectors form a permutation of the $U^Q$ charge-sectors contribute. For each of these surviving combinations of charge-sectors, denote by $n_\alpha$ the number of times that charge-sector $\alpha$ appears in $U^Q$, and choose permutation elements $\sigma,\tau\in S_Q$ that sort the $Q$ replicas into groups of the same charge. 
 \begin{widetext}
 Then we can write:
\begin{align}
\mathbb{E}_U\left[U^Q\otimes U^{*Q}\right] = \sum_{n_{1\dots R}:\sum_\alpha n_\alpha=Q}\sum_{\sigma,\tau \in \frac{S_Q}{S_{n_1}\times \dots S_{n_R}}} W_{\sigma,\tau} \bigotimes_{\alpha}\mathbb{E}_U\left(U_\alpha^{n_\alpha}P_\alpha^{n_\alpha}\otimes U^{*n_\alpha}_\alpha P^{*n_\alpha}_\alpha\right) W_{\sigma,\tau}^\dagger 
\end{align}
where $W_{\sigma,\tau}$ is the unitary acting on $\mathcal{H}^q\otimes \mathcal{H}^{*q}$ that permutes the $Q$ copies of $U$ by $\sigma$ and the $Q$ copies of $U^*$ by $\tau$, and the permutation elements $\sigma,\tau$ range over the quotient group $\frac{S_Q}{S_{n_1}\times \dots S_{n_R}}$ to avoid over counting equivalent permutations that cycle identical replicas with the same charge-sector. Here we have labeled projectors acting on $\mathcal{H}^*$ with a $^*$ simply for readability, and this mark carries no mathematical content.

The Haar average of each charge-sector-group is: $\mathbb{E}_U\left(U_\alpha^{n_\alpha}\otimes U^{*n_\alpha}_\alpha\right) = \sum_{\sigma_\alpha,\tau_\alpha \in S_{n_\alpha}} \text{Wg}_{Dd^2}(\sigma_\alpha^{-1}\tau_\alpha;n_\alpha) |\sigma_\alpha\rrangle\llangle\tau_\alpha|$, where Wg is the Weingarten function, $D_\alpha$ is the number of states in the charge-sector $\alpha$, and $|\sigma\rrangle$ denotes the operator which permutes the input legs of $U$ by $\sigma$, and contracts them with the corresponding legs of $U^*$ (and similarly for $\llangle \sigma|$ acting on the output legs). For our purposes, we will only need the large-d limit: \begin{align}
    \lim_{d\rightarrow \infty} \text{Wg}_{d}(\sigma^{-1}\tau;Q)\sim \frac{1}{d^{Q}}\delta_{\sigma,\tau}.
\end{align}
The sum over the total-charge sector permutations $\sigma_\alpha$, can now be combined with the quotient-group permutations $\sigma,\tau$ to yield a simpler sum over $S_Q$ permutations:
\begin{align}
\mathbb{E}_U\left[U^Q\otimes U^{*Q}\right] =
\frac{1}{d^{2Q}}\sum_{\alpha_1\dots \alpha_Q}\prod_i D_{\alpha_i}^{-1}\sum_{\sigma \in S_Q} P_{\alpha_1}\otimes \dots P_{\alpha_Q}\otimes P^*_{\alpha_{\sigma(1)}}\otimes \dots P^*_{\alpha_{\sigma(Q)}} |\sigma\rrangle\llangle \sigma| P_{\alpha_1}\otimes \dots P_{\alpha_Q}\otimes P^*_{\alpha_{\sigma(1)}}\otimes \dots P^*_{\alpha_{\sigma(Q)}}.
\label{eq:generalrule}
\end{align}
Note that, the two sets of projectors are partly redundant since $P^2_\alpha=P_\alpha$, and since $P_\alpha\otimes \mathbbm{1}|\sigma\rrangle = \mathbbm{1}\otimes P_{\alpha_{\sigma(i)}}|\sigma\rrangle$, but are written in this way to emphasize that charge is separately conserved in each replica, and also that the charge-sector labels on the $\mathcal{H}^*$ spaces are related to those in the $\mathcal{H}$ spaces by the permutation element $\sigma$.
\end{widetext}

Though complicated in appearance, Eq.~\ref{eq:generalrule} has a simple interpretation: each gate becomes a vertex in the statistical mechanics model labeled by i) a permutation element $\sigma \in S_Q$, and ii) $Q$-different total charge-sector labels (those of the conjugate copies are related by permutation), which can be conveniently associated with the $Q$ different individual link charges. After averaging, the indices of the gate input and output are unrelated, except by total charge conservation in each replica. Hence any input charge configuration can be transferred with equal weight $\sim 1/D_\alpha$ to any outgoing charge configuration with the same total charge $\alpha$. The remaining rules for domain wall and measurement constraints closely parallel those of the U(1) model described above.

\paragraph{Example: $\Z_2$ Symmetric Monitored Circuits}
As an example, consider a random monitored circuit ensemble with symmetry group $G=\Z_2$, consisting of qubits with $\Z_2$ symmetry charge $\mathfrak{q}_i = \frac{1+\sigma^x_i}{2} \in \{0,1\}$, and charge addition rule $\mathcal{Q} = \sum_{i} \mathfrak{q}_i~\text{mod}~2$, (each accompanied by large dimension qudits that transform trivially under the symmetry). The effective statistical mechanics model in the $d\rightarrow\infty$ limit is an ``8-vertex" model, which has an additional two vertices compared to the U(1)/6-vertex case that respectively correspond to pair creation and annihilation of charges, and differs also in that all 8 vertices come with weight $v=1/2$.

\paragraph{Example: $\Z_N$ Symmetric Monitored Circuits}
As a second example, we can consider models with symmetry group $G=\Z_N$ for general $N$ consisting of ``quNits" with charge basis states $\{|0\>,\dots |N-1\>\}$ having symmetry charge $\mathfrak{q}=0\dots N-1$ (again, each accompanied by large-d qudits). The resulting statistical mechanics model would be an $N^3$-vertex model, with $N$ different groups of vertices corresponding to the $N$-different total charges, and each group has $N$ different ways to apportion the incoming charge between the two input legs, and $N$ different ways to apportion it between the output legs, each weighted by a factor of $v=1/N$.

From these examples, one can readily generalize to arbitrary finite Abelian groups for which $G$ can be written as a product of different $\Z_N$ factors.

\subsection{Dual entangling phases classical and quantum orders}
In principle, the statistical mechanics models sketched above can be simulated by sign-problem-free Monte Carlo for arbitrary Abelian symmetry groups and dimensions.
Though we have only performed systematic numerics on the U$(1)$ symmetric models in one-dimension, we hypothesize that the resulting physics and phase diagram is similar for other Abelian symmetry groups and higher dimensions. Assuming this hypothesis holds, then standard duality transformations would relate these simple symmetric monitored random circuit models, to those with more complex types of phases including symmetry-protected topological (SPT) phases, and discrete gauge theories.

 In the following, we denote a system with $n$ spatial dimensions and one time dimension as $(n+1)d$, where $d$ should not be confused with the dimension of the qudits. In all the examples we discuss here, there will only be a single type of measurement. Generalizing the statistical mechanics framework to multiple types of incompatible measurements remains an open challenge for future work, and to date has only been done for Clifford circuit models~\cite{PhysRevResearch.2.023288,Lavasani2020,Ippoliti2020,Sang2020}. The $1+1d$ examples with $\Z_2$ symmetries were extensively explored in Ref.~\cite{2021arXiv210209164B} from a related but different perspective, and many aspects of the physics we discuss below echos that work (though we do not consider possible phases arising from additional replica permutation symmetries as it is not clear whether these can be realized in the physical replica limit).

\subsubsection{$1+1d$: Symmetry Protected Topology (SPT)}
In $1+1d$, discrete symmetry groups of the form $\Z_m\times \Z_n$ have non-trivial projective representations and can protect topological phases~\cite{PhysRevB.87.155114,Chen1604} when $m$ and $n$ have a nontrivial greatest common divisor. These SPT phases can be mapped (at the level of local operators) onto trivial paramagnets by a ``decorated domain wall" mapping~\cite{DecoratedDW}, which attaches conjugate operators that add (remove) a $\Z_n$ charge to the left (right) of the $\Z_m$ symmetry generator. For example, with $G=\Z^A_2\times\Z^B_2$ represented on a dimerized chain of spins-1/2s with the two $\Z_2$ factors acting on the A and B sublattices respectively, the duality mapping is generated by unitary transformation $U_\text{dual}=e^{-i\pi/4\sum_i \sigma^z_{i}\left(\sigma^z_{i-1}+\sigma^z_{i+1}\right)}$ (which commutes with the symmetry, but is not generated by a symmetric Hamiltonian). $U_\text{dual}$ interchanges the stabilizers of a trivial paramagnet with those of the cluster state SPT: $\sigma^x_{i}\rightarrow \sigma^z_{i-1}\sigma^x_{i}\sigma^z_{i+1}$.

How would the charge- sharp and fuzzy phases of the $\Z_2$-symmetry circuit model transform under this duality? To answer this question, note that the total charge in a region of length $\ell$ maps onto the string-order parameter of length $\ell$ for the SPT phase. For example, in the $G=\Z_2^A\times \Z_2^B$ example, the $\Z_2^A$ interval-charge maps onto $\sigma^z_{2i-1}\left(\prod_{i=1}^\ell \sigma^x_{2i}\right)\sigma^z_{2(i+\ell)+1}$. In the charge sharp phase, the interval charge at fixed time in a single trajectory has non-zero expectation value that is asymptotically independent of $\ell$ (i.e. charge fluctuations satisfy an area law). Correspondingly, in the dual model, there is a long-range string-order parameter in each trajectory. We note that, while the string order has a non-vanishing expectation value at any fixed time slice in the dual-charge-sharp phase, this expectation value will vary from time slice to time slice due to intermediate evolution causing short-range charge fluctuations across the end of the string.

While it is perhaps not surprising that in the area law phase with SPT order can be stabilized at large measurement rates (in this dual mapping the original charge measurements become projections onto the stabilizers of the SPT phase), the possibility of a volume law entangled phase with SPT order in a fixed trajectory does not have any precedent in equilibrium, and represents a fundamentally new type of phenomena that relies essentially on both non-equilibrium and non-unitary evolution.

\subsubsection{$1+1d$: Spontaneous Symmetry Breaking}
By a closely related reasoning to the section above, applying standard Kramers Wannier (KW) duality (which exchanges the paramagnetic and spontaneous symmetry broken phases) to a $1+1d$ $\Z_N$ symmetric random circuit model relates a charge-sharp phase to a dual phase with long-range symmetry breaking order in a given trajectory. I.e. the KW duality exchanges sharp and fuzzy phases. E.g. for a $\Z_2$ symmetric circuit duality (KW) mapping exchanges $\sigma_i^x\leftrightarrow \sigma^z_{i}\sigma^z_{i+1}$. 

\subsubsection{$2+1d$: Discrete Gauge Theories}
As a final application, we consider KW dualities in $2+1d$ systems, which map models with global symmetries onto dual ``pure" gauge theories that have gauge magnetic-flux degrees of freedom but lack dynamical electrical charge. In this mapping, the original site degrees of freedom, which live on vertices of a square lattice and transform under a $\Z_N$ symmetry, map to dual link variables residing on a dual lattice whose sites are centered on plaquettes of the original lattice. Symmetry domain wall operators map onto $\Z_N$ electric field operators living on links, and local symmetry generators map onto the $\Z_N$ flux through the dual plaquette. We define the duality mapping to act trivially on the large-d qudits, which in the dual lattice now reside at the center of plaquettes.

This duality mapping interchanges symmetry charge in the original variables with gauge magnetic flux in the dual variables. Hence, the charge-sharp and fuzzy phases respectively map onto confined and deconfined phases of the gauge theory. Namely, in the charge-sharp phase, the total charge in a region $\Sigma$, $\mathcal{Q}_\Sigma$, has fluctuations only from local quantum fluctuations near the boundary, $\d \Sigma$. In the dual mapping, $\mathcal{Q}_\Sigma$, transforms into a Wilson loop around the boundary, which we write schematically as $W_\Sigma = e^{i\oint_{\d \Sigma} \vec{A}\cdot d\vec{\ell}}$ (where $e^{iA}$ is the conjugate variable to the $\Z_N$ electric field, and all continuum notations should be appropriately interpreted as lattice sums). Thus in a given trajectory at late times, the dual of the charge- sharp and fuzzy phases are characterized by area- or volume- law scaling of Wilson loops respectively:
\begin{align}
\lim_{t\rightarrow \infty} \left[ |\<\psi_{\bf m}(t)|W_\Sigma|\psi_{\bf m}(t)\>|\right] \sim \begin{cases}
e^{-|\d \Sigma|} & \text{dual-charge-sharp} \\
e^{-|\Sigma|} & \text{dual-charge-fuzzy}
\end{cases},
\end{align}
i.e. correspond to confined and deconfined phases respectively. Note here, as for spin-glass-like order parameters, it is important to take absolute values before averaging to obtain the area law scaling. 

Again, the prospect of a volume-law entangled, but trajectory-deconfined gauge theory represents a new possibility not present in equilibrium settings or closed-system dynamics. We leave establishing the (non)existence of this phase, and the study of potential critical properties of putative unconventional volume-law entangled confinement transitions as a loose end for future study. Other potentially interesting generalizations include both gauge theories with electrically charged matter, which are not dual to systems with global symmetries, and continuous $U(1)$ gauge invariant random circuit dynamics, for which the duality omits monopole instantons that are known to confine equilibrium gauge theories in $2+1d$, but perhaps have a different fate in monitored random circuit dynamics.

\section{Additional numerical results}

\subsection{Entanglement dynamics in the statistical mechanics model}

\label{app:dynamics}

\begin{figure}
\includegraphics[width=0.45\textwidth]{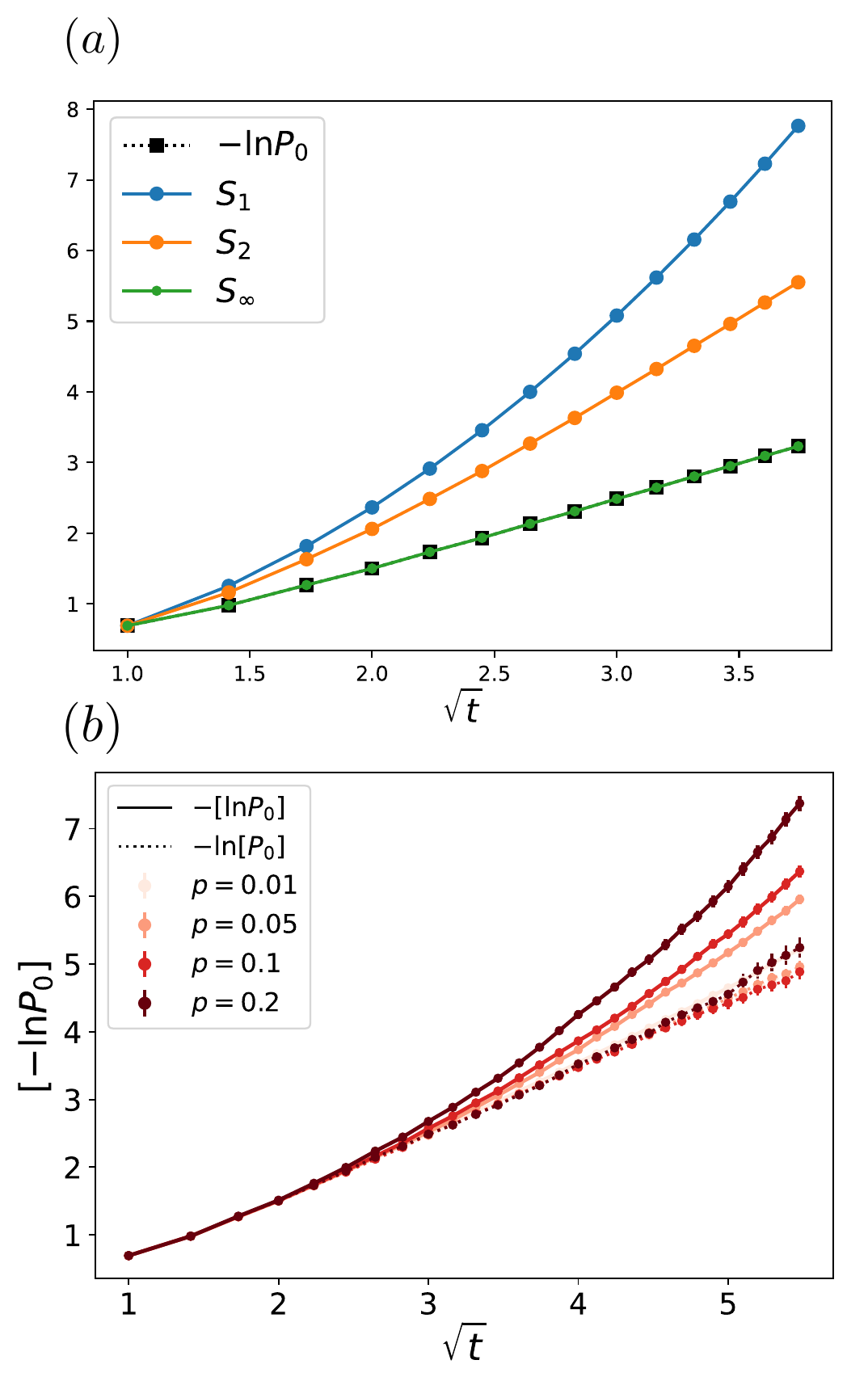}
 \caption{\textbf{Entanglement dynamics in the statistical mechanics model.} (a) Plot of $S_1$, $S_2$, $S_\infty$, and $-\ln P_0$ vs $\sqrt{t}$ for $p=0$ obtained using the statistical model with a fixed vertical minimal cut. We clearly see different growth of $S_1$ and $S_\infty$, $S_2$. The curve of $-\ln P_0$ exactly overlaps the $S_\infty$ curve as argued in the main text. (b) Plot showing $[ -\ln P_0 ]$ vs $\sqrt{t}$ for various $p$, and $L=12$. We find that this quantity grows linearly with time for any non-zero $p$. We also plot the average $-\ln [ P_0 ]$ for various $p$ (dashed curves), which grows as $\sqrt{t}$ independently of $p$. Error bars are obtained using standard deviations. }
\label{Fig. time evolution plot2}
\end{figure}

In this appendix, we present additional results on the entanglement dynamics obtained from the statistical mechanics model. 

We start with the $p=0$ case, and analyze how the argument for $\sqrt{t}$ growth translates to the statistical model language. We are working in the regime where $L\gg t$, so the minimal cut runs along the time direction.  
For simplicity of the argument, we assume that the cut does not fluctuate and is exactly vertical, that is, the cut passes through the same link at all times (we checked that our results are independent from averaging over fluctuations of the minimal cut). 
Using Eq.~\eqref{eq: final exp for s_n}, the R\'enyi entanglement entropies are related to the classical R\'enyi entropies for charge configurations along the minimal cut. Let us denote this distribution by $P_{\rm DW}$, with $S_\infty = \ln p_\text{max}$, where $p_\text{max}$ is the maximum of $P_{\rm DW}$. 
We find that $p_\text{max}$ is given by $P_{\rm DW}(0...0) \equiv P_0$. $P_0$ is the probability for all charges on the vertical cut to be equal to $0$ (equivalently, we could have also considered $P_{\rm DW}(1...1)\equiv P_1$). 
$P_0$ describes the part of the dynamics where there are no exchange of charge across the cut and is therefore dominated by dead regions in the initial state, which we know to be the source of the dominant contribution in the Schmidt values. As we discussed in Sec.~\ref{overview}, if the initial state has a dead region of size $\sqrt{t}$ centered at the entanglement cut, charges cannot diffuse to the cut until times of order $t$, so the configurational entropy of charge along the vertical minimal cut will remain zero. It follows that $P_0 \geq \exp(-\sqrt{Dt})$ and therefore that $S_n \leq \sqrt{Dt}$ for $n > 1$. 
Meanwhile, typical components of the initial wavefunction give rise to essentially random charge configurations on the minimal cut. Owing to the greater multiplicity of typical configurations they dominate $S_1$, which grows linearly in $t$. 
These expectations are borne out in Fig.~\ref{Fig. time evolution plot2}(a). 

We now turn to $p > 0$. For $p > 0$, the charge on measured sites in a given trajectory is constrained to match the measurement outcome. As we noted in Sec.~\ref{overview}, this suffices to eliminate dead regions in typical trajectories. To capture the effect of dead regions on the growth of $S_\infty$ for $p>0$, we calculate $ -\ln P_0$  as a proxy for entanglement entropies. For numerical convenience, we make two simplifying assumptions: (1) we ignore fluctuations of the minimal cut, (2) we do not perform measurements on links adjacent to the cut
(this avoids numerically expensive postselection procedures as the trajectories with non-zero $P_0$ quickly become rare as we increase $p$).

We plot $[ -\ln P_0 ]$ and $-\ln [ P_0 ]$ vs $\sqrt{t}$ in Fig. \ref{Fig. time evolution plot2}(b). We find that the quantity $-\ln [  P_0 ]$ grows as $\sqrt{t}$ for all $p$. This is expected because we are averaging over the trajectories before calculating the (pseudo)entropy; this is same as in the $p=0$ case where the unitary evolution can be seen as equivalent to doing the sum over all trajectories. We find that at low $p$, $[ -\ln P_0 ]$ stays closer to the $\sqrt{t}$ growth for longer times. At higher $p$ it diverges significantly and crosses over to linear growth. Though $[ -\ln P_0 ]$ is not exactly equal to the R\'enyi entropy, this transition from $\sqrt{t}$ to linear growth $\sim t$ is a generic phenomenon for all quantities where the survival of dead regions becomes a rare occurrence due to measurements, consistent with the general argument in Sec.~\ref{overview}, and the results of Sec.~\ref{statmech_numerics}.

\subsection{Charge sharpening dynamics in the fuzzy phase and near the charge-sharpening transition}
\label{app:chargesharpeningdynamics}

\begin{figure}
\includegraphics[scale=0.3]{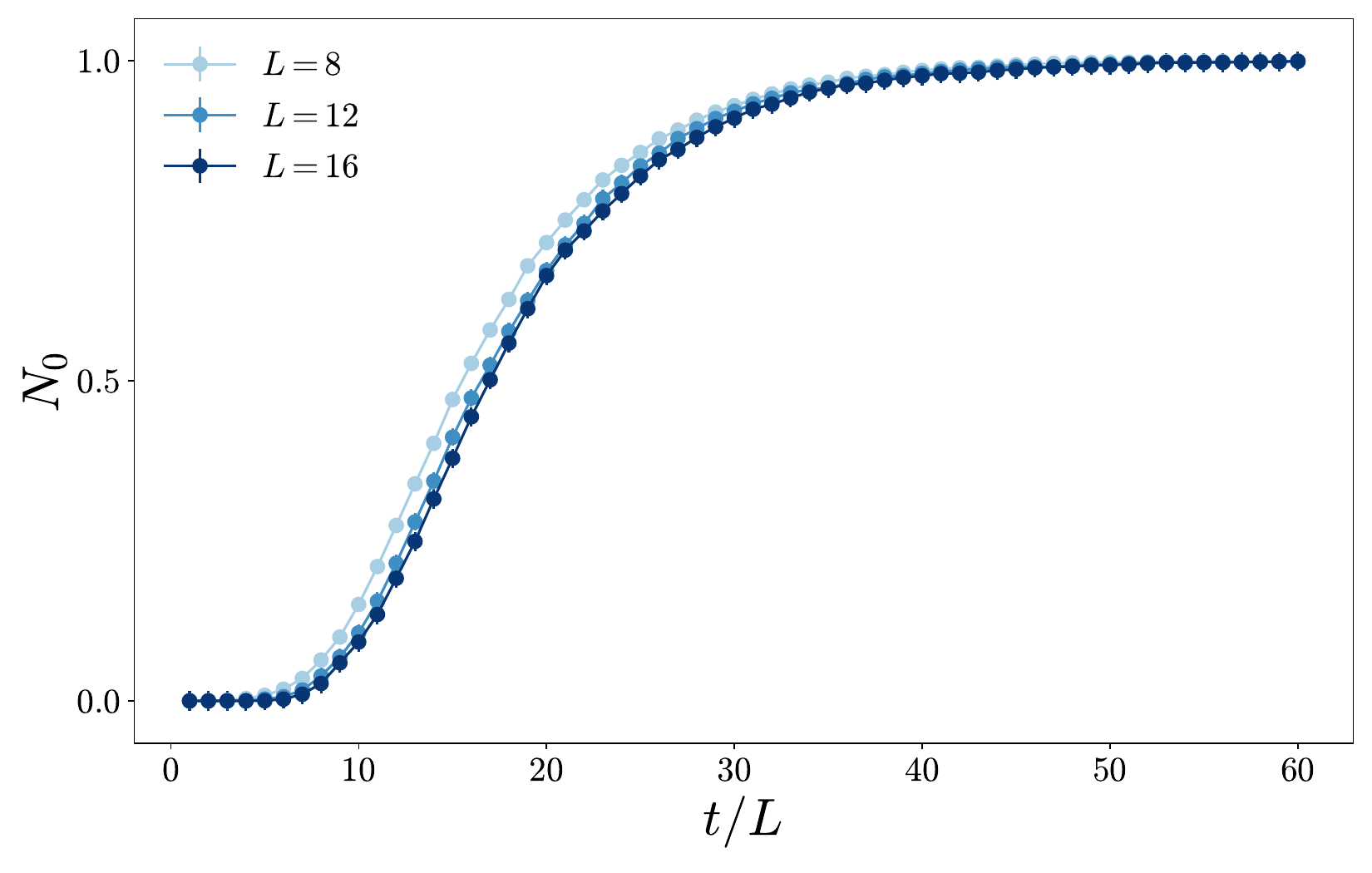}
\caption{{\bf Dynamics of charge sharpening in qubit chains.} Fraction of trajectories with $\delta \mathcal{Q}^2 < \epsilon$ with $\epsilon = 10^{-3}$ at $p=0.085$, inside the fuzzy phase. }\label{fig:ChargeSharpDynqubits}
\end{figure}

\begin{figure}
\includegraphics[scale=0.33]{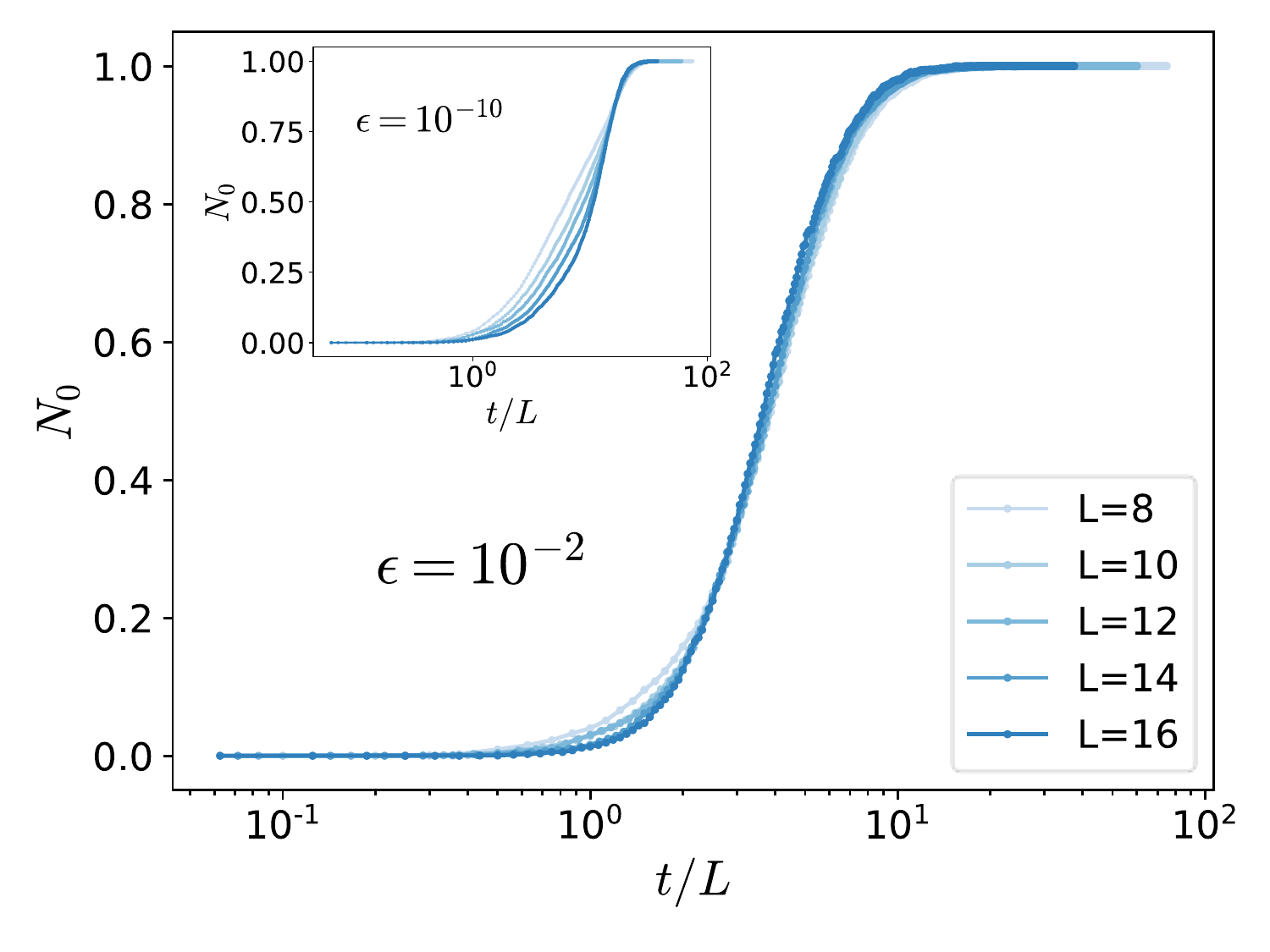}
\caption{{\bf Dynamics of charge sharpening in the statistical model.} Fraction of trajectories with $\delta \mathcal{Q}^2 < \epsilon$ with $\epsilon = 10^{-2}$ at $p=0.24$, inside the fuzzy phase. Inset: Different threshold $\epsilon=10^{-10}$, showing a similar scaling of the charge sharpening over a timescale $t \sim L$. }\label{fig:ChargeSharpDynStatmech}
\end{figure}

In this appendix, we present numerical evidence that the charge sharpens on a time scale $t_\# \sim L$, in agreement with the general argument of Sec.~\ref{overview}. We plot the fraction $N_0$ of trajectories with $\delta \mathcal{Q}^2 < \epsilon$ versus $t \sim L$, both in the qubit chain numerics (Fig.~\ref{fig:ChargeSharpDynqubits}) and in the statistical mechanics model (Fig.~\ref{fig:ChargeSharpDynStatmech}). We observe a clear crossover 
when $N_0$ lifts off from zero
on a time scale scaling linearly with $L$, as expected. Note that this sharpening time scale $t_\# \sim L/p$ is much smaller than the purification time scale $t_\pi \sim {\rm e}^L$~\cite{Gullans2019}.



\begin{figure*}
\includegraphics[width=\linewidth]{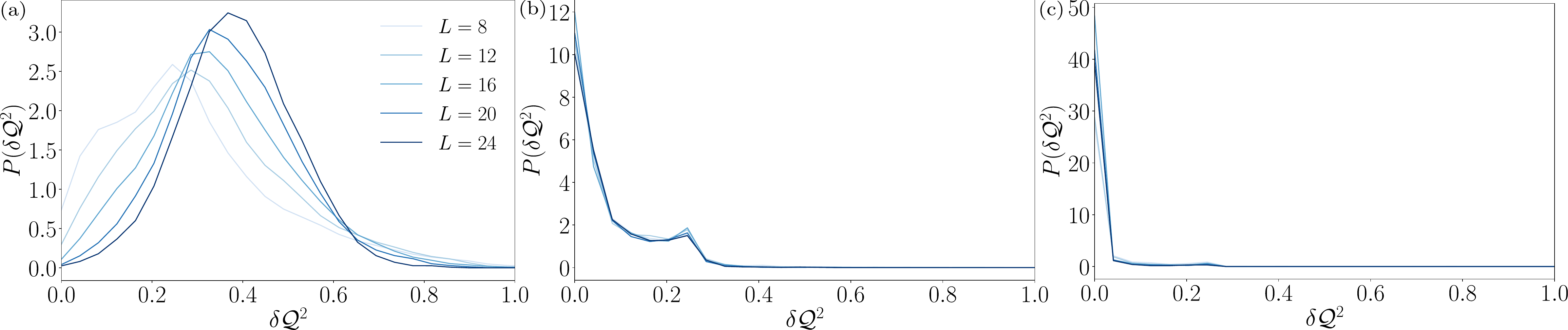}
\caption{{\bf{Charge variance distribution in the qubit model.}} Distribution of the charge variance at $t/L = 4$ (a) in the charge fuzzy phase $p = 0.05$, (b) near the critical point $p=0.10$, and (c) in the charge sharp phase $p = 0.14$.} \label{fig:DistributionsQubit}
\end{figure*}

We now turn to the critical dynamics near the charge sharpening phase transition in qubit chains. Before embarking into numerical details, we summarize the physics of the critical dynamics obtained from the simulations.

For generic initial states that mix multiple charge sectors, the charge sharpening happens in two stages: the measurement first sharpens the charge from multiple sectors to two consecutive sectors $(N, N+1)$ 
The measurement then further collapses the superposition of the two sectors $(N, N+1)$ to a unique charge (either $N$ or $N+1$).  The two stages are separated by the crossover time $t_\# \sim L/p$. 
At much later times $t\gg t_{\#}$ we find $N_0\rightarrow 1$ in a critical manner that we now turn to.

The numerical simulations of qubit chains suggest that the second stage is governed by charge-sharpening criticality. In the long-time limit, $t\gg t_\#$, we find that the universal scaling law for the critical dynamics is an exponential function,
\begin{equation}
\label{eq:critical_dynamcis_scaling}
    O(t, p) \sim A_O(x) e^{-t/\xi_t(x)}, \,\, \mathrm{where} \,\, x\equiv(p-p_\#)L^{1/\nu_\#},
\end{equation}
$L$ is the system size, and $O$ is an observable sensitive to the criticality (e.g. $N_0$ and $S_{1,\mathcal{Q}})$. The universal decay rate $\xi_t$ is a time scale that is different from the crossover time $t_\#$. Due to the space-time symmetry, the time scale $\xi_t$ follows a scaling law,
\begin{equation}
\label{eq:correlation_time}
    \xi_t(x)/L = B(x).
\end{equation}
Both universal scaling functions $A_O(x)$ and $B(x)$ are smooth in the critical regime. Right at the transition point, we obtain $\xi_t \approx 0.5 L/p_\#$. 

We now show our numerical evidence to support the above physical picture. 
We  present the charge variance distribution $\delta \mathcal{Q}^2$ in each phase and the vicinity of the charge sharpening critical point in Fig. \ref{fig:DistributionsQubit} to 
provide additional clarity on the nature of the charge sharpening dynamics.
More specifically, Fig.~\ref{fig:DistributionsQubit}(a) is deep in the charge fuzzy phase characteristic of the first stage of dynamics $t \ll t_\# $. It reveals a wide charge distribution, indicating that the quantum state at this stage is charge fuzzy and spread across multiple sectors. The middle panel depicts the second stage $t > t_\#$ with $p\approx p_{\#}$. The charge variance at this stage is peaked at zero and $0.25$, indicating the quantum state is either projected to a unique state or a superposition of two consecutive charge sectors $(N, N+1)$, respectively. The third panel is for late time dynamics $t \gg t_\#$ deep in the charge sharp phase. In this regime, only the peak near zero remains, indicating the long-time evolved quantum state has a unique sharp charge as expected.

\begin{figure}
\includegraphics[scale=0.3]{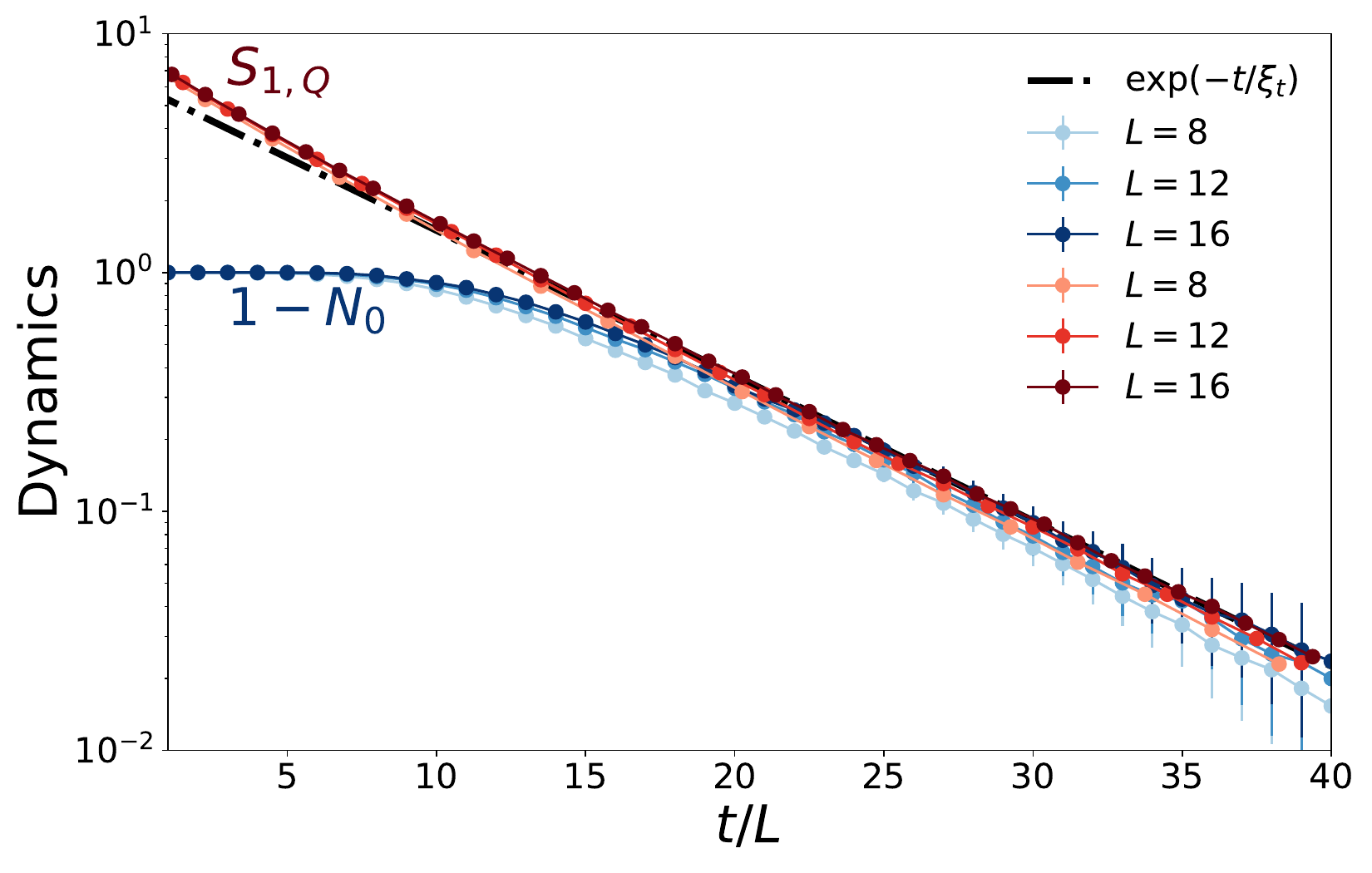}
\caption{{\bf Critical dynamics of charge-sharpening transition in qubit chains.} Comparison of the dynamics of the ancilla entanglement entropy $S_{1, Q}$ and the quantity $1-N_0$, where $N_0$ is the fraction of trajectories with $\delta \mathcal{Q}^2 < 0.01$ and the ancilla entanglement entropy $S_{1, Q}$. After rescaling the overall amplitude, both observables collapse to the same exponential function in the long-time limit.}\label{fig:critical_dynamics}
\end{figure}

We now focus on the  critical dynamics of the charge-sharpening phase transition. We first present a strong evidence to show that the critical dynamics is only about two consecutive sectors $(N, N+1)$. In Fig \ref{fig:critical_dynamics}, we compare the dynamics of $1-N_0$ to that of the ancilla-system entanglement entropy $S_{1, \mathcal{Q}}$ (see Sec. \ref{subsec:charge_sharpeing} for the definition) in the vicinity of the transition point. 
The former involves multiple charge sectors, while the latter only involves two consecutive sectors $(N, N+1)$ (with $N=L/2$). 
After rescaling the overall amplitude, the long-time dynamics of $S_{1,\mathcal{Q}}$ almost perfectly matches with the second stage dynamics of $1-N_0$, both going like $\sim e^{-t/\xi_t}$. Furthermore, due to the absence of other sectors, the ancilla probe saturates to the critical behavior much earlier than $1-N_0$. We thus conclude that the long-time critical dynamics only involves two consecutive sectors. In the long-time limit $t>t_\#$, 
both observables decay with the same exponential function, suggesting that the critical dynamics is exponential as in Eq.\eqref{eq:critical_dynamcis_scaling} and Eq.\eqref{eq:correlation_time}.

\begin{figure}
\includegraphics[scale=0.5]{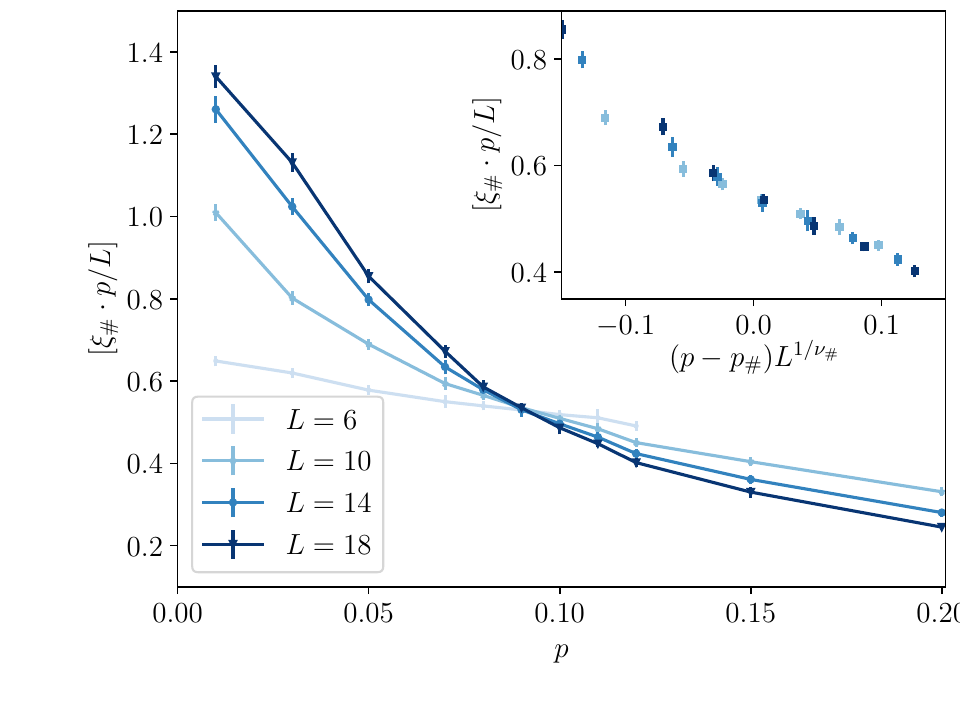}
\caption{{\bf Universal decay rate of charge-sharpening transition in qubit chains.} Data and collapse of the decay rate in the vicinity of the charge-sharpening phase transition. The transition point $p=0.088$ and the critical exponent $\nu=2.15$ established in the main text is used to collapse the curves.}\label{fig:universal_decay_rate}
\end{figure}

We use the ancilla dynamics to establish the universal scaling functions $A_O(x)$ and $B(x)$ in Eq.\eqref{eq:critical_dynamcis_scaling} and Eq.\eqref{eq:correlation_time}. The scaling function for the amplitude $A_O(x)$ depends on the choice of observable. It has been extracted for $N_0$ in Fig. \ref{fig:number-sharpening}(a) and for the ancilla probe $E_{1, Q}$ in Fig. \ref{fig:number-sharpening}(b). In this section, we extract the scaling of $B(x)$ for the universal decay rate. In Fig.\ref{fig:universal_decay_rate}, we calculate the long-time dynamics of $S_{1, \mathcal{Q}}$ for various $p$ and system sizes $L$, then fit the tail to extract the decay rate $\xi_t$. We find $\xi_t(p, L)$ cross at the transition point $p=p_\#$, indicating the existence of a scaling function $\xi_t(x)/L=B(x)$. To extract this function, we collapse the curves for different $L$ using the transition point $p_{\#}=0.088$ and the critical exponent $\nu_{\#}=2.15$ established in Fig.\ref{fig:number-sharpening}(b). 

\section{Finite size scaling}
In the case of qubit chains ($d=1$), the numerical simulations are performed for system sizes $L\le 24$. We rely on finite-size scaling protocols to extract the critical properties in the thermodynamic limit. We briefly explain the protocols in this section.

The quantities we studied in the main text, including the tripartite mutual information, the probability of a trajectory with certain charge variance and the ancilla probes shown in Figs.~\ref{fig:entanglement-transition} and~\ref{fig:number-sharpening}, all have zero scaling dimension. In the vicinity of the transition point, they share the same finite size scaling ansatz,
\begin{equation}
\label{eq:scaling_ansatz}
    R(p, L) = f\left((p-p_0)L^{1/\nu}, vL^{-\omega} \right)+ ...
\end{equation}
where $L$ is the system size, the measurement probability $p$ is the relevant scaling field with a critical exponent $\nu>0$. The transition point $p_0$ is either the entanglement phase transition $p_c$ or the charge-sharpening phase transition $p_{\#}$. 

To make the analysis systematic, we keep the leading irrelevant scaling variable $v$ in the above scaling ansatz. In thermodynamic limit $L\rightarrow \infty$, it is suppressed by a non-negative exponent $\omega$. In finite size scaling, however, this field may play an important role. Numerically, we find that the variable $v$ is significant for the ancilla probes while negligible for other quantities. We therefore have to use a more involved finite size scaling protocol as explained below to analyze the ancilla probes \cite{PhysRevE.68.036125}.

Since our numerics indicates that the scaling function $f(x, y)$ in Eq.\eqref{eq:scaling_ansatz} is analytic for both $x=(p-p_0)L^{1/\nu}$ and $y=vL^{-\omega}$, one can approximate $f$ with its Taylor expansion near the critical point,
\begin{equation}
\label{eq:taylor}
     R(p, L) = a_R+b_R(p-p_0)L^{1/\nu}+c_R(p-p_0)^2L^{2/\nu}+d_R/L^{\omega}+...
\end{equation}
where we assume $p$ is close to the critical point $p_0$ and the system size is sufficiently large so that both $x=(p-p_0)L^{1/\nu}$ and $y=vL^{-\omega}$ are small. We also redefine the Taylor coefficient $d_R$ to absorb the unknown amplitude $v$ of the irrelevant variable. In the ideal case $d_R=0$, $R(p_0, L)$ collapses to the universal constant $a_R$ at the critical point. It indicates that the curves $R(p, L)$ for different system sizes perfectly cross at $p=p_0$. However, in realistic models, $d_R$ is non-zero. The irrelevant field then shifts the crossing points as the system size increases. 

The ansatz Eq.\eqref{eq:taylor} allows us to extract the phase transition point $p_0$  and the critical exponent $\nu$ with the presence of a non-negligible irrelevant scaling variable. In practice, we collect dozens of data points for different $p$ and system size $L$ in the vicinity of the critical points. We then perform a non-linear fitting with the ansatz Eq.~\eqref{eq:taylor} by taking the coefficients $a, b, c, d$ and critical properties $p_0$ and $\nu$ as the fitting parameters. {Their error bars are defined as the standard error of the mean, which is approximately equal to a confidence interval of $67\%$}. We then try to drop some of the parameters to make sure the fitting is robust and the error bars are reliable.


\end{appendix}

\newpage
\bibliography{references_updated}

\begin{thebibliography}{80}%
\makeatletter
\providecommand \@ifxundefined [1]{%
 \@ifx{#1\undefined}
}%
\providecommand \@ifnum [1]{%
 \ifnum #1\expandafter \@firstoftwo
 \else \expandafter \@secondoftwo
 \fi
}%
\providecommand \@ifx [1]{%
 \ifx #1\expandafter \@firstoftwo
 \else \expandafter \@secondoftwo
 \fi
}%
\providecommand \natexlab [1]{#1}%
\providecommand \enquote  [1]{``#1''}%
\providecommand \bibnamefont  [1]{#1}%
\providecommand \bibfnamefont [1]{#1}%
\providecommand \citenamefont [1]{#1}%
\providecommand \href@noop [0]{\@secondoftwo}%
\providecommand \href [0]{\begingroup \@sanitize@url \@href}%
\providecommand \@href[1]{\@@startlink{#1}\@@href}%
\providecommand \@@href[1]{\endgroup#1\@@endlink}%
\providecommand \@sanitize@url [0]{\catcode `\\12\catcode `\$12\catcode
  `\&12\catcode `\#12\catcode `\^12\catcode `\_12\catcode `\%12\relax}%
\providecommand \@@startlink[1]{}%
\providecommand \@@endlink[0]{}%
\providecommand \url  [0]{\begingroup\@sanitize@url \@url }%
\providecommand \@url [1]{\endgroup\@href {#1}{\urlprefix }}%
\providecommand \urlprefix  [0]{URL }%
\providecommand \Eprint [0]{\href }%
\providecommand \doibase [0]{http://dx.doi.org/}%
\providecommand \selectlanguage [0]{\@gobble}%
\providecommand \bibinfo  [0]{\@secondoftwo}%
\providecommand \bibfield  [0]{\@secondoftwo}%
\providecommand \translation [1]{[#1]}%
\providecommand \BibitemOpen [0]{}%
\providecommand \bibitemStop [0]{}%
\providecommand \bibitemNoStop [0]{.\EOS\space}%
\providecommand \EOS [0]{\spacefactor3000\relax}%
\providecommand \BibitemShut  [1]{\csname bibitem#1\endcsname}%
\let\auto@bib@innerbib\@empty
\bibitem [{\citenamefont {Amico}\ \emph {et~al.}(2008)\citenamefont {Amico},
  \citenamefont {Fazio}, \citenamefont {Osterloh},\ and\ \citenamefont
  {Vedral}}]{RevModPhys.80.517}%
  \BibitemOpen
  \bibfield  {author} {\bibinfo {author} {\bibfnamefont {Luigi}\ \bibnamefont
  {Amico}}, \bibinfo {author} {\bibfnamefont {Rosario}\ \bibnamefont {Fazio}},
  \bibinfo {author} {\bibfnamefont {Andreas}\ \bibnamefont {Osterloh}}, \ and\
  \bibinfo {author} {\bibfnamefont {Vlatko}\ \bibnamefont {Vedral}},\
  }\bibfield  {title} {\enquote {\bibinfo {title} {{Entanglement in many-body
  systems}},}\ }\href {\doibase 10.1103/RevModPhys.80.517} {\bibfield
  {journal} {\bibinfo  {journal} {Reviews of Modern Physics}\ }\textbf
  {\bibinfo {volume} {80}},\ \bibinfo {pages} {517--576} (\bibinfo {year}
  {2008})},\ \Eprint {http://arxiv.org/abs/0703044} {arXiv:0703044 [quant-ph]}
  \BibitemShut {NoStop}%
\bibitem [{\citenamefont {Horodecki}\ \emph {et~al.}(2009)\citenamefont
  {Horodecki}, \citenamefont {Horodecki}, \citenamefont {Horodecki},\ and\
  \citenamefont {Horodecki}}]{RevModPhys.81.865}%
  \BibitemOpen
  \bibfield  {author} {\bibinfo {author} {\bibfnamefont {Ryszard}\ \bibnamefont
  {Horodecki}}, \bibinfo {author} {\bibfnamefont {Pawe{\l}}\ \bibnamefont
  {Horodecki}}, \bibinfo {author} {\bibfnamefont {Micha{\l}}\ \bibnamefont
  {Horodecki}}, \ and\ \bibinfo {author} {\bibfnamefont {Karol}\ \bibnamefont
  {Horodecki}},\ }\bibfield  {title} {\enquote {\bibinfo {title} {{Quantum
  entanglement}},}\ }\href {\doibase 10.1103/RevModPhys.81.865} {\bibfield
  {journal} {\bibinfo  {journal} {Reviews of Modern Physics}\ }\textbf
  {\bibinfo {volume} {81}},\ \bibinfo {pages} {865--942} (\bibinfo {year}
  {2009})},\ \Eprint {http://arxiv.org/abs/0702225} {arXiv:0702225 [quant-ph]}
  \BibitemShut {NoStop}%
\bibitem [{\citenamefont {Eisert}\ \emph {et~al.}(2010)\citenamefont {Eisert},
  \citenamefont {Cramer},\ and\ \citenamefont {Plenio}}]{RevModPhys.82.277}%
  \BibitemOpen
  \bibfield  {author} {\bibinfo {author} {\bibfnamefont {J.}~\bibnamefont
  {Eisert}}, \bibinfo {author} {\bibfnamefont {M.}~\bibnamefont {Cramer}}, \
  and\ \bibinfo {author} {\bibfnamefont {M.~B.}\ \bibnamefont {Plenio}},\
  }\bibfield  {title} {\enquote {\bibinfo {title} {{Colloquium: Area laws for
  the entanglement entropy}},}\ }\href {\doibase 10.1103/RevModPhys.82.277}
  {\bibfield  {journal} {\bibinfo  {journal} {Reviews of Modern Physics}\
  }\textbf {\bibinfo {volume} {82}},\ \bibinfo {pages} {277--306} (\bibinfo
  {year} {2010})}\BibitemShut {NoStop}%
\bibitem [{\citenamefont {Calabrese}\ and\ \citenamefont
  {Cardy}(2009)}]{Calabrese_2009}%
  \BibitemOpen
  \bibfield  {author} {\bibinfo {author} {\bibfnamefont {Pasquale}\
  \bibnamefont {Calabrese}}\ and\ \bibinfo {author} {\bibfnamefont {John}\
  \bibnamefont {Cardy}},\ }\bibfield  {title} {\enquote {\bibinfo {title}
  {{Entanglement entropy and conformal field theory}},}\ }\href {\doibase
  10.1088/1751-8113/42/50/504005} {\bibfield  {journal} {\bibinfo  {journal}
  {Journal of Physics A: Mathematical and Theoretical}\ }\textbf {\bibinfo
  {volume} {42}},\ \bibinfo {pages} {504005} (\bibinfo {year} {2009})},\
  \Eprint {http://arxiv.org/abs/0905.4013} {arXiv:0905.4013} \BibitemShut
  {NoStop}%
\bibitem [{\citenamefont {Laflorencie}(2016)}]{LAFLORENCIE20161}%
  \BibitemOpen
  \bibfield  {author} {\bibinfo {author} {\bibfnamefont {Nicolas}\ \bibnamefont
  {Laflorencie}},\ }\bibfield  {title} {\enquote {\bibinfo {title} {{Quantum
  entanglement in condensed matter systems}},}\ }\href {\doibase
  10.1016/j.physrep.2016.06.008} {\bibfield  {journal} {\bibinfo  {journal}
  {Physics Reports}\ }\textbf {\bibinfo {volume} {646}},\ \bibinfo {pages}
  {1--59} (\bibinfo {year} {2016})},\ \Eprint {http://arxiv.org/abs/1512.03388}
  {arXiv:1512.03388} \BibitemShut {NoStop}%
\bibitem [{\citenamefont {Preskill}(2018)}]{Preskill2018quantumcomputingin}%
  \BibitemOpen
  \bibfield  {author} {\bibinfo {author} {\bibfnamefont {John}\ \bibnamefont
  {Preskill}},\ }\bibfield  {title} {\enquote {\bibinfo {title} {{Quantum
  {\{}C{\}}omputing in the {\{}NISQ{\}} era and beyond}},}\ }\href {\doibase
  10.22331/q-2018-08-06-79} {\bibfield  {journal} {\bibinfo  {journal}
  {Quantum}\ }\textbf {\bibinfo {volume} {2}},\ \bibinfo {pages} {79} (\bibinfo
  {year} {2018})}\BibitemShut {NoStop}%
\bibitem [{\citenamefont {Nahum}\ \emph {et~al.}(2017)\citenamefont {Nahum},
  \citenamefont {Ruhman}, \citenamefont {Vijay},\ and\ \citenamefont
  {Haah}}]{PhysRevX.7.031016}%
  \BibitemOpen
  \bibfield  {author} {\bibinfo {author} {\bibfnamefont {Adam}\ \bibnamefont
  {Nahum}}, \bibinfo {author} {\bibfnamefont {Jonathan}\ \bibnamefont
  {Ruhman}}, \bibinfo {author} {\bibfnamefont {Sagar}\ \bibnamefont {Vijay}}, \
  and\ \bibinfo {author} {\bibfnamefont {Jeongwan}\ \bibnamefont {Haah}},\
  }\bibfield  {title} {\enquote {\bibinfo {title} {{Quantum entanglement growth
  under random unitary dynamics}},}\ }\href {\doibase
  10.1103/PhysRevX.7.031016} {\bibfield  {journal} {\bibinfo  {journal}
  {Physical Review X}\ }\textbf {\bibinfo {volume} {7}},\ \bibinfo {pages}
  {31016} (\bibinfo {year} {2017})},\ \Eprint {http://arxiv.org/abs/1608.06950}
  {arXiv:1608.06950} \BibitemShut {NoStop}%
\bibitem [{\citenamefont {Nahum}\ \emph {et~al.}(2018)\citenamefont {Nahum},
  \citenamefont {Vijay},\ and\ \citenamefont {Haah}}]{Nahum2018}%
  \BibitemOpen
  \bibfield  {author} {\bibinfo {author} {\bibfnamefont {Adam}\ \bibnamefont
  {Nahum}}, \bibinfo {author} {\bibfnamefont {Sagar}\ \bibnamefont {Vijay}}, \
  and\ \bibinfo {author} {\bibfnamefont {Jeongwan}\ \bibnamefont {Haah}},\
  }\bibfield  {title} {\enquote {\bibinfo {title} {{Operator Spreading in
  Random Unitary Circuits}},}\ }\href {\doibase 10.1103/PhysRevX.8.021014}
  {\bibfield  {journal} {\bibinfo  {journal} {Physical Review X}\ }\textbf
  {\bibinfo {volume} {8}},\ \bibinfo {pages} {21014} (\bibinfo {year}
  {2018})},\ \Eprint {http://arxiv.org/abs/1705.08975} {arXiv:1705.08975}
  \BibitemShut {NoStop}%
\bibitem [{\citenamefont {{Von Keyserlingk}}\ \emph {et~al.}(2018)\citenamefont
  {{Von Keyserlingk}}, \citenamefont {Rakovszky}, \citenamefont {Pollmann},\
  and\ \citenamefont {Sondhi}}]{PhysRevX.8.021013}%
  \BibitemOpen
  \bibfield  {author} {\bibinfo {author} {\bibfnamefont {C.~W.}\ \bibnamefont
  {{Von Keyserlingk}}}, \bibinfo {author} {\bibfnamefont {Tibor}\ \bibnamefont
  {Rakovszky}}, \bibinfo {author} {\bibfnamefont {Frank}\ \bibnamefont
  {Pollmann}}, \ and\ \bibinfo {author} {\bibfnamefont {S.~L.}\ \bibnamefont
  {Sondhi}},\ }\bibfield  {title} {\enquote {\bibinfo {title} {{Operator
  Hydrodynamics, OTOCs, and Entanglement Growth in Systems without Conservation
  Laws}},}\ }\href {\doibase 10.1103/PhysRevX.8.021013} {\bibfield  {journal}
  {\bibinfo  {journal} {Physical Review X}\ }\textbf {\bibinfo {volume} {8}},\
  \bibinfo {pages} {021013} (\bibinfo {year} {2018})},\ \Eprint
  {http://arxiv.org/abs/1705.08910} {arXiv:1705.08910} \BibitemShut {NoStop}%
\bibitem [{\citenamefont {Zhou}\ and\ \citenamefont {Nahum}(2019)}]{Zhou2019}%
  \BibitemOpen
  \bibfield  {author} {\bibinfo {author} {\bibfnamefont {Tianci}\ \bibnamefont
  {Zhou}}\ and\ \bibinfo {author} {\bibfnamefont {Adam}\ \bibnamefont
  {Nahum}},\ }\bibfield  {title} {\enquote {\bibinfo {title} {{Emergent
  statistical mechanics of entanglement in random unitary circuits}},}\ }\href
  {\doibase 10.1103/PhysRevB.99.174205} {\bibfield  {journal} {\bibinfo
  {journal} {Physical Review B}\ }\textbf {\bibinfo {volume} {99}},\ \bibinfo
  {pages} {174205} (\bibinfo {year} {2019})},\ \Eprint
  {http://arxiv.org/abs/1804.09737} {arXiv:1804.09737} \BibitemShut {NoStop}%
\bibitem [{\citenamefont {Rakovszky}\ \emph {et~al.}(2018)\citenamefont
  {Rakovszky}, \citenamefont {Pollmann},\ and\ \citenamefont {{Von
  Keyserlingk}}}]{PhysRevX.8.031058}%
  \BibitemOpen
  \bibfield  {author} {\bibinfo {author} {\bibfnamefont {Tibor}\ \bibnamefont
  {Rakovszky}}, \bibinfo {author} {\bibfnamefont {Frank}\ \bibnamefont
  {Pollmann}}, \ and\ \bibinfo {author} {\bibfnamefont {C.~W.}\ \bibnamefont
  {{Von Keyserlingk}}},\ }\bibfield  {title} {\enquote {\bibinfo {title}
  {{Diffusive Hydrodynamics of Out-of-Time-Ordered Correlators with Charge
  Conservation}},}\ }\href {\doibase 10.1103/PhysRevX.8.031058} {\bibfield
  {journal} {\bibinfo  {journal} {Physical Review X}\ }\textbf {\bibinfo
  {volume} {8}},\ \bibinfo {pages} {031058} (\bibinfo {year} {2018})},\ \Eprint
  {http://arxiv.org/abs/1710.09827} {arXiv:1710.09827} \BibitemShut {NoStop}%
\bibitem [{\citenamefont {Khemani}\ \emph {et~al.}(2018)\citenamefont
  {Khemani}, \citenamefont {Vishwanath},\ and\ \citenamefont
  {Huse}}]{PhysRevX.8.031057}%
  \BibitemOpen
  \bibfield  {author} {\bibinfo {author} {\bibfnamefont {Vedika}\ \bibnamefont
  {Khemani}}, \bibinfo {author} {\bibfnamefont {Ashvin}\ \bibnamefont
  {Vishwanath}}, \ and\ \bibinfo {author} {\bibfnamefont {David~A.}\
  \bibnamefont {Huse}},\ }\bibfield  {title} {\enquote {\bibinfo {title}
  {{Operator Spreading and the Emergence of Dissipative Hydrodynamics under
  Unitary Evolution with Conservation Laws}},}\ }\href {\doibase
  10.1103/PhysRevX.8.031057} {\bibfield  {journal} {\bibinfo  {journal}
  {Physical Review X}\ }\textbf {\bibinfo {volume} {8}},\ \bibinfo {pages}
  {031057} (\bibinfo {year} {2018})},\ \Eprint
  {http://arxiv.org/abs/1710.09835} {arXiv:1710.09835} \BibitemShut {NoStop}%
\bibitem [{\citenamefont {Li}\ \emph {et~al.}(2018)\citenamefont {Li},
  \citenamefont {Chen},\ and\ \citenamefont {Fisher}}]{PhysRevB.98.205136}%
  \BibitemOpen
  \bibfield  {author} {\bibinfo {author} {\bibfnamefont {Yaodong}\ \bibnamefont
  {Li}}, \bibinfo {author} {\bibfnamefont {Xiao}\ \bibnamefont {Chen}}, \ and\
  \bibinfo {author} {\bibfnamefont {Matthew~P.A.}\ \bibnamefont {Fisher}},\
  }\bibfield  {title} {\enquote {\bibinfo {title} {{Quantum Zeno effect and the
  many-body entanglement transition}},}\ }\href {\doibase
  10.1103/PhysRevB.98.205136} {\bibfield  {journal} {\bibinfo  {journal}
  {Physical Review B}\ }\textbf {\bibinfo {volume} {98}},\ \bibinfo {pages}
  {205136} (\bibinfo {year} {2018})},\ \Eprint
  {http://arxiv.org/abs/1808.06134} {arXiv:1808.06134} \BibitemShut {NoStop}%
\bibitem [{\citenamefont {Skinner}\ \emph {et~al.}(2019)\citenamefont
  {Skinner}, \citenamefont {Ruhman},\ and\ \citenamefont
  {Nahum}}]{Skinner2019}%
  \BibitemOpen
  \bibfield  {author} {\bibinfo {author} {\bibfnamefont {Brian}\ \bibnamefont
  {Skinner}}, \bibinfo {author} {\bibfnamefont {Jonathan}\ \bibnamefont
  {Ruhman}}, \ and\ \bibinfo {author} {\bibfnamefont {Adam}\ \bibnamefont
  {Nahum}},\ }\bibfield  {title} {\enquote {\bibinfo {title}
  {{Measurement-Induced Phase Transitions in the Dynamics of Entanglement}},}\
  }\href {\doibase 10.1103/PhysRevX.9.031009} {\bibfield  {journal} {\bibinfo
  {journal} {Physical Review X}\ }\textbf {\bibinfo {volume} {9}},\ \bibinfo
  {pages} {031009} (\bibinfo {year} {2019})},\ \Eprint
  {http://arxiv.org/abs/1808.05953} {arXiv:1808.05953} \BibitemShut {NoStop}%
\bibitem [{\citenamefont {Choi}\ \emph {et~al.}(2020)\citenamefont {Choi},
  \citenamefont {Bao}, \citenamefont {Qi},\ and\ \citenamefont
  {Altman}}]{Choi2020}%
  \BibitemOpen
  \bibfield  {author} {\bibinfo {author} {\bibfnamefont {Soonwon}\ \bibnamefont
  {Choi}}, \bibinfo {author} {\bibfnamefont {Yimu}\ \bibnamefont {Bao}},
  \bibinfo {author} {\bibfnamefont {Xiao~Liang}\ \bibnamefont {Qi}}, \ and\
  \bibinfo {author} {\bibfnamefont {Ehud}\ \bibnamefont {Altman}},\ }\bibfield
  {title} {\enquote {\bibinfo {title} {{Quantum Error Correction in Scrambling
  Dynamics and Measurement-Induced Phase Transition}},}\ }\href {\doibase
  10.1103/PhysRevLett.125.030505} {\bibfield  {journal} {\bibinfo  {journal}
  {Physical Review Letters}\ }\textbf {\bibinfo {volume} {125}},\ \bibinfo
  {pages} {030505} (\bibinfo {year} {2020})},\ \Eprint
  {http://arxiv.org/abs/1903.05124} {arXiv:1903.05124} \BibitemShut {NoStop}%
\bibitem [{\citenamefont {Gullans}\ and\ \citenamefont
  {Huse}(2020{\natexlab{a}})}]{Gullans2019}%
  \BibitemOpen
  \bibfield  {author} {\bibinfo {author} {\bibfnamefont {Michael~J.}\
  \bibnamefont {Gullans}}\ and\ \bibinfo {author} {\bibfnamefont {David~A.}\
  \bibnamefont {Huse}},\ }\bibfield  {title} {\enquote {\bibinfo {title}
  {{Dynamical Purification Phase Transition Induced by Quantum
  Measurements}},}\ }\href {\doibase 10.1103/PhysRevX.10.041020} {\bibfield
  {journal} {\bibinfo  {journal} {Physical Review X}\ }\textbf {\bibinfo
  {volume} {10}},\ \bibinfo {pages} {41020} (\bibinfo {year}
  {2020}{\natexlab{a}})},\ \Eprint {http://arxiv.org/abs/1905.05195}
  {arXiv:1905.05195} \BibitemShut {NoStop}%
\bibitem [{\citenamefont {Li}\ and\ \citenamefont {Fisher}(2021)}]{Li2020b}%
  \BibitemOpen
  \bibfield  {author} {\bibinfo {author} {\bibfnamefont {Yaodong}\ \bibnamefont
  {Li}}\ and\ \bibinfo {author} {\bibfnamefont {Matthew~P.A.}\ \bibnamefont
  {Fisher}},\ }\bibfield  {title} {\enquote {\bibinfo {title} {{Statistical
  mechanics of quantum error correcting codes}},}\ }\href {\doibase
  10.1103/PhysRevB.103.104306} {\bibfield  {journal} {\bibinfo  {journal}
  {Physical Review B}\ }\textbf {\bibinfo {volume} {103}},\ \bibinfo {pages}
  {104306} (\bibinfo {year} {2021})},\ \Eprint
  {http://arxiv.org/abs/2007.03822} {arXiv:2007.03822} \BibitemShut {NoStop}%
\bibitem [{\citenamefont {Fan}\ \emph {et~al.}(2021)\citenamefont {Fan},
  \citenamefont {Vijay}, \citenamefont {Vishwanath},\ and\ \citenamefont
  {You}}]{Fan2020}%
  \BibitemOpen
  \bibfield  {author} {\bibinfo {author} {\bibfnamefont {Ruihua}\ \bibnamefont
  {Fan}}, \bibinfo {author} {\bibfnamefont {Sagar}\ \bibnamefont {Vijay}},
  \bibinfo {author} {\bibfnamefont {Ashvin}\ \bibnamefont {Vishwanath}}, \ and\
  \bibinfo {author} {\bibfnamefont {Yi~Zhuang}\ \bibnamefont {You}},\
  }\bibfield  {title} {\enquote {\bibinfo {title} {{Self-organized error
  correction in random unitary circuits with measurement}},}\ }\href {\doibase
  10.1103/PhysRevB.103.174309} {\bibfield  {journal} {\bibinfo  {journal}
  {Physical Review B}\ }\textbf {\bibinfo {volume} {103}},\ \bibinfo {pages}
  {174309} (\bibinfo {year} {2021})},\ \Eprint
  {http://arxiv.org/abs/2002.12385} {arXiv:2002.12385} \BibitemShut {NoStop}%
\bibitem [{\citenamefont {Bao}\ \emph {et~al.}(2020)\citenamefont {Bao},
  \citenamefont {Choi},\ and\ \citenamefont {Altman}}]{Bao2020}%
  \BibitemOpen
  \bibfield  {author} {\bibinfo {author} {\bibfnamefont {Yimu}\ \bibnamefont
  {Bao}}, \bibinfo {author} {\bibfnamefont {Soonwon}\ \bibnamefont {Choi}}, \
  and\ \bibinfo {author} {\bibfnamefont {Ehud}\ \bibnamefont {Altman}},\
  }\bibfield  {title} {\enquote {\bibinfo {title} {{Theory of the phase
  transition in random unitary circuits with measurements}},}\ }\href {\doibase
  10.1103/PhysRevB.101.104301} {\bibfield  {journal} {\bibinfo  {journal}
  {Physical Review B}\ }\textbf {\bibinfo {volume} {101}},\ \bibinfo {pages}
  {104301} (\bibinfo {year} {2020})},\ \Eprint
  {http://arxiv.org/abs/1908.04305} {arXiv:1908.04305} \BibitemShut {NoStop}%
\bibitem [{\citenamefont {Jian}\ \emph
  {et~al.}(2020{\natexlab{a}})\citenamefont {Jian}, \citenamefont {You},
  \citenamefont {Vasseur},\ and\ \citenamefont {Ludwig}}]{Jian2020}%
  \BibitemOpen
  \bibfield  {author} {\bibinfo {author} {\bibfnamefont {Chao~Ming}\
  \bibnamefont {Jian}}, \bibinfo {author} {\bibfnamefont {Yi~Zhuang}\
  \bibnamefont {You}}, \bibinfo {author} {\bibfnamefont {Romain}\ \bibnamefont
  {Vasseur}}, \ and\ \bibinfo {author} {\bibfnamefont {Andreas~W.W.}\
  \bibnamefont {Ludwig}},\ }\bibfield  {title} {\enquote {\bibinfo {title}
  {{Measurement-induced criticality in random quantum circuits}},}\ }\href
  {\doibase 10.1103/PhysRevB.101.104302} {\bibfield  {journal} {\bibinfo
  {journal} {Physical Review B}\ }\textbf {\bibinfo {volume} {101}},\ \bibinfo
  {pages} {104302} (\bibinfo {year} {2020}{\natexlab{a}})},\ \Eprint
  {http://arxiv.org/abs/1908.08051} {arXiv:1908.08051} \BibitemShut {NoStop}%
\bibitem [{\citenamefont {Li}\ \emph {et~al.}(2021)\citenamefont {Li},
  \citenamefont {Chen}, \citenamefont {Ludwig},\ and\ \citenamefont
  {Fisher}}]{Li2020a}%
  \BibitemOpen
  \bibfield  {author} {\bibinfo {author} {\bibfnamefont {Yaodong}\ \bibnamefont
  {Li}}, \bibinfo {author} {\bibfnamefont {Xiao}\ \bibnamefont {Chen}},
  \bibinfo {author} {\bibfnamefont {Andreas~W.W.}\ \bibnamefont {Ludwig}}, \
  and\ \bibinfo {author} {\bibfnamefont {Matthew~P.A.}\ \bibnamefont
  {Fisher}},\ }\bibfield  {title} {\enquote {\bibinfo {title} {{Conformal
  invariance and quantum nonlocality in critical hybrid circuits}},}\ }\href
  {\doibase 10.1103/PhysRevB.104.104305} {\bibfield  {journal} {\bibinfo
  {journal} {Physical Review B}\ }\textbf {\bibinfo {volume} {104}},\ \bibinfo
  {pages} {104305} (\bibinfo {year} {2021})},\ \Eprint
  {http://arxiv.org/abs/2003.12721} {arXiv:2003.12721} \BibitemShut {NoStop}%
\bibitem [{\citenamefont {Zabalo}\ \emph {et~al.}(2022)\citenamefont {Zabalo},
  \citenamefont {Gullans}, \citenamefont {Wilson}, \citenamefont {Vasseur},
  \citenamefont {Ludwig}, \citenamefont {Gopalakrishnan}, \citenamefont
  {Huse},\ and\ \citenamefont {Pixley}}]{2021arXiv210703393Z}%
  \BibitemOpen
  \bibfield  {author} {\bibinfo {author} {\bibfnamefont {A.}~\bibnamefont
  {Zabalo}}, \bibinfo {author} {\bibfnamefont {M.~J.}\ \bibnamefont {Gullans}},
  \bibinfo {author} {\bibfnamefont {J.~H.}\ \bibnamefont {Wilson}}, \bibinfo
  {author} {\bibfnamefont {R.}~\bibnamefont {Vasseur}}, \bibinfo {author}
  {\bibfnamefont {A.~W.W.}\ \bibnamefont {Ludwig}}, \bibinfo {author}
  {\bibfnamefont {S.}~\bibnamefont {Gopalakrishnan}}, \bibinfo {author}
  {\bibfnamefont {David~A.}\ \bibnamefont {Huse}}, \ and\ \bibinfo {author}
  {\bibfnamefont {J.~H.}\ \bibnamefont {Pixley}},\ }\bibfield  {title}
  {\enquote {\bibinfo {title} {{Operator Scaling Dimensions and Multifractality
  at Measurement-Induced Transitions}},}\ }\href {\doibase
  10.1103/PhysRevLett.128.050602} {\bibfield  {journal} {\bibinfo  {journal}
  {Physical Review Letters}\ }\textbf {\bibinfo {volume} {128}},\ \bibinfo
  {pages} {050602} (\bibinfo {year} {2022})},\ \Eprint
  {http://arxiv.org/abs/2107.03393} {arXiv:2107.03393} \BibitemShut {NoStop}%
\bibitem [{\citenamefont {Li}\ \emph {et~al.}(2019)\citenamefont {Li},
  \citenamefont {Chen},\ and\ \citenamefont {Fisher}}]{Li2019}%
  \BibitemOpen
  \bibfield  {author} {\bibinfo {author} {\bibfnamefont {Yaodong}\ \bibnamefont
  {Li}}, \bibinfo {author} {\bibfnamefont {Xiao}\ \bibnamefont {Chen}}, \ and\
  \bibinfo {author} {\bibfnamefont {Matthew~P.A.}\ \bibnamefont {Fisher}},\
  }\bibfield  {title} {\enquote {\bibinfo {title} {{Measurement-driven
  entanglement transition in hybrid quantum circuits}},}\ }\href {\doibase
  10.1103/PhysRevB.100.134306} {\bibfield  {journal} {\bibinfo  {journal}
  {Physical Review B}\ }\textbf {\bibinfo {volume} {100}},\ \bibinfo {pages}
  {134306} (\bibinfo {year} {2019})},\ \Eprint
  {http://arxiv.org/abs/1901.08092} {arXiv:1901.08092} \BibitemShut {NoStop}%
\bibitem [{\citenamefont {Chan}\ \emph {et~al.}(2019)\citenamefont {Chan},
  \citenamefont {Nandkishore}, \citenamefont {Pretko},\ and\ \citenamefont
  {Smith}}]{PhysRevB.99.224307}%
  \BibitemOpen
  \bibfield  {author} {\bibinfo {author} {\bibfnamefont {Amos}\ \bibnamefont
  {Chan}}, \bibinfo {author} {\bibfnamefont {Rahul~M.}\ \bibnamefont
  {Nandkishore}}, \bibinfo {author} {\bibfnamefont {Michael}\ \bibnamefont
  {Pretko}}, \ and\ \bibinfo {author} {\bibfnamefont {Graeme}\ \bibnamefont
  {Smith}},\ }\bibfield  {title} {\enquote {\bibinfo {title}
  {{Unitary-projective entanglement dynamics}},}\ }\href {\doibase
  10.1103/PhysRevB.99.224307} {\bibfield  {journal} {\bibinfo  {journal}
  {Physical Review B}\ }\textbf {\bibinfo {volume} {99}},\ \bibinfo {pages}
  {224307} (\bibinfo {year} {2019})},\ \Eprint
  {http://arxiv.org/abs/1808.05949} {arXiv:1808.05949} \BibitemShut {NoStop}%
\bibitem [{\citenamefont {Cao}\ \emph {et~al.}(2019)\citenamefont {Cao},
  \citenamefont {Tilloy},\ and\ \citenamefont
  {de~Luca}}]{10.21468/SciPostPhys.7.2.024}%
  \BibitemOpen
  \bibfield  {author} {\bibinfo {author} {\bibfnamefont {Xiangyu}\ \bibnamefont
  {Cao}}, \bibinfo {author} {\bibfnamefont {Antoine}\ \bibnamefont {Tilloy}}, \
  and\ \bibinfo {author} {\bibfnamefont {Andrea}\ \bibnamefont {de~Luca}},\
  }\bibfield  {title} {\enquote {\bibinfo {title} {{Entanglement in a free
  fermion chain under continuous monitoring}},}\ }\href {\doibase
  10.21468/SciPostPhys.7.2.024} {\bibfield  {journal} {\bibinfo  {journal}
  {SciPost Physics}\ }\textbf {\bibinfo {volume} {7}},\ \bibinfo {pages} {24}
  (\bibinfo {year} {2019})},\ \Eprint {http://arxiv.org/abs/1804.04638}
  {arXiv:1804.04638} \BibitemShut {NoStop}%
\bibitem [{\citenamefont {Szyniszewski}\ \emph {et~al.}(2019)\citenamefont
  {Szyniszewski}, \citenamefont {Romito},\ and\ \citenamefont
  {Schomerus}}]{Szyniszewski2019}%
  \BibitemOpen
  \bibfield  {author} {\bibinfo {author} {\bibfnamefont {M.}~\bibnamefont
  {Szyniszewski}}, \bibinfo {author} {\bibfnamefont {A.}~\bibnamefont
  {Romito}}, \ and\ \bibinfo {author} {\bibfnamefont {H.}~\bibnamefont
  {Schomerus}},\ }\bibfield  {title} {\enquote {\bibinfo {title} {{Entanglement
  transition from variable-strength weak measurements}},}\ }\href {\doibase
  10.1103/PhysRevB.100.064204} {\bibfield  {journal} {\bibinfo  {journal}
  {Physical Review B}\ }\textbf {\bibinfo {volume} {100}},\ \bibinfo {pages}
  {064204} (\bibinfo {year} {2019})},\ \Eprint
  {http://arxiv.org/abs/1903.05452} {arXiv:1903.05452} \BibitemShut {NoStop}%
\bibitem [{\citenamefont {Gullans}\ and\ \citenamefont
  {Huse}(2020{\natexlab{b}})}]{Gullans2020}%
  \BibitemOpen
  \bibfield  {author} {\bibinfo {author} {\bibfnamefont {Michael~J.}\
  \bibnamefont {Gullans}}\ and\ \bibinfo {author} {\bibfnamefont {David~A.}\
  \bibnamefont {Huse}},\ }\bibfield  {title} {\enquote {\bibinfo {title}
  {{Scalable Probes of Measurement-Induced Criticality}},}\ }\href {\doibase
  10.1103/PhysRevLett.125.070606} {\bibfield  {journal} {\bibinfo  {journal}
  {Physical Review Letters}\ }\textbf {\bibinfo {volume} {125}},\ \bibinfo
  {pages} {070606} (\bibinfo {year} {2020}{\natexlab{b}})},\ \Eprint
  {http://arxiv.org/abs/1910.00020} {arXiv:1910.00020} \BibitemShut {NoStop}%
\bibitem [{\citenamefont {Zabalo}\ \emph {et~al.}(2020)\citenamefont {Zabalo},
  \citenamefont {Gullans}, \citenamefont {Wilson}, \citenamefont
  {Gopalakrishnan}, \citenamefont {Huse},\ and\ \citenamefont
  {Pixley}}]{Zabalo2020}%
  \BibitemOpen
  \bibfield  {author} {\bibinfo {author} {\bibfnamefont {Aidan}\ \bibnamefont
  {Zabalo}}, \bibinfo {author} {\bibfnamefont {Michael~J.}\ \bibnamefont
  {Gullans}}, \bibinfo {author} {\bibfnamefont {Justin~H.}\ \bibnamefont
  {Wilson}}, \bibinfo {author} {\bibfnamefont {Sarang}\ \bibnamefont
  {Gopalakrishnan}}, \bibinfo {author} {\bibfnamefont {David~A.}\ \bibnamefont
  {Huse}}, \ and\ \bibinfo {author} {\bibfnamefont {J.~H.}\ \bibnamefont
  {Pixley}},\ }\bibfield  {title} {\enquote {\bibinfo {title} {{Critical
  properties of the measurement-induced transition in random quantum
  circuits}},}\ }\href {\doibase 10.1103/PhysRevB.101.060301} {\bibfield
  {journal} {\bibinfo  {journal} {Physical Review B}\ }\textbf {\bibinfo
  {volume} {101}},\ \bibinfo {pages} {060301} (\bibinfo {year} {2020})},\
  \Eprint {http://arxiv.org/abs/1911.00008} {arXiv:1911.00008} \BibitemShut
  {NoStop}%
\bibitem [{\citenamefont {Nahum}\ and\ \citenamefont
  {Skinner}(2020)}]{PhysRevResearch.2.023288}%
  \BibitemOpen
  \bibfield  {author} {\bibinfo {author} {\bibfnamefont {Adam}\ \bibnamefont
  {Nahum}}\ and\ \bibinfo {author} {\bibfnamefont {Brian}\ \bibnamefont
  {Skinner}},\ }\bibfield  {title} {\enquote {\bibinfo {title} {{Entanglement
  and dynamics of diffusion-annihilation processes with Majorana defects}},}\
  }\href {\doibase 10.1103/PhysRevResearch.2.023288} {\bibfield  {journal}
  {\bibinfo  {journal} {Physical Review Research}\ }\textbf {\bibinfo {volume}
  {2}},\ \bibinfo {pages} {23288} (\bibinfo {year} {2020})},\ \Eprint
  {http://arxiv.org/abs/1911.11169} {arXiv:1911.11169} \BibitemShut {NoStop}%
\bibitem [{\citenamefont {Ippoliti}\ \emph {et~al.}(2021)\citenamefont
  {Ippoliti}, \citenamefont {Gullans}, \citenamefont {Gopalakrishnan},
  \citenamefont {Huse},\ and\ \citenamefont {Khemani}}]{Ippoliti2020}%
  \BibitemOpen
  \bibfield  {author} {\bibinfo {author} {\bibfnamefont {Matteo}\ \bibnamefont
  {Ippoliti}}, \bibinfo {author} {\bibfnamefont {Michael~J.}\ \bibnamefont
  {Gullans}}, \bibinfo {author} {\bibfnamefont {Sarang}\ \bibnamefont
  {Gopalakrishnan}}, \bibinfo {author} {\bibfnamefont {David~A.}\ \bibnamefont
  {Huse}}, \ and\ \bibinfo {author} {\bibfnamefont {Vedika}\ \bibnamefont
  {Khemani}},\ }\bibfield  {title} {\enquote {\bibinfo {title} {{Entanglement
  Phase Transitions in Measurement-Only Dynamics}},}\ }\href {\doibase
  10.1103/PhysRevX.11.011030} {\bibfield  {journal} {\bibinfo  {journal}
  {Physical Review X}\ }\textbf {\bibinfo {volume} {11}},\ \bibinfo {pages}
  {011030} (\bibinfo {year} {2021})},\ \Eprint
  {http://arxiv.org/abs/2004.09560} {arXiv:2004.09560} \BibitemShut {NoStop}%
\bibitem [{\citenamefont {Lavasani}\ \emph {et~al.}(2021)\citenamefont
  {Lavasani}, \citenamefont {Alavirad},\ and\ \citenamefont
  {Barkeshli}}]{Lavasani2020}%
  \BibitemOpen
  \bibfield  {author} {\bibinfo {author} {\bibfnamefont {Ali}\ \bibnamefont
  {Lavasani}}, \bibinfo {author} {\bibfnamefont {Yahya}\ \bibnamefont
  {Alavirad}}, \ and\ \bibinfo {author} {\bibfnamefont {Maissam}\ \bibnamefont
  {Barkeshli}},\ }\bibfield  {title} {\enquote {\bibinfo {title}
  {{Measurement-induced topological entanglement transitions in symmetric
  random quantum circuits}},}\ }\href {\doibase 10.1038/s41567-020-01112-z}
  {\bibfield  {journal} {\bibinfo  {journal} {Nature Physics}\ }\textbf
  {\bibinfo {volume} {17}},\ \bibinfo {pages} {342--347} (\bibinfo {year}
  {2021})},\ \Eprint {http://arxiv.org/abs/2004.07243} {arXiv:2004.07243}
  \BibitemShut {NoStop}%
\bibitem [{\citenamefont {Sang}\ and\ \citenamefont {Hsieh}(2021)}]{Sang2020}%
  \BibitemOpen
  \bibfield  {author} {\bibinfo {author} {\bibfnamefont {Shengqi}\ \bibnamefont
  {Sang}}\ and\ \bibinfo {author} {\bibfnamefont {Timothy~H.}\ \bibnamefont
  {Hsieh}},\ }\bibfield  {title} {\enquote {\bibinfo {title}
  {{Measurement-protected quantum phases}},}\ }\href {\doibase
  10.1103/PhysRevResearch.3.023200} {\bibfield  {journal} {\bibinfo  {journal}
  {Physical Review Research}\ }\textbf {\bibinfo {volume} {3}},\ \bibinfo
  {pages} {023200} (\bibinfo {year} {2021})},\ \Eprint
  {http://arxiv.org/abs/2004.09509} {arXiv:2004.09509} \BibitemShut {NoStop}%
\bibitem [{\citenamefont {Tang}\ and\ \citenamefont
  {Zhu}(2020)}]{PhysRevResearch.2.013022}%
  \BibitemOpen
  \bibfield  {author} {\bibinfo {author} {\bibfnamefont {Qicheng}\ \bibnamefont
  {Tang}}\ and\ \bibinfo {author} {\bibfnamefont {W.}~\bibnamefont {Zhu}},\
  }\bibfield  {title} {\enquote {\bibinfo {title} {{Measurement-induced phase
  transition: A case study in the nonintegrable model by density-matrix
  renormalization group calculations}},}\ }\href {\doibase
  10.1103/PhysRevResearch.2.013022} {\bibfield  {journal} {\bibinfo  {journal}
  {Physical Review Research}\ }\textbf {\bibinfo {volume} {2}},\ \bibinfo
  {pages} {13022} (\bibinfo {year} {2020})},\ \Eprint
  {http://arxiv.org/abs/1908.11253} {arXiv:1908.11253} \BibitemShut {NoStop}%
\bibitem [{\citenamefont {Lopez-Piqueres}\ \emph {et~al.}(2020)\citenamefont
  {Lopez-Piqueres}, \citenamefont {Ware},\ and\ \citenamefont
  {Vasseur}}]{PhysRevB.102.064202}%
  \BibitemOpen
  \bibfield  {author} {\bibinfo {author} {\bibfnamefont {Javier}\ \bibnamefont
  {Lopez-Piqueres}}, \bibinfo {author} {\bibfnamefont {Brayden}\ \bibnamefont
  {Ware}}, \ and\ \bibinfo {author} {\bibfnamefont {Romain}\ \bibnamefont
  {Vasseur}},\ }\bibfield  {title} {\enquote {\bibinfo {title} {{Mean-field
  entanglement transitions in random tree tensor networks}},}\ }\href {\doibase
  10.1103/PhysRevB.102.064202} {\bibfield  {journal} {\bibinfo  {journal}
  {Physical Review B}\ }\textbf {\bibinfo {volume} {102}},\ \bibinfo {pages}
  {64202} (\bibinfo {year} {2020})},\ \Eprint {http://arxiv.org/abs/2003.01138}
  {arXiv:2003.01138} \BibitemShut {NoStop}%
\bibitem [{\citenamefont {Nahum}\ \emph {et~al.}(2021)\citenamefont {Nahum},
  \citenamefont {Roy}, \citenamefont {Skinner},\ and\ \citenamefont
  {Ruhman}}]{Nahum2020}%
  \BibitemOpen
  \bibfield  {author} {\bibinfo {author} {\bibfnamefont {Adam}\ \bibnamefont
  {Nahum}}, \bibinfo {author} {\bibfnamefont {Sthitadhi}\ \bibnamefont {Roy}},
  \bibinfo {author} {\bibfnamefont {Brian}\ \bibnamefont {Skinner}}, \ and\
  \bibinfo {author} {\bibfnamefont {Jonathan}\ \bibnamefont {Ruhman}},\
  }\bibfield  {title} {\enquote {\bibinfo {title} {{Measurement and
  Entanglement Phase Transitions in All-To-All Quantum Circuits, on Quantum
  Trees, and in Landau-Ginsburg Theory}},}\ }\href {\doibase
  10.1103/PRXQuantum.2.010352} {\bibfield  {journal} {\bibinfo  {journal} {PRX
  Quantum}\ }\textbf {\bibinfo {volume} {2}},\ \bibinfo {pages} {10352}
  (\bibinfo {year} {2021})},\ \Eprint {http://arxiv.org/abs/2009.11311}
  {arXiv:2009.11311} \BibitemShut {NoStop}%
\bibitem [{\citenamefont {Turkeshi}\ \emph {et~al.}(2020)\citenamefont
  {Turkeshi}, \citenamefont {Fazio},\ and\ \citenamefont
  {Dalmonte}}]{Turkeshi2020}%
  \BibitemOpen
  \bibfield  {author} {\bibinfo {author} {\bibfnamefont {Xhek}\ \bibnamefont
  {Turkeshi}}, \bibinfo {author} {\bibfnamefont {Rosario}\ \bibnamefont
  {Fazio}}, \ and\ \bibinfo {author} {\bibfnamefont {Marcello}\ \bibnamefont
  {Dalmonte}},\ }\bibfield  {title} {\enquote {\bibinfo {title}
  {{Measurement-induced criticality in (2+1) -dimensional hybrid quantum
  circuits}},}\ }\href {\doibase 10.1103/PhysRevB.102.014315} {\bibfield
  {journal} {\bibinfo  {journal} {Physical Review B}\ }\textbf {\bibinfo
  {volume} {102}},\ \bibinfo {pages} {014315} (\bibinfo {year} {2020})},\
  \Eprint {http://arxiv.org/abs/2007.02970} {arXiv:2007.02970} \BibitemShut
  {NoStop}%
\bibitem [{\citenamefont {Fuji}\ and\ \citenamefont {Ashida}(2020)}]{Fuji2020}%
  \BibitemOpen
  \bibfield  {author} {\bibinfo {author} {\bibfnamefont {Yohei}\ \bibnamefont
  {Fuji}}\ and\ \bibinfo {author} {\bibfnamefont {Yuto}\ \bibnamefont
  {Ashida}},\ }\bibfield  {title} {\enquote {\bibinfo {title}
  {{Measurement-induced quantum criticality under continuous monitoring}},}\
  }\href {\doibase 10.1103/PhysRevB.102.054302} {\bibfield  {journal} {\bibinfo
   {journal} {Physical Review B}\ }\textbf {\bibinfo {volume} {102}},\ \bibinfo
  {pages} {054302} (\bibinfo {year} {2020})},\ \Eprint
  {http://arxiv.org/abs/2004.11957} {arXiv:2004.11957} \BibitemShut {NoStop}%
\bibitem [{\citenamefont {Lunt}\ \emph {et~al.}(2021)\citenamefont {Lunt},
  \citenamefont {Szyniszewski},\ and\ \citenamefont {Pal}}]{Lunt}%
  \BibitemOpen
  \bibfield  {author} {\bibinfo {author} {\bibfnamefont {Oliver}\ \bibnamefont
  {Lunt}}, \bibinfo {author} {\bibfnamefont {Marcin}\ \bibnamefont
  {Szyniszewski}}, \ and\ \bibinfo {author} {\bibfnamefont {Arijeet}\
  \bibnamefont {Pal}},\ }\bibfield  {title} {\enquote {\bibinfo {title}
  {{Measurement-induced criticality and entanglement clusters: A study of
  one-dimensional and two-dimensional Clifford circuits}},}\ }\href {\doibase
  10.1103/PhysRevB.104.155111} {\bibfield  {journal} {\bibinfo  {journal}
  {Physical Review B}\ }\textbf {\bibinfo {volume} {104}},\ \bibinfo {pages}
  {155111} (\bibinfo {year} {2021})},\ \Eprint
  {http://arxiv.org/abs/2012.03857} {arXiv:2012.03857} \BibitemShut {NoStop}%
\bibitem [{\citenamefont {Lunt}\ and\ \citenamefont {Pal}(2020)}]{Lunt2020}%
  \BibitemOpen
  \bibfield  {author} {\bibinfo {author} {\bibfnamefont {Oliver}\ \bibnamefont
  {Lunt}}\ and\ \bibinfo {author} {\bibfnamefont {Arijeet}\ \bibnamefont
  {Pal}},\ }\bibfield  {title} {\enquote {\bibinfo {title}
  {{Measurement-induced entanglement transitions in many-body localized
  systems}},}\ }\href {\doibase 10.1103/PhysRevResearch.2.043072} {\bibfield
  {journal} {\bibinfo  {journal} {Physical Review Research}\ }\textbf {\bibinfo
  {volume} {2}},\ \bibinfo {pages} {043072} (\bibinfo {year} {2020})},\ \Eprint
  {http://arxiv.org/abs/2005.13603} {arXiv:2005.13603} \BibitemShut {NoStop}%
\bibitem [{\citenamefont {Vijay}(2020)}]{2020arXiv200503052V}%
  \BibitemOpen
  \bibfield  {author} {\bibinfo {author} {\bibfnamefont {Sagar}\ \bibnamefont
  {Vijay}},\ }\bibfield  {title} {\enquote {\bibinfo {title}
  {{Measurement-Driven Phase Transition within a Volume-Law Entangled
  Phase}},}\ }\href {http://arxiv.org/abs/2005.03052} {\bibfield  {journal}
  {\bibinfo  {journal} {arXiv e-prints}\ ,\ \bibinfo {pages}
  {arXiv:2005.03052}} (\bibinfo {year} {2020})},\ \Eprint
  {http://arxiv.org/abs/2005.03052} {arXiv:2005.03052} \BibitemShut {NoStop}%
\bibitem [{\citenamefont {Turkeshi}\ \emph {et~al.}(2021)\citenamefont
  {Turkeshi}, \citenamefont {Biella}, \citenamefont {Fazio}, \citenamefont
  {Dalmonte},\ and\ \citenamefont {Schir{\'{o}}}}]{PhysRevB.103.224210}%
  \BibitemOpen
  \bibfield  {author} {\bibinfo {author} {\bibfnamefont {Xhek}\ \bibnamefont
  {Turkeshi}}, \bibinfo {author} {\bibfnamefont {Alberto}\ \bibnamefont
  {Biella}}, \bibinfo {author} {\bibfnamefont {Rosario}\ \bibnamefont {Fazio}},
  \bibinfo {author} {\bibfnamefont {Marcello}\ \bibnamefont {Dalmonte}}, \ and\
  \bibinfo {author} {\bibfnamefont {Marco}\ \bibnamefont {Schir{\'{o}}}},\
  }\bibfield  {title} {\enquote {\bibinfo {title} {{Measurement-induced
  entanglement transitions in the quantum Ising chain: From infinite to zero
  clicks}},}\ }\href {\doibase 10.1103/PhysRevB.103.224210} {\bibfield
  {journal} {\bibinfo  {journal} {Physical Review B}\ }\textbf {\bibinfo
  {volume} {103}},\ \bibinfo {pages} {224210} (\bibinfo {year} {2021})},\
  \Eprint {http://arxiv.org/abs/2103.09138} {arXiv:2103.09138} \BibitemShut
  {NoStop}%
\bibitem [{\citenamefont {Ippoliti}\ and\ \citenamefont
  {Khemani}(2021)}]{PhysRevLett.126.060501}%
  \BibitemOpen
  \bibfield  {author} {\bibinfo {author} {\bibfnamefont {Matteo}\ \bibnamefont
  {Ippoliti}}\ and\ \bibinfo {author} {\bibfnamefont {Vedika}\ \bibnamefont
  {Khemani}},\ }\bibfield  {title} {\enquote {\bibinfo {title}
  {{Postselection-Free Entanglement Dynamics via Spacetime Duality}},}\ }\href
  {\doibase 10.1103/PhysRevLett.126.060501} {\bibfield  {journal} {\bibinfo
  {journal} {Physical Review Letters}\ }\textbf {\bibinfo {volume} {126}},\
  \bibinfo {pages} {060501} (\bibinfo {year} {2021})},\ \Eprint
  {http://arxiv.org/abs/2010.15840} {arXiv:2010.15840} \BibitemShut {NoStop}%
\bibitem [{\citenamefont {Lu}\ and\ \citenamefont
  {Grover}(2021)}]{2021arXiv210306356L}%
  \BibitemOpen
  \bibfield  {author} {\bibinfo {author} {\bibfnamefont {Tsung~Cheng}\
  \bibnamefont {Lu}}\ and\ \bibinfo {author} {\bibfnamefont {Tarun}\
  \bibnamefont {Grover}},\ }\bibfield  {title} {\enquote {\bibinfo {title}
  {{Spacetime duality between localization transitions and measurement-induced
  transitions}},}\ }\href {\doibase 10.1103/PRXQuantum.2.040319} {\bibfield
  {journal} {\bibinfo  {journal} {PRX Quantum}\ }\textbf {\bibinfo {volume}
  {2}},\ \bibinfo {pages} {1--16} (\bibinfo {year} {2021})},\ \Eprint
  {http://arxiv.org/abs/2103.06356} {arXiv:2103.06356} \BibitemShut {NoStop}%
\bibitem [{\citenamefont {Jian}\ \emph
  {et~al.}(2020{\natexlab{b}})\citenamefont {Jian}, \citenamefont {Bauer},
  \citenamefont {Keselman},\ and\ \citenamefont
  {Ludwig}}]{2020arXiv201204666J}%
  \BibitemOpen
  \bibfield  {author} {\bibinfo {author} {\bibfnamefont {Chao-Ming}\
  \bibnamefont {Jian}}, \bibinfo {author} {\bibfnamefont {Bela}\ \bibnamefont
  {Bauer}}, \bibinfo {author} {\bibfnamefont {Anna}\ \bibnamefont {Keselman}},
  \ and\ \bibinfo {author} {\bibfnamefont {Andreas W.~W.}\ \bibnamefont
  {Ludwig}},\ }\bibfield  {title} {\enquote {\bibinfo {title} {{Criticality and
  entanglement in non-unitary quantum circuits and tensor networks of
  non-interacting fermions}},}\ }\href {http://arxiv.org/abs/2012.04666} {\
  (\bibinfo {year} {2020}{\natexlab{b}})},\ \Eprint
  {http://arxiv.org/abs/2012.04666} {arXiv:2012.04666} \BibitemShut {NoStop}%
\bibitem [{\citenamefont {Gopalakrishnan}\ and\ \citenamefont
  {Gullans}(2021)}]{PhysRevLett.126.170503}%
  \BibitemOpen
  \bibfield  {author} {\bibinfo {author} {\bibfnamefont {Sarang}\ \bibnamefont
  {Gopalakrishnan}}\ and\ \bibinfo {author} {\bibfnamefont {Michael~J.}\
  \bibnamefont {Gullans}},\ }\bibfield  {title} {\enquote {\bibinfo {title}
  {{Entanglement and Purification Transitions in Non-Hermitian Quantum
  Mechanics}},}\ }\href {\doibase 10.1103/PhysRevLett.126.170503} {\bibfield
  {journal} {\bibinfo  {journal} {Physical Review Letters}\ }\textbf {\bibinfo
  {volume} {126}},\ \bibinfo {pages} {170503} (\bibinfo {year} {2021})},\
  \Eprint {http://arxiv.org/abs/2012.01435} {arXiv:2012.01435} \BibitemShut
  {NoStop}%
\bibitem [{\citenamefont {Turkeshi}(2021)}]{2021arXiv210106245T}%
  \BibitemOpen
  \bibfield  {author} {\bibinfo {author} {\bibfnamefont {Xhek}\ \bibnamefont
  {Turkeshi}},\ }\bibfield  {title} {\enquote {\bibinfo {title}
  {{Measurement-induced criticality as a data-structure transition}},}\ }\href
  {http://arxiv.org/abs/2101.06245} {\bibfield  {journal} {\bibinfo  {journal}
  {arXiv e-prints}\ ,\ \bibinfo {pages} {arXiv:2101.06245}} (\bibinfo {year}
  {2021})},\ \Eprint {http://arxiv.org/abs/2101.06245} {arXiv:2101.06245}
  \BibitemShut {NoStop}%
\bibitem [{\citenamefont {Bao}\ \emph {et~al.}(2021)\citenamefont {Bao},
  \citenamefont {Choi},\ and\ \citenamefont {Altman}}]{2021arXiv210209164B}%
  \BibitemOpen
  \bibfield  {author} {\bibinfo {author} {\bibfnamefont {Yimu}\ \bibnamefont
  {Bao}}, \bibinfo {author} {\bibfnamefont {Soonwon}\ \bibnamefont {Choi}}, \
  and\ \bibinfo {author} {\bibfnamefont {Ehud}\ \bibnamefont {Altman}},\
  }\bibfield  {title} {\enquote {\bibinfo {title} {{Symmetry enriched phases of
  quantum circuits}},}\ }\href {\doibase 10.1016/j.aop.2021.168618} {\bibfield
  {journal} {\bibinfo  {journal} {Annals of Physics}\ }\textbf {\bibinfo
  {volume} {435}},\ \bibinfo {pages} {168618} (\bibinfo {year} {2021})},\
  \Eprint {http://arxiv.org/abs/2102.09164} {arXiv:2102.09164} \BibitemShut
  {NoStop}%
\bibitem [{\citenamefont {Block}\ \emph {et~al.}(2022)\citenamefont {Block},
  \citenamefont {Bao}, \citenamefont {Choi}, \citenamefont {Altman},\ and\
  \citenamefont {Yao}}]{2021arXiv210413372B}%
  \BibitemOpen
  \bibfield  {author} {\bibinfo {author} {\bibfnamefont {Maxwell}\ \bibnamefont
  {Block}}, \bibinfo {author} {\bibfnamefont {Yimu}\ \bibnamefont {Bao}},
  \bibinfo {author} {\bibfnamefont {Soonwon}\ \bibnamefont {Choi}}, \bibinfo
  {author} {\bibfnamefont {Ehud}\ \bibnamefont {Altman}}, \ and\ \bibinfo
  {author} {\bibfnamefont {Norman~Y.}\ \bibnamefont {Yao}},\ }\bibfield
  {title} {\enquote {\bibinfo {title} {{Measurement-Induced Transition in
  Long-Range Interacting Quantum Circuits}},}\ }\href {\doibase
  10.1103/PhysRevLett.128.010604} {\bibfield  {journal} {\bibinfo  {journal}
  {Physical Review Letters}\ }\textbf {\bibinfo {volume} {128}},\ \bibinfo
  {pages} {010604} (\bibinfo {year} {2022})},\ \Eprint
  {http://arxiv.org/abs/2104.13372} {arXiv:2104.13372} \BibitemShut {NoStop}%
\bibitem [{\citenamefont {Bentsen}\ \emph {et~al.}(2021)\citenamefont
  {Bentsen}, \citenamefont {Sahu},\ and\ \citenamefont
  {Swingle}}]{2021arXiv210407688B}%
  \BibitemOpen
  \bibfield  {author} {\bibinfo {author} {\bibfnamefont {Gregory~S.}\
  \bibnamefont {Bentsen}}, \bibinfo {author} {\bibfnamefont {Subhayan}\
  \bibnamefont {Sahu}}, \ and\ \bibinfo {author} {\bibfnamefont {Brian}\
  \bibnamefont {Swingle}},\ }\bibfield  {title} {\enquote {\bibinfo {title}
  {{Measurement-induced purification in large- N hybrid Brownian circuits}},}\
  }\href {\doibase 10.1103/PhysRevB.104.094304} {\bibfield  {journal} {\bibinfo
   {journal} {Physical Review B}\ }\textbf {\bibinfo {volume} {104}},\ \bibinfo
  {pages} {094304} (\bibinfo {year} {2021})},\ \Eprint
  {http://arxiv.org/abs/2104.07688} {arXiv:2104.07688} \BibitemShut {NoStop}%
\bibitem [{\citenamefont {Noel}\ \emph {et~al.}(2022)\citenamefont {Noel},
  \citenamefont {Niroula}, \citenamefont {Zhu}, \citenamefont {Risinger},
  \citenamefont {Egan}, \citenamefont {Biswas}, \citenamefont {Cetina},
  \citenamefont {Gorshkov}, \citenamefont {Gullans}, \citenamefont {Huse},\
  and\ \citenamefont {Monroe}}]{2021arXiv210605881N}%
  \BibitemOpen
  \bibfield  {author} {\bibinfo {author} {\bibfnamefont {Crystal}\ \bibnamefont
  {Noel}}, \bibinfo {author} {\bibfnamefont {Pradeep}\ \bibnamefont {Niroula}},
  \bibinfo {author} {\bibfnamefont {Daiwei}\ \bibnamefont {Zhu}}, \bibinfo
  {author} {\bibfnamefont {Andrew}\ \bibnamefont {Risinger}}, \bibinfo {author}
  {\bibfnamefont {Laird}\ \bibnamefont {Egan}}, \bibinfo {author}
  {\bibfnamefont {Debopriyo}\ \bibnamefont {Biswas}}, \bibinfo {author}
  {\bibfnamefont {Marko}\ \bibnamefont {Cetina}}, \bibinfo {author}
  {\bibfnamefont {Alexey~V.}\ \bibnamefont {Gorshkov}}, \bibinfo {author}
  {\bibfnamefont {Michael~J.}\ \bibnamefont {Gullans}}, \bibinfo {author}
  {\bibfnamefont {David~A.}\ \bibnamefont {Huse}}, \ and\ \bibinfo {author}
  {\bibfnamefont {Christopher}\ \bibnamefont {Monroe}},\ }\bibfield  {title}
  {\enquote {\bibinfo {title} {{Measurement-induced quantum phases realized in
  a trapped-ion quantum computer}},}\ }\href {\doibase
  10.1038/s41567-022-01619-7} {\bibfield  {journal} {\bibinfo  {journal}
  {Nature Physics}\ ,\ \bibinfo {pages} {760--764}} (\bibinfo {year} {2022})},\
  \Eprint {http://arxiv.org/abs/2106.05881} {arXiv:2106.05881 [quant-ph]}
  \BibitemShut {NoStop}%
\bibitem [{\citenamefont {Friedman}\ \emph {et~al.}(2019)\citenamefont
  {Friedman}, \citenamefont {Chan}, \citenamefont {{De Luca}},\ and\
  \citenamefont {Chalker}}]{Friedman2019}%
  \BibitemOpen
  \bibfield  {author} {\bibinfo {author} {\bibfnamefont {Aaron~J.}\
  \bibnamefont {Friedman}}, \bibinfo {author} {\bibfnamefont {Amos}\
  \bibnamefont {Chan}}, \bibinfo {author} {\bibfnamefont {Andrea}\ \bibnamefont
  {{De Luca}}}, \ and\ \bibinfo {author} {\bibfnamefont {J.~T.}\ \bibnamefont
  {Chalker}},\ }\bibfield  {title} {\enquote {\bibinfo {title} {{Spectral
  Statistics and Many-Body Quantum Chaos with Conserved Charge}},}\ }\href
  {\doibase 10.1103/PhysRevLett.123.210603} {\bibfield  {journal} {\bibinfo
  {journal} {Physical Review Letters}\ }\textbf {\bibinfo {volume} {123}},\
  \bibinfo {pages} {210603} (\bibinfo {year} {2019})},\ \Eprint
  {http://arxiv.org/abs/1906.07736} {arXiv:1906.07736} \BibitemShut {NoStop}%
\bibitem [{\citenamefont {Rakovszky}\ \emph {et~al.}(2019)\citenamefont
  {Rakovszky}, \citenamefont {Pollmann},\ and\ \citenamefont {{Von
  Keyserlingk}}}]{Rakovszky2019}%
  \BibitemOpen
  \bibfield  {author} {\bibinfo {author} {\bibfnamefont {Tibor}\ \bibnamefont
  {Rakovszky}}, \bibinfo {author} {\bibfnamefont {Frank}\ \bibnamefont
  {Pollmann}}, \ and\ \bibinfo {author} {\bibfnamefont {C.~W.}\ \bibnamefont
  {{Von Keyserlingk}}},\ }\bibfield  {title} {\enquote {\bibinfo {title}
  {{Sub-ballistic Growth of R{\'{e}}nyi Entropies due to Diffusion}},}\ }\href
  {\doibase 10.1103/PhysRevLett.122.250602} {\bibfield  {journal} {\bibinfo
  {journal} {Physical Review Letters}\ }\textbf {\bibinfo {volume} {122}},\
  \bibinfo {pages} {250602} (\bibinfo {year} {2019})},\ \Eprint
  {http://arxiv.org/abs/1901.10502} {arXiv:1901.10502} \BibitemShut {NoStop}%
\bibitem [{\citenamefont {Huang}(2020{\natexlab{a}})}]{Huang2019}%
  \BibitemOpen
  \bibfield  {author} {\bibinfo {author} {\bibfnamefont {Yichen}\ \bibnamefont
  {Huang}},\ }\bibfield  {title} {\enquote {\bibinfo {title} {{Dynamics of
  Renyi entanglement entropy in local quantum circuits with charge
  conservation}},}\ }\href {http://arxiv.org/abs/1902.00977} {\bibfield
  {journal} {\bibinfo  {journal} {IOPSciNotes}\ }\textbf {\bibinfo {volume}
  {1}},\ \bibinfo {pages} {035205} (\bibinfo {year} {2020}{\natexlab{a}})},\
  \Eprint {http://arxiv.org/abs/1902.00977} {arXiv:1902.00977} \BibitemShut
  {NoStop}%
\bibitem [{\citenamefont {Zhou}\ and\ \citenamefont {Ludwig}(2020)}]{Zhou2020}%
  \BibitemOpen
  \bibfield  {author} {\bibinfo {author} {\bibfnamefont {Tianci}\ \bibnamefont
  {Zhou}}\ and\ \bibinfo {author} {\bibfnamefont {Andreas~W.W.}\ \bibnamefont
  {Ludwig}},\ }\bibfield  {title} {\enquote {\bibinfo {title} {{Diffusive
  scaling of R{\'{e}}nyi entanglement entropy}},}\ }\href {\doibase
  10.1103/PhysRevResearch.2.033020} {\bibfield  {journal} {\bibinfo  {journal}
  {Physical Review Research}\ }\textbf {\bibinfo {volume} {2}},\ \bibinfo
  {pages} {33020} (\bibinfo {year} {2020})},\ \Eprint
  {http://arxiv.org/abs/1911.12384} {arXiv:1911.12384} \BibitemShut {NoStop}%
\bibitem [{\citenamefont {Rakovszky}\ \emph {et~al.}(2020)\citenamefont
  {Rakovszky}, \citenamefont {Pollmann},\ and\ \citenamefont {von
  Keyserlingk}}]{Rakovszky2020}%
  \BibitemOpen
  \bibfield  {author} {\bibinfo {author} {\bibfnamefont {Tibor}\ \bibnamefont
  {Rakovszky}}, \bibinfo {author} {\bibfnamefont {Frank}\ \bibnamefont
  {Pollmann}}, \ and\ \bibinfo {author} {\bibfnamefont {C~W}\ \bibnamefont {von
  Keyserlingk}},\ }\bibfield  {title} {\enquote {\bibinfo {title} {{Comment on
  "Entanglement growth in diffusive systems"}},}\ }\href
  {http://arxiv.org/abs/2010.07969} {\  (\bibinfo {year} {2020})},\ \Eprint
  {http://arxiv.org/abs/2010.07969} {arXiv:2010.07969} \BibitemShut {NoStop}%
\bibitem [{\citenamefont {{\v{Z}}nidari{\v{c}}}(2020)}]{Znidaric2020}%
  \BibitemOpen
  \bibfield  {author} {\bibinfo {author} {\bibfnamefont {Marko}\ \bibnamefont
  {{\v{Z}}nidari{\v{c}}}},\ }\bibfield  {title} {\enquote {\bibinfo {title}
  {{Entanglement growth in diffusive systems}},}\ }\href {\doibase
  10.1038/s42005-020-0366-7} {\bibfield  {journal} {\bibinfo  {journal}
  {Communications Physics}\ }\textbf {\bibinfo {volume} {3}},\ \bibinfo {pages}
  {1--9} (\bibinfo {year} {2020})},\ \Eprint {http://arxiv.org/abs/1912.03645}
  {arXiv:1912.03645} \BibitemShut {NoStop}%
\bibitem [{\citenamefont {Barratt}\ \emph {et~al.}(2021)\citenamefont
  {Barratt}, \citenamefont {Agrawal}, \citenamefont {Gopalakrishnan},
  \citenamefont {Huse}, \citenamefont {Vasseur},\ and\ \citenamefont
  {Potter}}]{FieldTheorySharpening}%
  \BibitemOpen
  \bibfield  {author} {\bibinfo {author} {\bibfnamefont {Fergus}\ \bibnamefont
  {Barratt}}, \bibinfo {author} {\bibfnamefont {Utkarsh}\ \bibnamefont
  {Agrawal}}, \bibinfo {author} {\bibfnamefont {Sarang}\ \bibnamefont
  {Gopalakrishnan}}, \bibinfo {author} {\bibfnamefont {David~A.}\ \bibnamefont
  {Huse}}, \bibinfo {author} {\bibfnamefont {Romain}\ \bibnamefont {Vasseur}},
  \ and\ \bibinfo {author} {\bibfnamefont {Andrew~C.}\ \bibnamefont {Potter}},\
  }\bibfield  {title} {\enquote {\bibinfo {title} {{Field theory of charge
  sharpening in symmetric monitored quantum circuits}},}\ }\href
  {http://arxiv.org/abs/2111.09336} {\bibfield  {journal} {\bibinfo  {journal}
  {arXiv e-prints}\ ,\ \bibinfo {pages} {arXiv:2111.09336}} (\bibinfo {year}
  {2021})},\ \Eprint {http://arxiv.org/abs/2111.09336} {arXiv:2111.09336}
  \BibitemShut {NoStop}%
\bibitem [{Note1()}]{Note1}%
  \BibitemOpen
  \bibinfo {note} {Since the unitaries acting on the qudits are random, the
  randomizing measurement basis is superfluous.}\BibitemShut {Stop}%
\bibitem [{\citenamefont {Blais}\ \emph {et~al.}(2021)\citenamefont {Blais},
  \citenamefont {Grimsmo}, \citenamefont {Girvin},\ and\ \citenamefont
  {Wallraff}}]{blais2021circuit}%
  \BibitemOpen
  \bibfield  {author} {\bibinfo {author} {\bibfnamefont {Alexandre}\
  \bibnamefont {Blais}}, \bibinfo {author} {\bibfnamefont {Arne~L.}\
  \bibnamefont {Grimsmo}}, \bibinfo {author} {\bibfnamefont {S.~M.}\
  \bibnamefont {Girvin}}, \ and\ \bibinfo {author} {\bibfnamefont {Andreas}\
  \bibnamefont {Wallraff}},\ }\bibfield  {title} {\enquote {\bibinfo {title}
  {{Circuit quantum electrodynamics}},}\ }\href {\doibase
  10.1103/RevModPhys.93.025005} {\bibfield  {journal} {\bibinfo  {journal}
  {Reviews of Modern Physics}\ }\textbf {\bibinfo {volume} {93}},\ \bibinfo
  {pages} {25005} (\bibinfo {year} {2021})},\ \Eprint
  {http://arxiv.org/abs/2005.12667} {arXiv:2005.12667} \BibitemShut {NoStop}%
\bibitem [{Note2()}]{Note2}%
  \BibitemOpen
  \bibinfo {note} {The ensemble-averaged $\rho (t)$ resulting from
  maximally-random, local, open-systems dynamics is indistinguishable from an
  infinite temperature state over distances $\sim t$.}\BibitemShut {Stop}%
\bibitem [{Note3()}]{Note3}%
  \BibitemOpen
  \bibinfo {note} {This can seen by the central limit theorem as follows:
  assuming each measurement outcome to be independent, the statistical error in
  the outcomes of the measurement of charge density goes as $1/\protect \sqrt
  {\protect \mathcal {N}_M}$. To distinguish the states with global charge $N$
  and $N-1$, we require this error to become smaller than $\sim 1/L$. This
  gives $\protect \mathcal {N}_M \sim L^2$.}\BibitemShut {Stop}%
\bibitem [{\citenamefont {Huang}(2020{\natexlab{b}})}]{Huang2019a}%
  \BibitemOpen
  \bibfield  {author} {\bibinfo {author} {\bibfnamefont {Yichen}\ \bibnamefont
  {Huang}},\ }\bibfield  {title} {\enquote {\bibinfo {title} {{Dynamics of
  R{\'{e}}nyi entanglement entropy in diffusive qudit systems}},}\ }\href
  {\doibase 10.1088/2633-1357/abd1e2} {\bibfield  {journal} {\bibinfo
  {journal} {IOP SciNotes}\ }\textbf {\bibinfo {volume} {1}},\ \bibinfo {pages}
  {035205} (\bibinfo {year} {2020}{\natexlab{b}})},\ \Eprint
  {http://arxiv.org/abs/2008.00944} {arXiv:2008.00944} \BibitemShut {NoStop}%
\bibitem [{Note4()}]{Note4}%
  \BibitemOpen
  \bibinfo {note} {Though other quantities such as $\protect \qopname \relax
  o{log}[e^{-S_n}]$ are dominated by rare dead-region contributions and do
  exhibit $\protect \sqrt {t}$ growth due to rare dilute measurement locations.
  The parametrically strong discrepancy between the average purity and the
  average R\'enyi entropies is also seen numerically in our statistical model
  approach (Sec.~\ref {statmech_numerics}).}\BibitemShut {Stop}%
\bibitem [{Note5()}]{Note5}%
  \BibitemOpen
  \bibinfo {note} {To obtain a uniform distribution over $U(2)$ we must pick
  $\alpha ,\psi ,\chi \in \left [0,2\pi \right )$ and $\xi \in \left [0,1\right
  ]$ uniformly and compute $\phi = \protect \qopname \relax o{arcsin}\protect
  \sqrt {\xi }$.}\BibitemShut {Stop}%
\bibitem [{Note6()}]{Note6}%
  \BibitemOpen
  \bibinfo {note} {In the absence of a conservation law, it has been shown that
  the probes we have used for the entanglement transition ($\protect \mathcal
  {I}_{3,n}$ and $[S_{1,E}]$) have weak finite size drifts in Clifford
  circuits~\cite {Zabalo2020} by examining small and large system
  sizes.}\BibitemShut {Stop}%
\bibitem [{\citenamefont {Hayden}\ \emph {et~al.}(2016)\citenamefont {Hayden},
  \citenamefont {Nezami}, \citenamefont {Qi}, \citenamefont {Thomas},
  \citenamefont {Walter},\ and\ \citenamefont {Yang}}]{RTN}%
  \BibitemOpen
  \bibfield  {author} {\bibinfo {author} {\bibfnamefont {Patrick}\ \bibnamefont
  {Hayden}}, \bibinfo {author} {\bibfnamefont {Sepehr}\ \bibnamefont {Nezami}},
  \bibinfo {author} {\bibfnamefont {Xiao~Liang}\ \bibnamefont {Qi}}, \bibinfo
  {author} {\bibfnamefont {Nathaniel}\ \bibnamefont {Thomas}}, \bibinfo
  {author} {\bibfnamefont {Michael}\ \bibnamefont {Walter}}, \ and\ \bibinfo
  {author} {\bibfnamefont {Zhao}\ \bibnamefont {Yang}},\ }\bibfield  {title}
  {\enquote {\bibinfo {title} {{Holographic duality from random tensor
  networks}},}\ }\href {\doibase 10.1007/JHEP11(2016)009} {\bibfield  {journal}
  {\bibinfo  {journal} {Journal of High Energy Physics}\ }\textbf {\bibinfo
  {volume} {2016}},\ \bibinfo {pages} {9} (\bibinfo {year} {2016})},\ \Eprint
  {http://arxiv.org/abs/1601.01694} {arXiv:1601.01694} \BibitemShut {NoStop}%
\bibitem [{\citenamefont {Vasseur}\ \emph {et~al.}(2019)\citenamefont
  {Vasseur}, \citenamefont {Potter}, \citenamefont {You},\ and\ \citenamefont
  {Ludwig}}]{PhysRevB.100.134203}%
  \BibitemOpen
  \bibfield  {author} {\bibinfo {author} {\bibfnamefont {Romain}\ \bibnamefont
  {Vasseur}}, \bibinfo {author} {\bibfnamefont {Andrew~C.}\ \bibnamefont
  {Potter}}, \bibinfo {author} {\bibfnamefont {Yi~Zhuang}\ \bibnamefont {You}},
  \ and\ \bibinfo {author} {\bibfnamefont {Andreas~W.W.}\ \bibnamefont
  {Ludwig}},\ }\bibfield  {title} {\enquote {\bibinfo {title} {{Entanglement
  transitions from holographic random tensor networks}},}\ }\href {\doibase
  10.1103/PhysRevB.100.134203} {\bibfield  {journal} {\bibinfo  {journal}
  {Physical Review B}\ }\textbf {\bibinfo {volume} {100}},\ \bibinfo {pages}
  {134203} (\bibinfo {year} {2019})},\ \Eprint
  {http://arxiv.org/abs/1807.07082} {arXiv:1807.07082} \BibitemShut {NoStop}%
\bibitem [{Note7()}]{Note7}%
  \BibitemOpen
  \bibinfo {note} {DWs are restricted by unitarity to only make certain
  ``turns'' (See \cite {Zhou2019} for details). E.g, for $p=0$, this leads to a
  unique DW where the DW follows a ``light cone''. For $p>0$ one can still have
  many degenerate paths ~\cite {Li2020b}.}\BibitemShut {Stop}%
\bibitem [{Note8()}]{Note8}%
  \BibitemOpen
  \bibinfo {note} {Any difference in the vertex elements will lead to the
  creation of DW which are suppressed as ${\protect \cal O}(1/d)$ and whose
  contribution goes to zero as $d\rightarrow \infty $.}\BibitemShut {Stop}%
\bibitem [{Note9()}]{Note9}%
  \BibitemOpen
  \bibinfo {note} {Note that the DW permutation element $(1...n)^{\otimes k}$
  has $k+1$ cycles and each cycle can follow an independent path (to the
  leading order)~\cite {Zhou2019}. However, this subtlety will not change the
  final result about the minimal cut, but only leads to fluctuations
  contributing sub-leading logarithmic corrections (for $p>0$).}\BibitemShut
  {Stop}%
\bibitem [{Note10()}]{Note10}%
  \BibitemOpen
  \bibinfo {note} {Note that this description of the $d \to \infty $ differs
  from that of Ref.~\cite {Jian2020}. There the measurement locations $\DOTSB
  \sum@ \slimits@ _{\protect \mathbf {X}}$ were averaged over directly in the
  partition function, in an annealed way, while we chose here to keep the
  measurement locations as quenched disorder. Our approach predicts a minimal
  cut picture consistent with Ref.~\cite {Skinner2019}. We leave a discussion
  of the validity of the replica trick in this limit to future
  work.}\BibitemShut {Stop}%
\bibitem [{\citenamefont {Schollw{\"{o}}ck}(2011)}]{SCHOLLWOCK201196}%
  \BibitemOpen
  \bibfield  {author} {\bibinfo {author} {\bibfnamefont {Ulrich}\ \bibnamefont
  {Schollw{\"{o}}ck}},\ }\bibfield  {title} {\enquote {\bibinfo {title} {{The
  density-matrix renormalization group in the age of matrix product states}},}\
  }\href {\doibase 10.1016/j.aop.2010.09.012} {\bibfield  {journal} {\bibinfo
  {journal} {Annals of Physics}\ }\textbf {\bibinfo {volume} {326}},\ \bibinfo
  {pages} {96--192} (\bibinfo {year} {2011})},\ \Eprint
  {http://arxiv.org/abs/1008.3477} {arXiv:1008.3477} \BibitemShut {NoStop}%
\bibitem [{Note11()}]{Note11}%
  \BibitemOpen
  \bibinfo {note} {Frozen in the sense that there is not much quantum
  superposition of different charge configurations.}\BibitemShut {Stop}%
\bibitem [{\citenamefont {Weinstein}\ \emph {et~al.}(2022)\citenamefont
  {Weinstein}, \citenamefont {Bao},\ and\ \citenamefont
  {Altman}}]{Weinstein2022}%
  \BibitemOpen
  \bibfield  {author} {\bibinfo {author} {\bibfnamefont {Zack}\ \bibnamefont
  {Weinstein}}, \bibinfo {author} {\bibfnamefont {Yimu}\ \bibnamefont {Bao}}, \
  and\ \bibinfo {author} {\bibfnamefont {Ehud}\ \bibnamefont {Altman}},\ }\href
  {\doibase 10.48550/ARXIV.2202.12905} {\enquote {\bibinfo {title}
  {{Measurement-induced power law negativity in an open monitored quantum
  circuit}},}\ } (\bibinfo {year} {2022})\BibitemShut {NoStop}%
\bibitem [{\citenamefont {Fisher}(2004)}]{fisher2004duality}%
  \BibitemOpen
  \bibfield  {author} {\bibinfo {author} {\bibfnamefont {Matthew}\ \bibnamefont
  {Fisher}},\ }\bibfield  {title} {\enquote {\bibinfo {title} {{Duality in low
  dimensional quantum field theories}},}\ }in\ \href@noop {} {\emph {\bibinfo
  {booktitle} {Strong interactions in low dimensions}}}\ (\bibinfo  {publisher}
  {Springer},\ \bibinfo {year} {2004})\ pp.\ \bibinfo {pages}
  {419--438}\BibitemShut {NoStop}%
\bibitem [{\citenamefont {Potter}\ and\ \citenamefont
  {Vasseur}(2016)}]{PhysRevB.94.224206}%
  \BibitemOpen
  \bibfield  {author} {\bibinfo {author} {\bibfnamefont {Andrew~C.}\
  \bibnamefont {Potter}}\ and\ \bibinfo {author} {\bibfnamefont {Romain}\
  \bibnamefont {Vasseur}},\ }\bibfield  {title} {\enquote {\bibinfo {title}
  {{Symmetry constraints on many-body localization}},}\ }\href {\doibase
  10.1103/PhysRevB.94.224206} {\bibfield  {journal} {\bibinfo  {journal}
  {Physical Review B}\ }\textbf {\bibinfo {volume} {94}},\ \bibinfo {pages}
  {224206} (\bibinfo {year} {2016})},\ \Eprint
  {http://arxiv.org/abs/1605.03601} {arXiv:1605.03601} \BibitemShut {NoStop}%
\bibitem [{\citenamefont {Chen}\ \emph {et~al.}(2013)\citenamefont {Chen},
  \citenamefont {Gu}, \citenamefont {Liu},\ and\ \citenamefont
  {Wen}}]{PhysRevB.87.155114}%
  \BibitemOpen
  \bibfield  {author} {\bibinfo {author} {\bibfnamefont {Xie}\ \bibnamefont
  {Chen}}, \bibinfo {author} {\bibfnamefont {Zheng~Cheng}\ \bibnamefont {Gu}},
  \bibinfo {author} {\bibfnamefont {Zheng~Xin}\ \bibnamefont {Liu}}, \ and\
  \bibinfo {author} {\bibfnamefont {Xiao~Gang}\ \bibnamefont {Wen}},\
  }\bibfield  {title} {\enquote {\bibinfo {title} {{Symmetry protected
  topological orders and the group cohomology of their symmetry group}},}\
  }\href {\doibase 10.1103/PhysRevB.87.155114} {\bibfield  {journal} {\bibinfo
  {journal} {Physical Review B - Condensed Matter and Materials Physics}\
  }\textbf {\bibinfo {volume} {87}},\ \bibinfo {pages} {155114} (\bibinfo
  {year} {2013})},\ \Eprint {http://arxiv.org/abs/1106.4772} {arXiv:1106.4772}
  \BibitemShut {NoStop}%
\bibitem [{\citenamefont {Chen}\ \emph {et~al.}(2012)\citenamefont {Chen},
  \citenamefont {Gu}, \citenamefont {Liu},\ and\ \citenamefont
  {Wen}}]{Chen1604}%
  \BibitemOpen
  \bibfield  {author} {\bibinfo {author} {\bibfnamefont {Xie}\ \bibnamefont
  {Chen}}, \bibinfo {author} {\bibfnamefont {Zheng~Cheng}\ \bibnamefont {Gu}},
  \bibinfo {author} {\bibfnamefont {Zheng~Xin}\ \bibnamefont {Liu}}, \ and\
  \bibinfo {author} {\bibfnamefont {Xiao~Gang}\ \bibnamefont {Wen}},\
  }\bibfield  {title} {\enquote {\bibinfo {title} {{Symmetry-protected
  topological orders in interacting bosonic systems}},}\ }\href {\doibase
  10.1126/science.1227224} {\bibfield  {journal} {\bibinfo  {journal}
  {Science}\ }\textbf {\bibinfo {volume} {338}},\ \bibinfo {pages} {1604--1606}
  (\bibinfo {year} {2012})},\ \Eprint {http://arxiv.org/abs/1301.0861}
  {arXiv:1301.0861} \BibitemShut {NoStop}%
\bibitem [{\citenamefont {Chen}\ \emph {et~al.}(2014)\citenamefont {Chen},
  \citenamefont {Lu},\ and\ \citenamefont {Vishwanath}}]{DecoratedDW}%
  \BibitemOpen
  \bibfield  {author} {\bibinfo {author} {\bibfnamefont {Xie}\ \bibnamefont
  {Chen}}, \bibinfo {author} {\bibfnamefont {Yuan~Ming}\ \bibnamefont {Lu}}, \
  and\ \bibinfo {author} {\bibfnamefont {Ashvin}\ \bibnamefont {Vishwanath}},\
  }\bibfield  {title} {\enquote {\bibinfo {title} {{Symmetry-protected
  topological phases from decorated domain walls}},}\ }\href {\doibase
  10.1038/ncomms4507} {\bibfield  {journal} {\bibinfo  {journal} {Nature
  Communications}\ }\textbf {\bibinfo {volume} {5}},\ \bibinfo {pages} {3507}
  (\bibinfo {year} {2014})},\ \Eprint {http://arxiv.org/abs/1303.4301}
  {arXiv:1303.4301} \BibitemShut {NoStop}%
\bibitem [{\citenamefont {Deng}\ and\ \citenamefont
  {Bl{\"{o}}te}(2003)}]{PhysRevE.68.036125}%
  \BibitemOpen
  \bibfield  {author} {\bibinfo {author} {\bibfnamefont {Youjin}\ \bibnamefont
  {Deng}}\ and\ \bibinfo {author} {\bibfnamefont {Henk~W.J.}\ \bibnamefont
  {Bl{\"{o}}te}},\ }\bibfield  {title} {\enquote {\bibinfo {title}
  {{Simultaneous analysis of several models in the three-dimensional Ising
  universality class}},}\ }\href {\doibase 10.1103/PhysRevE.68.036125}
  {\bibfield  {journal} {\bibinfo  {journal} {Physical Review E - Statistical
  Physics, Plasmas, Fluids, and Related Interdisciplinary Topics}\ }\textbf
  {\bibinfo {volume} {68}},\ \bibinfo {pages} {9} (\bibinfo {year}
  {2003})}\BibitemShut {NoStop}%
\end{thebibliography}%

\end{document}